\documentclass[twocolumn,appendixfloats,numberedappendix]{aastex631}

\usepackage{apjfonts}
\usepackage{times}
\usepackage{CJK}
\usepackage{amsmath}
\usepackage{hyperref}
\usepackage{cleveref}
\usepackage{natbib}
\usepackage{amssymb}
\usepackage{rotating}
\usepackage{lipsum} 
\usepackage{academicons}
\usepackage{graphicx}
\usepackage{booktabs}

\usepackage{color}

\accepted{for publication in the Astrophysical Journal}
\shorttitle{}
\shortauthors{Le \& Xue}

\newcommand{\Hb}{H{$\beta$}}
\newcommand{\Ha}{H{$\alpha$}}

\newcommand{\OIII}{[\ion{O}{3}]}

\newcommand{\NII}{[\ion{N}{2}]}

\def\gsim{\mathrel{\rlap{\lower4pt\hbox{\hskip1pt$\sim$}}
    \raise1pt\hbox{$>$}}}         

\def\lsim{\mathrel{\rlap{\lower4pt\hbox{\hskip1pt$\sim$}}
    \raise1pt\hbox{$<$}}}         

\newcommand{\ustc}{\affil{CAS Key Laboratory for Research in Galaxies and Cosmology, Department of Astronomy, University of Science and Technology of China, Hefei 230026, China; \href{mailto: lha@ustc.edu.cn}{lha@ustc.edu.cn}; \href{mailto: xuey@ustc.edu.cn}{xuey@ustc.edu.cn}}}
\newcommand{\sustc}{\affil{School of Astronomy and Space Science, University of Science and Technology of China, Hefei 230026, China}}

\begin{document}

\begin{CJK*}{UTF8}{gbsn}

\title{Nuclear and Star Formation Activities in Nearby Galaxies: Roles of Gas Supply and AGN Feedback}

\author[0000-0003-1270-9802]{Huynh Anh N. Le} \ustc \sustc
\author[0000-0002-1935-8104]{Yongquan Xue} \ustc \sustc


\begin{abstract} 

We analyzed a sample of $\sim$113,000 galaxies ($\rm z < 0.3$) from the Sloan Digital Sky Survey, divided into star-forming, composite, Seyfert, and LINER types, to explore the relationships between UV-to-optical colors ($\rm u-r$), star formation rates (SFRs), specific star formation rates (sSFRs), stellar velocity dispersions ($\rm \sigma_{*}$), mass accretion rates onto the black hole ($\rm L_{[OIII]}/\sigma_{*}^{4}$), and Eddington ratios. Star-forming galaxies predominantly feature young, blue stars along the main-sequence (MS) line, while composite, Seyfert, and LINER galaxies deviate from this line, displaying progressively older stellar populations and lower SFRs. $\rm L_{[OIII]}/\sigma_{*}^{4}$ and Eddington ratios are highest in Seyfert galaxies, moderate in composite galaxies, and lowest in LINERs, with higher ratios associated with bluer colors, indicating a younger stellar population and stronger active galactic nucleus (AGN) activity. These trends suggest a strong correlation between sSFRs and Eddington ratios, highlighting a close connection between AGN and star formation activities. These results may imply an evolutionary sequence where galaxies transition from blue star-forming galaxies to red LINERs, passing through composite and Seyfert phases, driven primarily by gas supply, with AGN feedback playing a secondary role. Radio excess shows low values on the MS line and higher values below the line, potentially indicating AGN and/or jet contributions suppressing SFRs. While both radio luminosities ($\rm L_{1.4GHz}$) and Eddington ratios correlate with SFRs, their trends differ on the SFR$-$stellar mass ($\rm M_{*}$) plane, with radio luminosities increasing with stellar mass along the MS line, and no direct connection between radio luminosities and Eddington ratios. These findings may provide new insights into the interplay between star formation, AGN activity, and radio emission in galaxies, shedding light on their evolutionary pathways.

\end{abstract}


\section{INTRODUCTION}\label{section:intro}

Over the past two decades, extensive research has delved into the relationship between black holes and the properties of their host galaxies, shedding light on the coevolution of supermassive black holes and their galactic environments \citep[e.g.,][]{Ferrarese+00, Gebhardt+00, McLure+02, Tremaine+02, Woo+06, Xue+10, Kormendy+13, Le+14, Le+20}. The relationship between black holes and their host galaxies, characterized by a remarkably low intrinsic scatter of $\sim$0.3 dex between the black hole and bugle mass (e.g., \citealp{Haring+04}), suggests a strong, mutual correlation between black holes and their host galaxies. Despite the breadth of these studies, the mechanisms that link active galactic nucleus (AGN) activity to star formation (SF) in host galaxies remain largely unclear. Theoretical models often propose that AGN activity plays a pivotal role in this relationship, both triggering and regulating star formation through AGN feedback \citep[e.g.,][]{Croton+06, Woo+16, Woo+20}. Nevertheless, critical questions persist: What specifically drives the connection between AGN activity and SF? How does AGN activity affect the large-scale interstellar medium (ISM) within galaxies? What impacts do AGN energy outputs, including ionized gas outflows and luminosities, have on SF? Does nuclear activity, such as gas outflows, suppress SF by depleting the gas supply, or does it enhance SF by compressing the gas? These unanswered questions continue to spark intense discussions in astrophysics \citep[e.g.,][]{Alexander+12, Harrison+24}.

Many observational studies and theoretical models have aimed to fully understand the link between AGN activity and the SF process in their host galaxies \citep[e.g.,][]{Silk+98, Croton+06, Fabian12, Le+17a, Le+19, Le+23, Le+24}. Two main opposing scenarios have been proposed to explain the observational results. Firstly, AGN feedback can suppress SF. In this scenario, SF activity is regulated by black hole accretion. The ISM surrounding the black hole will be heated or cleared by the winds or jets launched from the accretion disk of the black hole. The lack of cold gas in the host galaxy will lead to the quenching of SF activity (negative feedback, e.g., \citealp{Page+12}, \citealp{Shin+19}). Secondly, contrary to the first scenario, AGN feedback can compress the gas in the ISM, enhancing SF activity (positive feedback, e.g., \citealp{Woo+20}, \citealp{Zhuang+20}). Additionally, some studies have reported the observational signatures of both negative and positive feedback in individual galaxies \citep{Shin+19}. From the diverse observational results, it is clear that the nature of AGN feedback and the interaction between AGN output and the nucleus activity with the ISM surrounding is a much more complex process that cannot be explained by simple scenarios alone. Many more observations need to be conducted to gain further insights into the physical nature of the connection between AGN and SF activities. Positive and negative feedback on the surrounding gas may occur in individual AGNs. Therefore, understanding the global impact of multiple AGN events is challenging and requires comprehensive statistical studies.

Recently, \citet{Leslie+16}, utilizing a large sample from the Sloan Digital Sky Survey (SDSS) data, compared star formation rates (SFRs) as a function of stellar mass across different galaxy types, including star-forming, composite, Seyfert, and LINER galaxies. They proposed an evolutionary pathway where galaxies transition from active star-forming states to composite, Seyfert, and finally, LINER galaxies, illustrating a shift from \emph{blue to red (active to quenching)} phases. They emphasized the significant role of AGNs in quenching SF along this pathway, a conclusion consistent with earlier studies based on optical and X-ray selected samples \citep[e.g.,][]{Salim+07, Schawinski+07, Xue+10, Shimizu+15}. While the results of \citet{Leslie+16} are crucial for understanding galaxy evolution, a more detailed analysis incorporating critical physical parameters such as Eddington ratio, UV-to-optical colors ($\rm u-r$), and radio luminosity is needed to deepen our understanding of these processes.

In our earlier research \citep{Le+24b}, we analyzed a substantial sample of $\sim$113,000 galaxies ($z$ $<$ 0.3) with high signal-to-noise (S/N) ratios in both continuum and emission lines. These galaxies were the focus of a series of statistical studies on gas outflows in type 2 AGNs, specifically investigating the kinematics of the \OIII\ 5007\AA\ emission line \citep[]{Bae+14, Woo+16, Woo+17, Woo+20}. In this paper, by using this large and high-quality sample, we aim to conduct a detailed analysis of the relationships between UV-to-optical colors ($\rm u-r$), SFRs, Eddington ratios, and radio luminosities across various galaxy types. By examining these correlations in the $\rm SFR-M_{*}$ and $\rm (u-r)-M_{*}$ planes, we seek to explore the interplay between SF and AGN activities and to better understand the roles of these physical quantities in the evolutionary processes of galaxies.

We describe the sample selection in Section \ref{section:sample}. Section \ref{section:meas} provides information on the measurements of physical properties. The results are presented in Section \ref{section:result}. The discussions are shown in Section \ref{section:discuss}. Finally, the summary is presented in Section \ref{section:sum}. We adopted the \citet{Kroupa01} stellar initial mass function in this work. The following cosmological parameters are used throughout the paper: $H_0 = 70$~km~s$^{-1}$~Mpc$^{-1}$, $\Omega_{\rm m} = 0.30$, and $\Omega_{\Lambda} = 0.70$. 
  
\section{Sample Selection}\label{section:sample}

\begin{figure*}
\centering
	\includegraphics[width=0.245\textwidth]{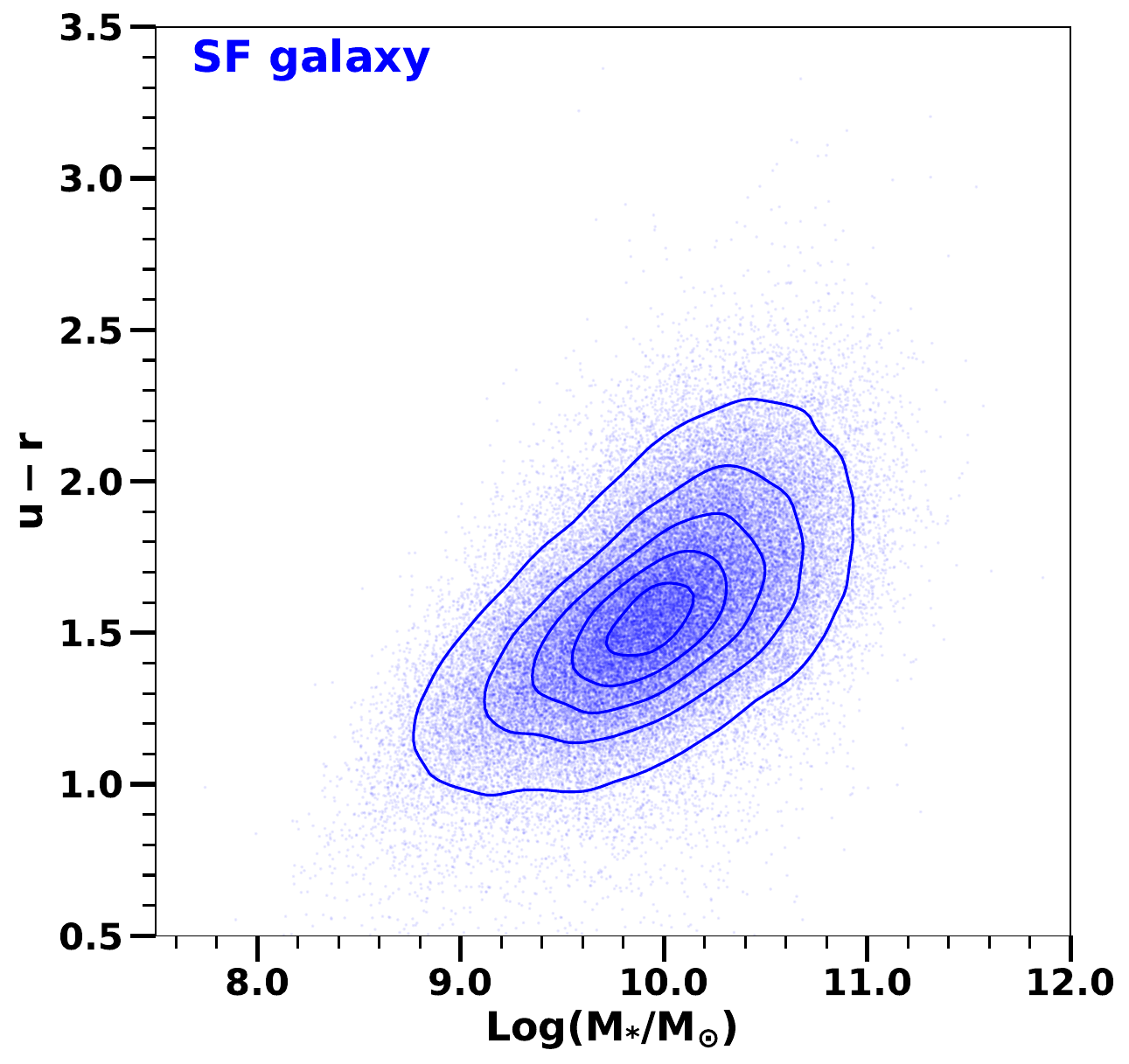}
	\includegraphics[width=0.245\textwidth]{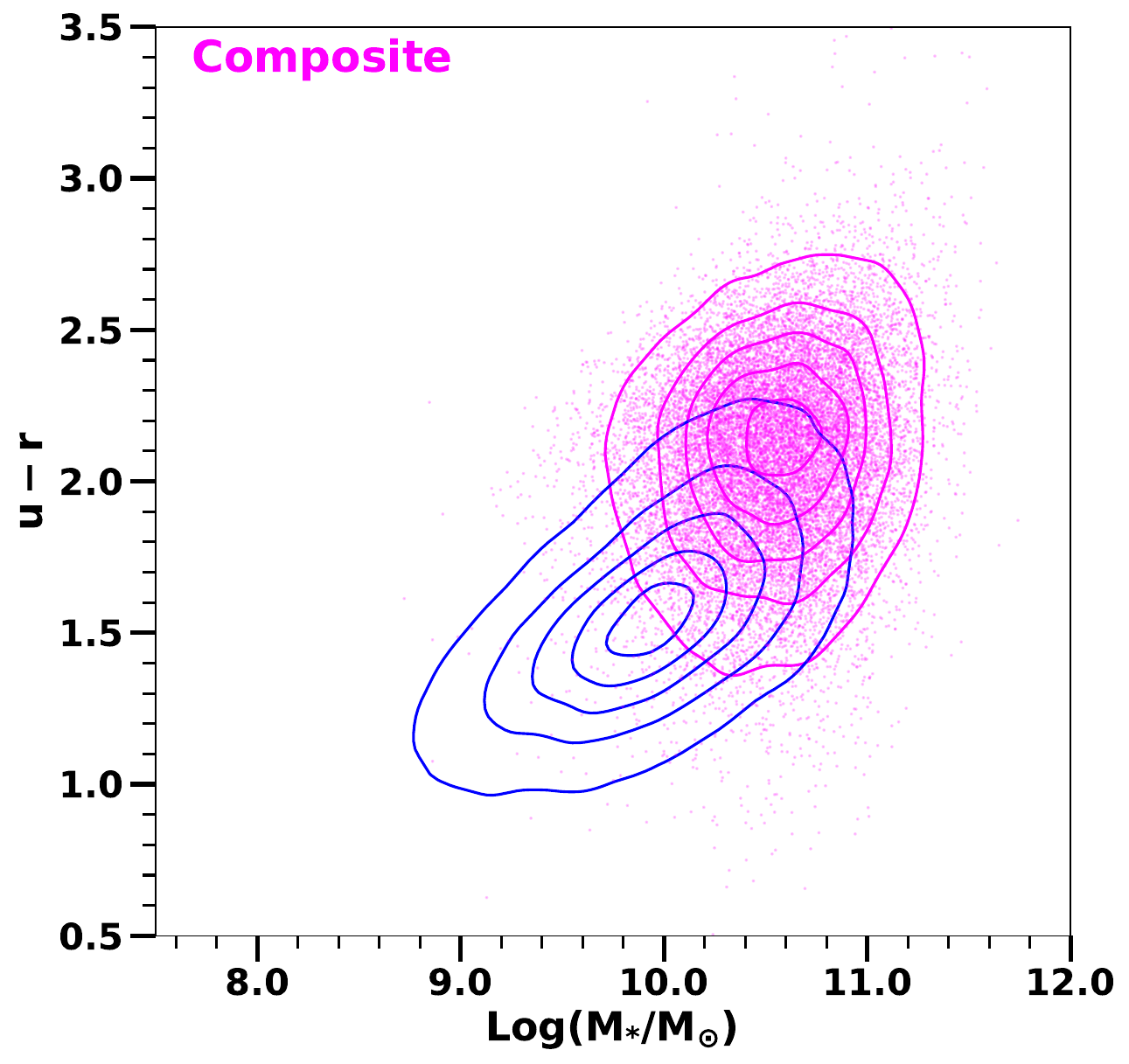}
	\includegraphics[width=0.245\textwidth]{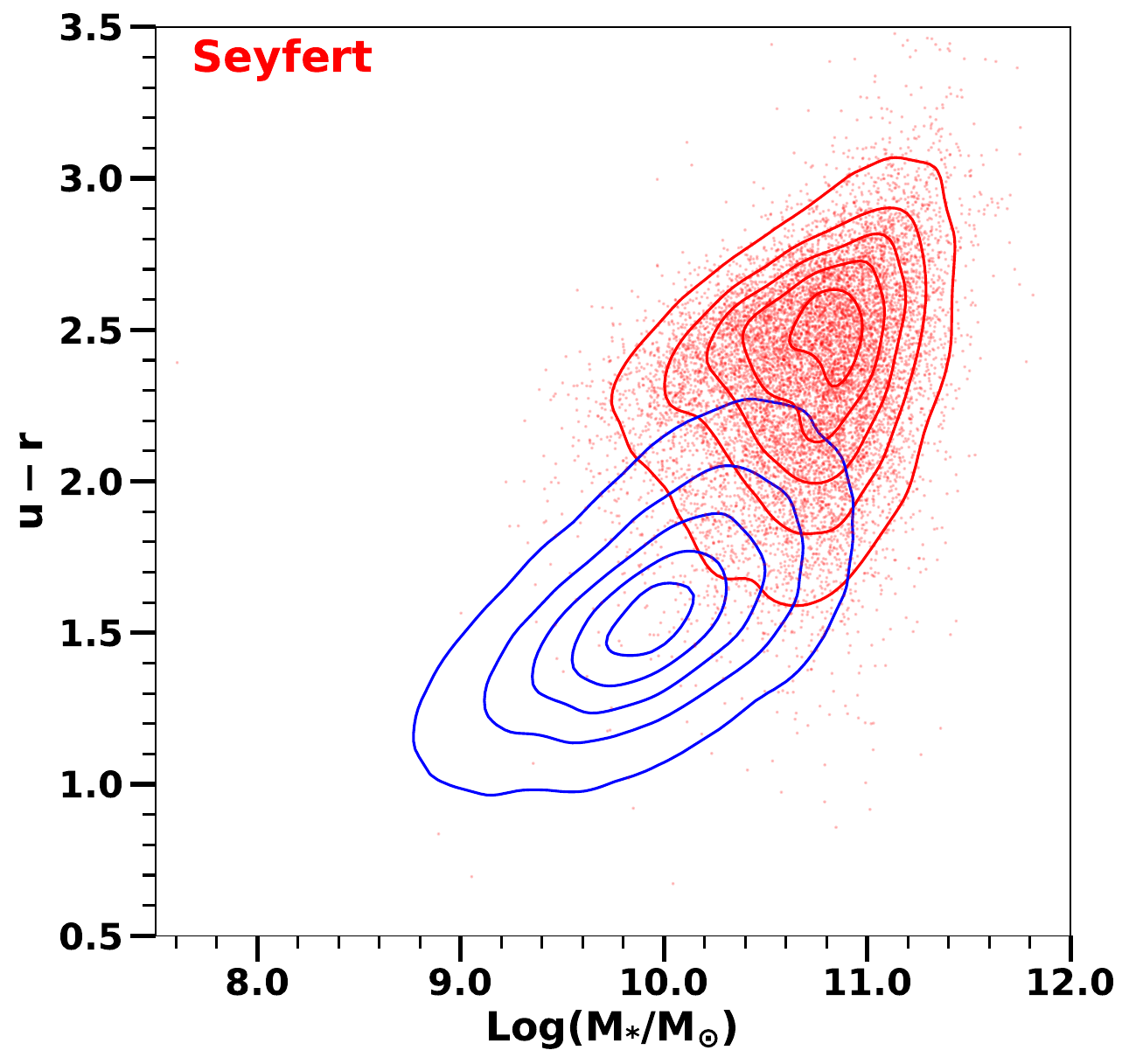}
	\includegraphics[width=0.245\textwidth]{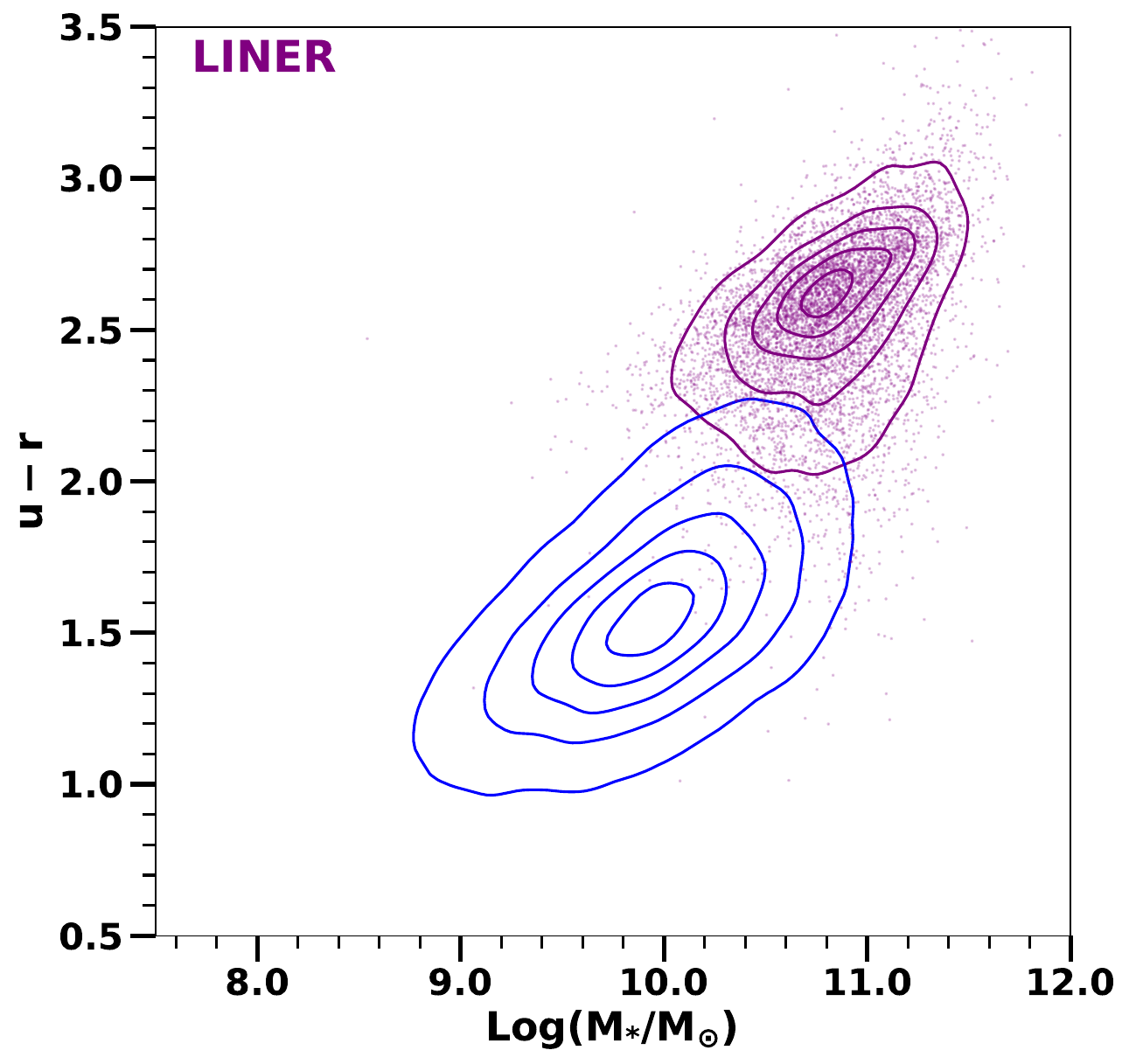}
	\caption{The Galactic extinction corrected UV-to-optical ($\rm u-r$) colors as a function of stellar mass. The $\rm (u-r)-M_{*}$ space is divided into a grid of 150x150 bins, and contour lines are drawn to represent 10, 30, 50, 70, and 90 percent of the maximum number density. Each panel displays the classifications among galaxies, such as SF (blue), composite (pink), Seyfert (red), and LINER (magenta), respectively. The blue contour lines of the SF galaxies are overplotted in other panels for a proper comparison. The stellar mass is adopted from the MPA-JHU catalog. 
		\label{fig:color_contour}}
\end{figure*}

The sample used in this paper was initially selected based on our work \citep{Le+24b}, in that we aim to compare the discrepancy of various SFR tracers in different observational bands and study the relation between SFRs and AGN activities (indicated by, e.g., Eddington ratio). In this paper, by matching our initial sample with the available photometry data of the SDSS catalog, we investigate the relations of UV-to-optical colors, SFRs, and AGN strength activities. We aim to examine how SF and Eddington ratios vary across the color-mass diagram and the SFR-stellar mass (SFR$-$$\rm M_{*}$) diagram across different galaxy types. The sample was initially selected with 235,922 emission line galaxies (z $<$ 0.3) from the Max Planck Institute for Astrophysics and Johns Hopkins University (MPA-JHU) catalog \footnote{http://www.mpa-garching.mpg.de/SDSS/} \citep{Brinchmann+04}, based on SDSS Data Release 7 \citep{Abazajian+09}. These galaxies met the criteria of having a continuum with S/N $\geq$ 10 and S/N $\geq$ 3 for the \Hb, \OIII\ 5007 \AA, \Ha, and \NII\ emission lines. We then narrowed the sample to 112,726 galaxies with well-defined emission line profiles with an amplitude peak to noise (A/N) ratio $>$ 5 for \OIII\ and \Ha. These were classified into 69,262 star-forming, 20,712 composite, 14,067 Seyfert, and 8,685 LINER galaxies using the Baldwin-Phillips-Terlevich (BPT) diagram \citep{Baldwin+81}. The classification of galaxies into star-forming, composite, Seyfert, and LINER categories is based on \citet{Kauffmann+03}, \citet{Kewley+06}, and \citet{Schawinski+07}, respectively. Figure \ref{fig:color_contour} presents the Galactic extinction corrected UV-to-optical ($\rm u-r$) colors as a function of the stellar mass of different galaxy types, such as SF, composite, Seyfert, and LINER, respectively.

\section{Measurements}\label{section:meas}


In this section, we outline the physical quantities measured and utilized in this paper. As mentioned in the sample selection, we cross-matched our sample with the available photometry data of the SDSS catalog to obtain their UV-to-optical colors ($\rm u-r$). We also adopted the stellar mass ($\rm M_{*}$) of our sample from the MPA-JHU catalog. 

We calculated the AGN bolometric luminosity, black hole mass, and Eddington ratio for each galaxy in our sample. To estimate the bolometric luminosity, we employed the extinction corrected \OIII\ luminosity as a proxy, following the relation $\rm L_{Bol} = 600 \times L_{[OIII]}$ \citep{Kauffmann09}. Dust extinction in the \OIII\ emission line was corrected using the observed Balmer decrements and the Milky Way extinction law from \citet{Cardelli+89}, adopting $\rm R_{V} = A_{V}/E(B-V) = 3.1$. Assuming electron temperatures of $\rm T_{e} = 10^4\ K$ and electron densities between $\rm n_{e} \approx 10^2$ and $\rm 10^4\ cm^{-3}$, the intrinsic \Ha/\Hb\ ratios are 2.86 for SF galaxies and 3.1 for AGNs \citep{Osterbrock+06}. The calculated bolometric luminosities for our sample range from $\rm 10^{40} - 10^{47}\ erg\ s^{-1}$. 

The estimation of AGN bolometric luminosity using the \OIII\ emission line can be biased because both star formation and AGN components can contribute to the \OIII\ line flux. Some earlier studies have calculated the contributions of SF and AGN fractions to the total \OIII\ luminosity of galaxies in the SDSS (e.g., \citealp{Kauffmann09}). Using the position of a galaxy on the BPT diagram, the AGN fraction contributing to the total \OIII\ luminosity can be estimated by comparing the position of the galaxy to the assumed \textit{pure-SF} and \textit{pure-AGN} templates on the diagram (for details, see Figure 3 in \citealp{Kauffmann09}). The assumption of \textit{pure} SF and AGN regions is based on averages derived from other galaxies, which may introduce biases in determining the \textit{true} AGN fraction. However, as argued by \citet{Kauffmann09}, this method is reasonable for statistical studies involving large samples. Therefore, we have adopted their approach in our analysis. In \citet{Le+24b}, we estimated the AGN contribution to the \OIII\ emission line for each source in our sample using the Python package Rainbow\footnote{https://gitlab.com/SPIrina/rainbow}. We applied these AGN fractions to the \OIII\ line flux to obtain more accurate AGN bolometric luminosities in this work. The Rainbow package is designed to quantify the ionization contributions in AGN host galaxies. It leverages the concept of a mixing sequence \textbf{\citep[e.g.,][]{Davies+16, Smirnova+22}}, modeling each position on the BPT diagram as a linear combination of foundational vectors that represent the ionization fractions contributed by AGN and star formation. 

The stellar velocity dispersion of our sample was measured by using the 47 MILES simple stellar population models with solar metallicity and ages ranging from 60 Myr to 12.6 Gyr \citep{Sanchez+06}. The black hole masse was determined using the relation between black hole mass and stellar mass described in \citet{Marconi03}, with values in the range of $\rm 10^{6} - 10^{9}\ M_{\odot}$. Finally, the Eddington ratios were calculated using the bolometric luminosities and the corresponding black hole masses.

Recently, \citet{Salim+16} and \citet{Salim+18} utilized UV/optical/mid-infrared spectral energy distribution (SED) fitting to determine SFRs for $\sim$700,000 galaxies at redshift z $<$ 0.3 in the GALEX-SDSS-WISE Legacy Catalog (GSWLC). We cross-matched our sample with the GSWLC-2 catalog, identifying 98,341 match sources, and used the provided SFRs for our study.

Additionally, we also cross-matched our sample with the catalog of the FIRST (Faint Images of the Radio Sky at Twenty Centimeters) survey at 1.4 GHz, which has a flux density limit of 1 mJy \citep{Becker+95} \footnote {http://sundog.stsci.edu/first/catalogs.html}. Using a matching radius of 5$\arcsec$, we found 4,877 sources in our sample, including 2,107 SF, 111 composite, 1,496 Seyfert, and 1163 LINER galaxies, which are available in the FIRST catalog. Based on the flux provided by the FIRST catalog, we calculated the K-corrected radio luminosity ($\rm L_{1.4GHz}$, \citealp[e.g.,][]{Ceraj+18, Wang+24}). Since most of the non-SF sources in our sample exhibit compact radio morphology, we assume a radio spectral index of $\rm \alpha = 0$ for the calculation. From the obtained $\rm L_{1.4GHz}$, we aim to study the relation between the Eddington ratio and radio luminosity in the SFR$-$$\rm M_{*}$ plane in our sample. 


\section{Results}\label{section:result}

In this section, we present our analysis of the relationships between UV-to-optical colors ($\rm u-r$), SFRs, Eddington ratios, and radio luminosities across various galaxy types. By examining these correlations in the $\rm SFR-M_{*}$ and $\rm (u-r)-M_{*}$ planes, we aim to study the roles of these physical quantities in different types of galaxies. 

\subsection{SFRs versus Stellar Mass}\label{subsec:sfr_mass}

\begin{figure*}
\centering
	\includegraphics[width=0.245\textwidth]{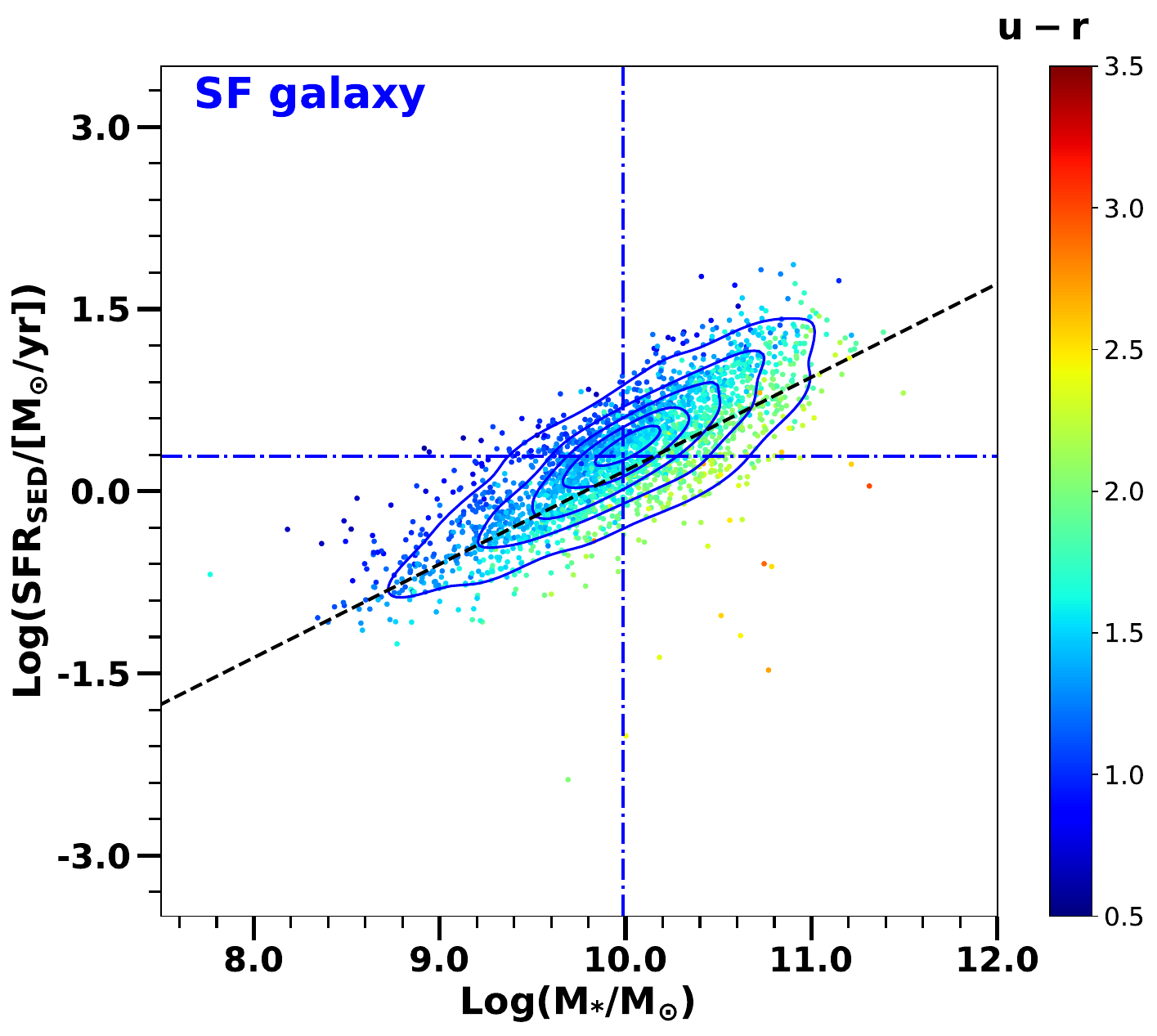}
	\includegraphics[width=0.245\textwidth]{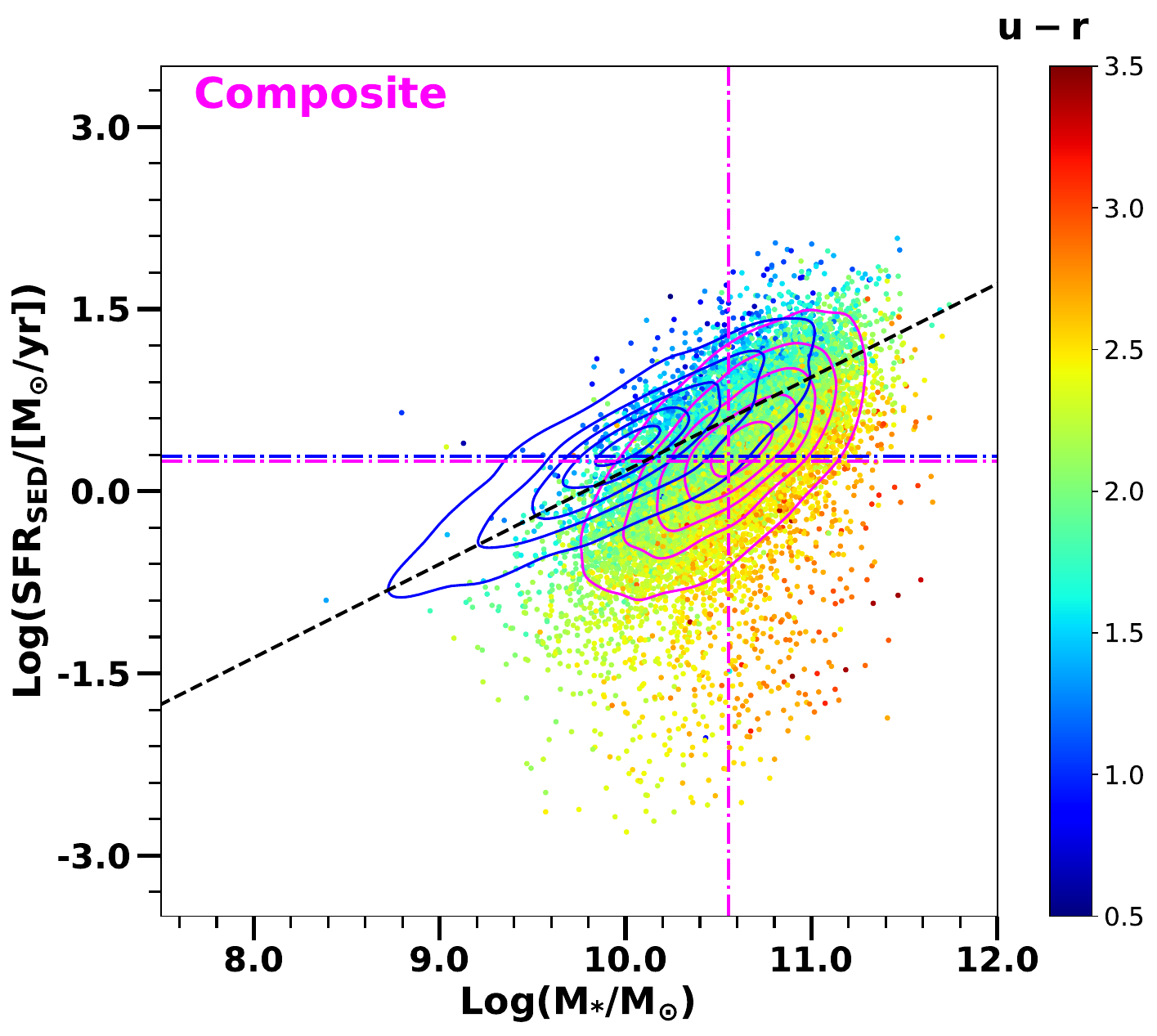}
	\includegraphics[width=0.245\textwidth]{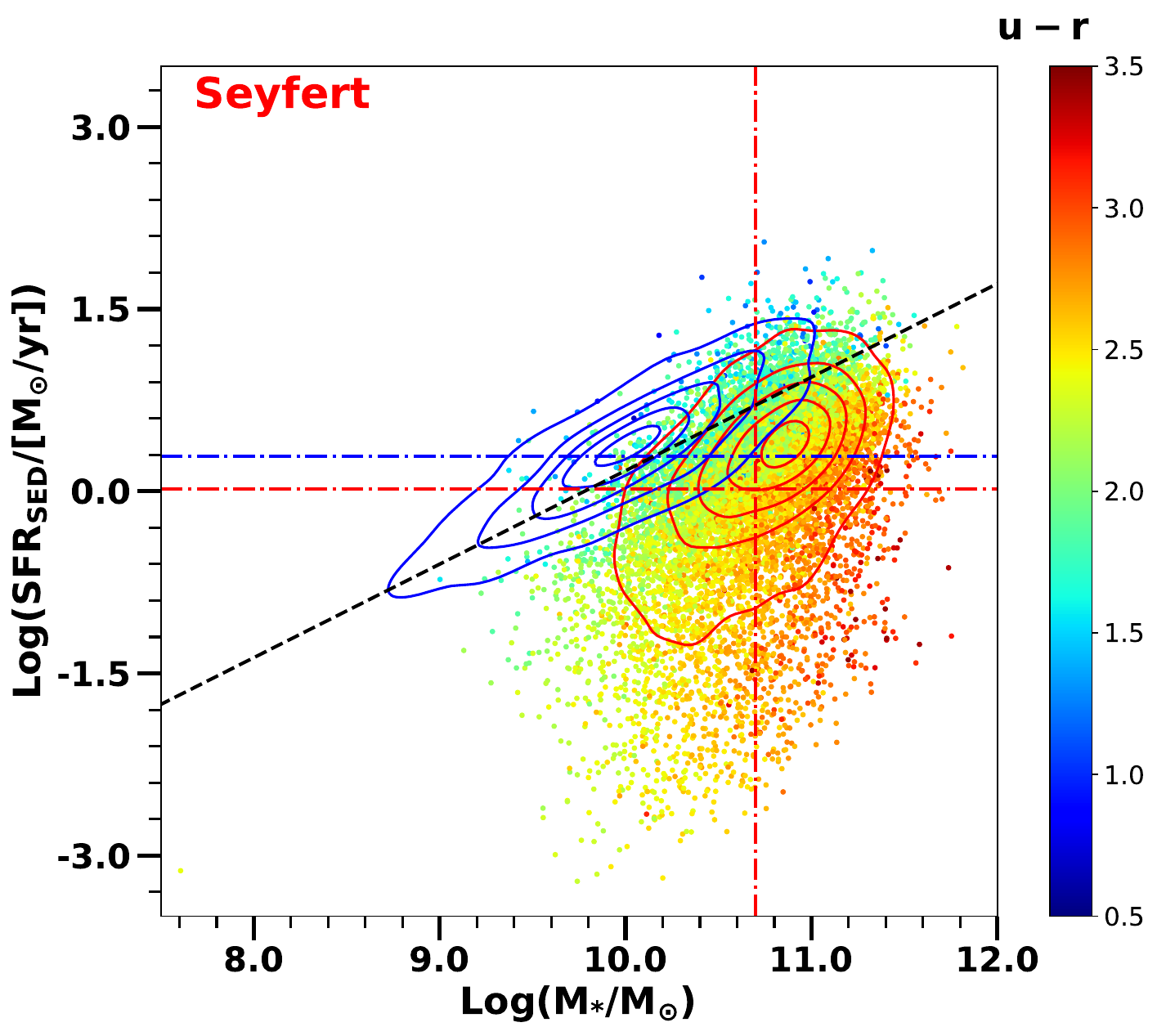}
	\includegraphics[width=0.245\textwidth]{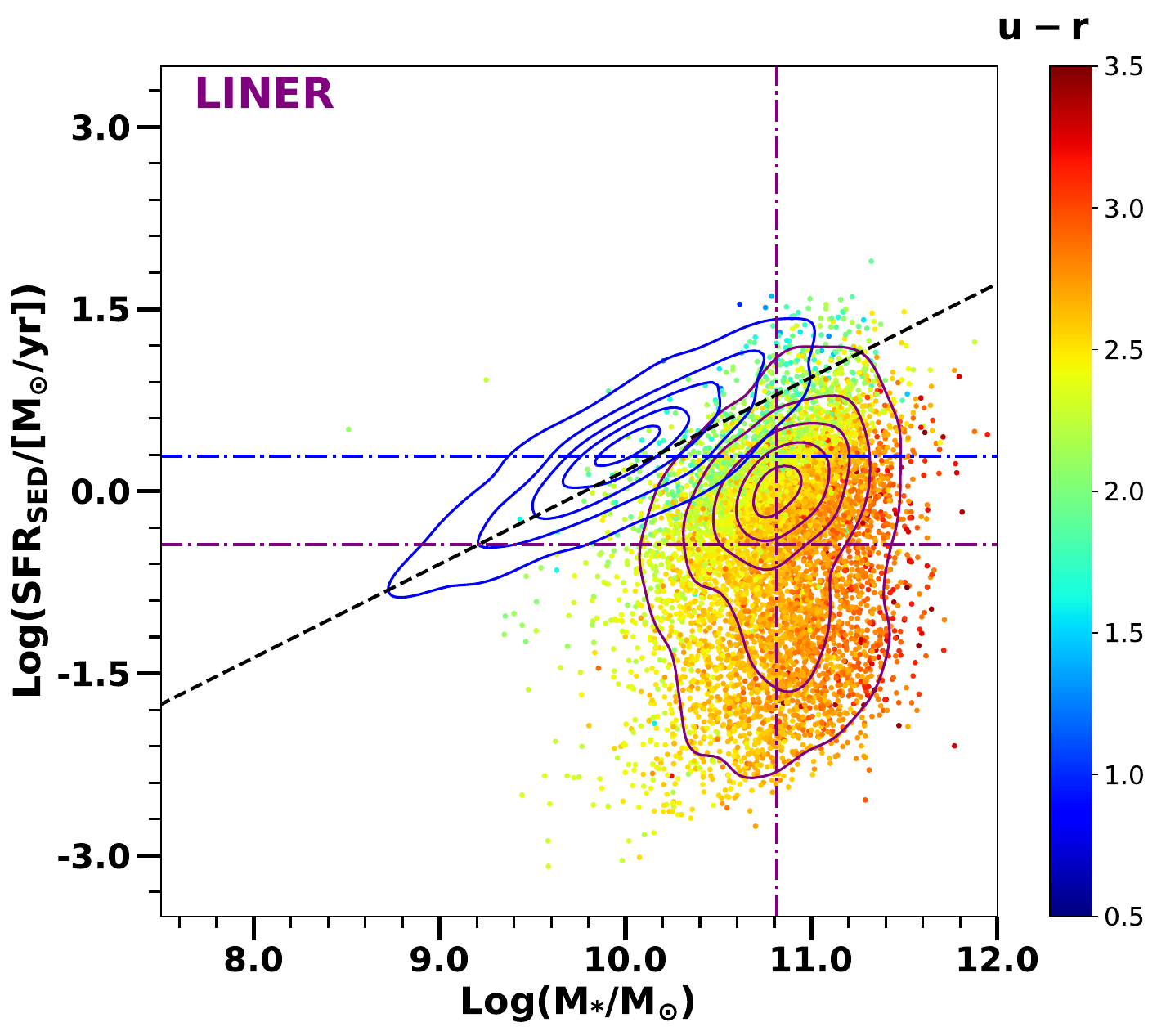}
	\caption{SFRs as a function of stellar mass. The $\rm SFR-M_{*}$ space is divided into a grid of 150x150 bins, and contour lines are drawn to represent 10, 30, 50, 70, and 90 percent of the maximum number density. Each panel displays the classifications among galaxies, such as SF (blue), composite (pink), Seyfert (red), and LINER (magenta), respectively. The blue contour lines of the SF galaxies are overplotted in other panels for a proper comparison. The black dashed lines show the local MS relation for blue galaxies determined by \citet{Elbaz+07}. The vertical and horizontal dashed-dotted lines display the medium values of $\rm Log(M_{*})$ and $\rm Log(SFR)$ of each galaxy type, respectively. The UV-to-optical ($\rm u-r$) color scale is displayed.
	\label{fig:sfr_mass_ur}}
\end{figure*}

We present the SFRs as a function of stellar mass in Figure \ref{fig:sfr_mass_ur}. In this $\rm SFR-M_{*}$ plane, as mentioned in the sample selection, we divided our sample into four galaxy types: SF, composite, Seyfert, and LINER. In each panel, we used the UV-to-optical color ($\rm u-r$) as a color scale to examine the variation of ($\rm u-r$) in the $\rm SFR-M_{*}$ plane for each galaxy type.

We found that in SF galaxies, most of the sources are young, exhibit blue colors ($\rm u-r$ $<$ 2.0), and lie on the main-sequence (MS) relation \citep{Elbaz+07}. In contrast, sources in composite, Seyfert, and LINER galaxies tend to deviate from the MS line, displaying decreasing trends in SFRs, with the highest SFRs in composite galaxies, lower in Seyfert galaxies, and lowest in LINER galaxies. These trends indicate that the stellar populations in composite, Seyfert, and LINER galaxies are generally older than those in SF galaxies. Similar trends are also observed for stellar mass, with average stellar mass increasing from SF to composite, Seyfert, and LINER galaxies. Furthermore, sources below the MS line typically exhibit red colors ($\rm u-r$ $>$ 2.5). The UV-to-optical colors ($\rm u-r$) become increasingly red from composite to Seyfert and finally to LINER galaxies, with the reddest sources observed in LINER galaxies.

\begin{figure*}
\centering
	\includegraphics[width=0.33\textwidth]{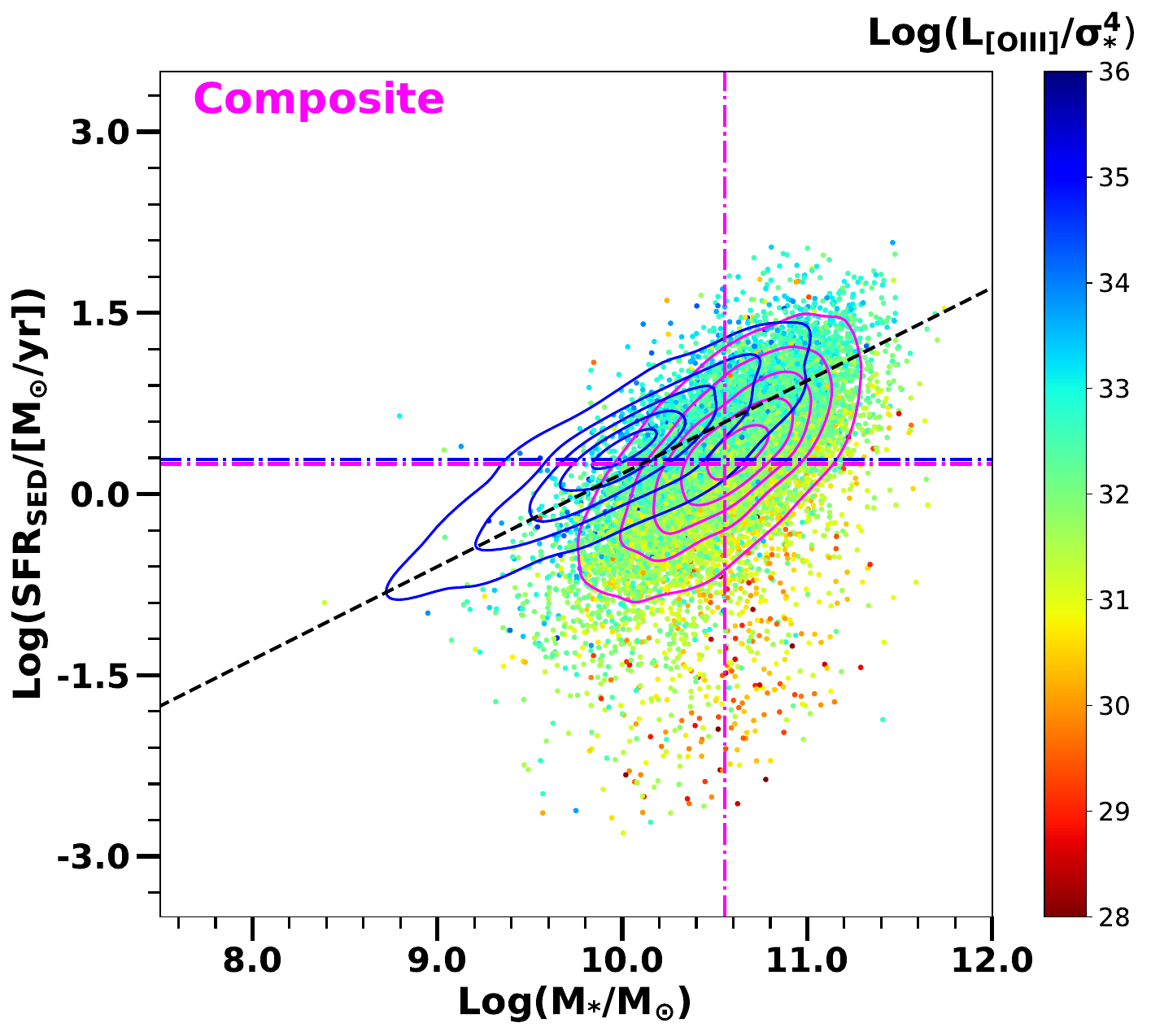}
	\includegraphics[width=0.33\textwidth]{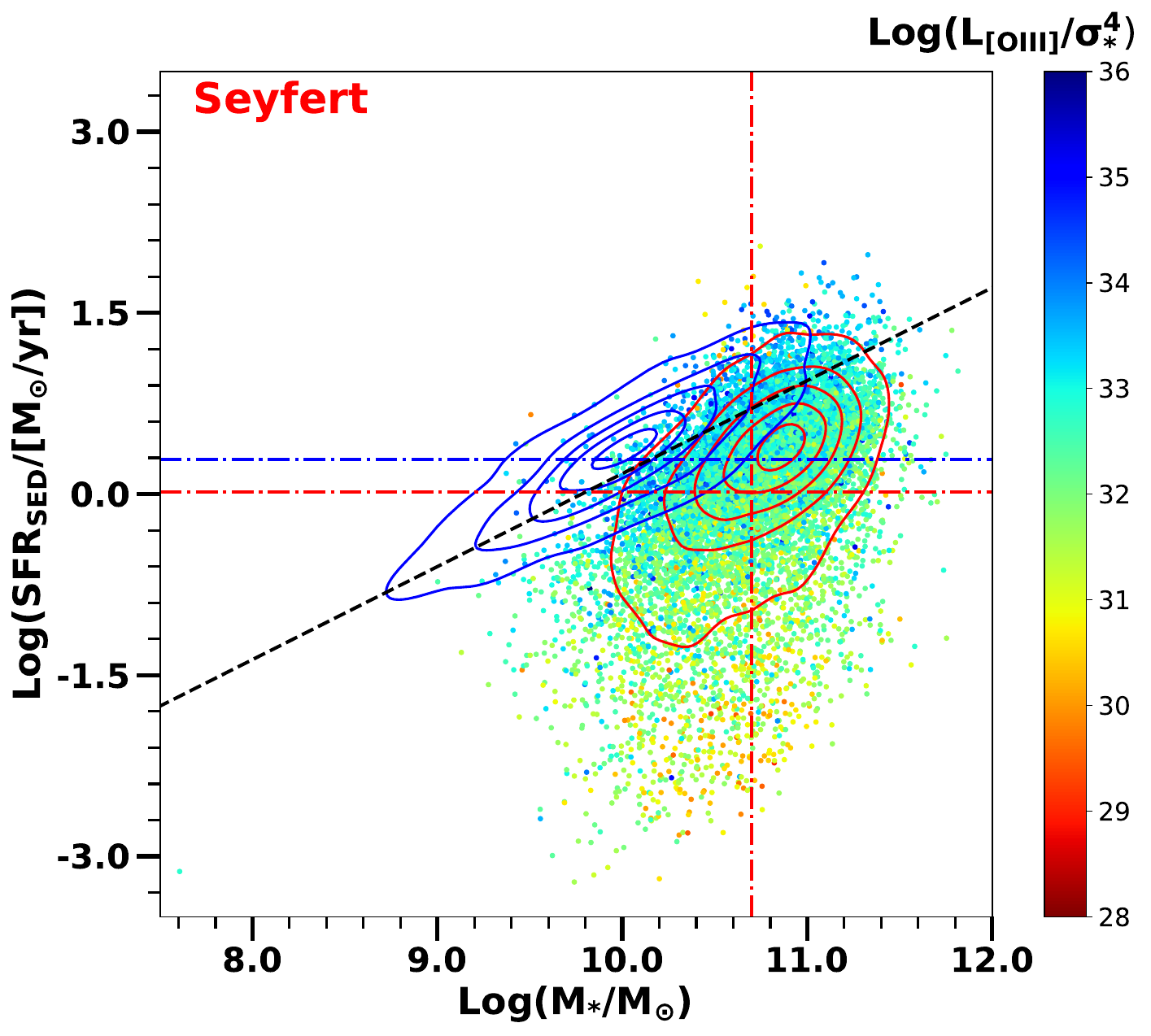}
	\includegraphics[width=0.33\textwidth]{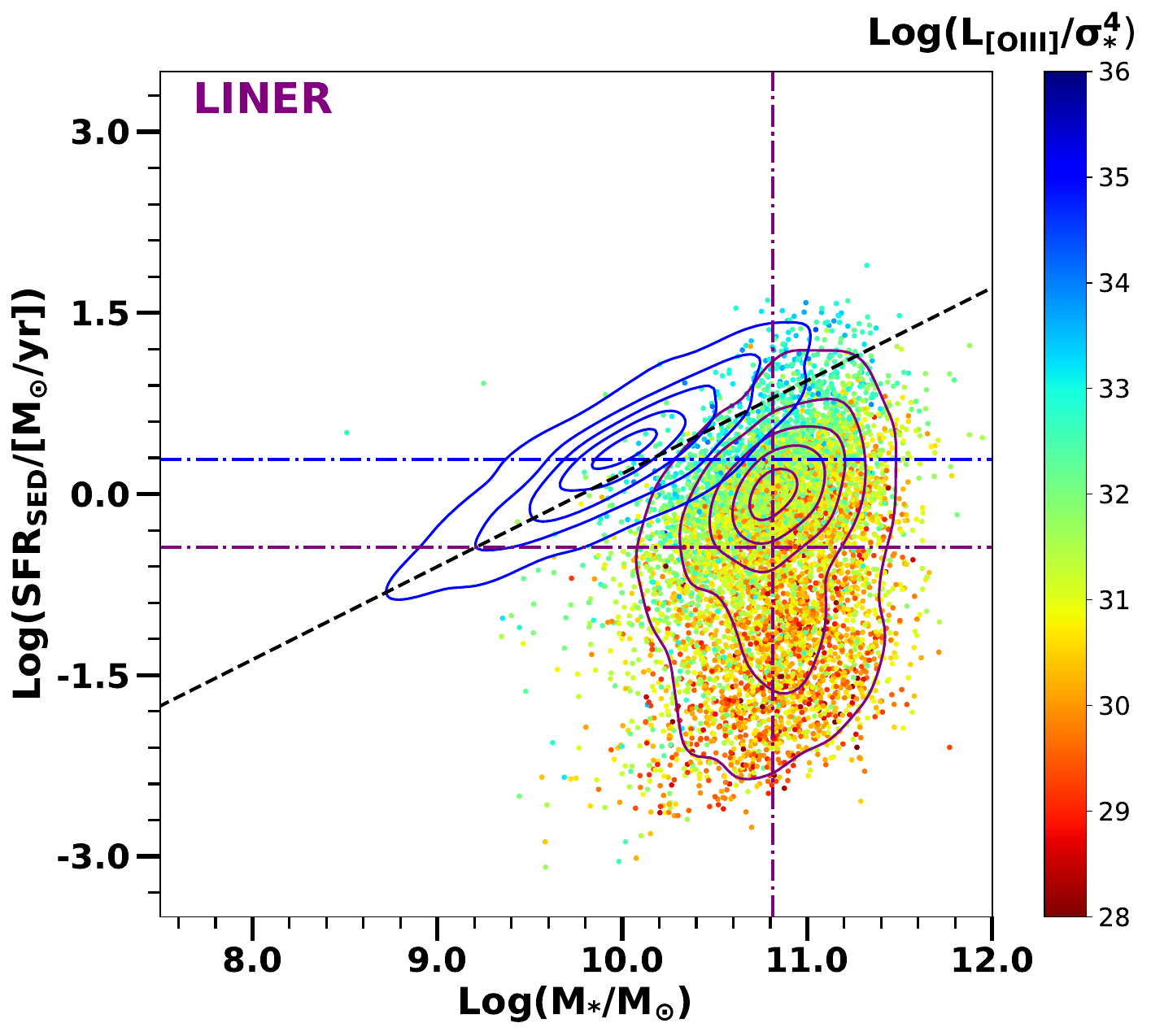}
	\caption{Similar to Figure \ref{fig:sfr_mass_ur}, but with color scales representing $\rm L_{[OIII]}/\sigma_{*}^{4}$. SF galaxies are not shown, as we assume no AGN contribution to their \OIII\ luminosities.
	\label{fig:sfr_mass_bhar}}
\end{figure*}

As discussed in \citet{Heckman+04} and \citet{Kewley+06}, the mass accretion rate onto the black hole is related to the bolometric luminosity, and the ratio $\rm L_{[OIII]}/\sigma_{*}^{4}$ is proportional to the Eddington ratio. In Figure \ref{fig:sfr_mass_bhar}, we present $\rm L_{[OIII]}/\sigma_{*}^{4}$, represented as color scales on the $\rm SFR-M_{*}$ plane. Interestingly, the mass accretion rate onto the black hole relative to the Eddington ratio exhibits decreasing trends similar to the UV-to-optical color ($\rm u-r$). Sources along the MS line show high $\rm L_{[OIII]}/\sigma_{*}^{4}$ values, while those with lower values tend to deviate from the MS line. The ratio $\rm L_{[OIII]}/\sigma_{*}^{4}$ decreases progressively, with the highest values in Seyfert galaxies, lower values in composite galaxies, and the lowest in LINER galaxies. The results may suggest a close connection between AGN activity and star formation processes, particularly during the Seyfert stage, where AGN feedback, along with the gas supply, begins to suppress SFRs. AGN activity is strongest in Seyfert galaxies and diminishes in LINER galaxies.

\subsection{UV-to-optical Colors versus Stellar Mass}\label{subsec:color_mass}

\begin{figure*}
\centering
	\includegraphics[width=0.33\textwidth]{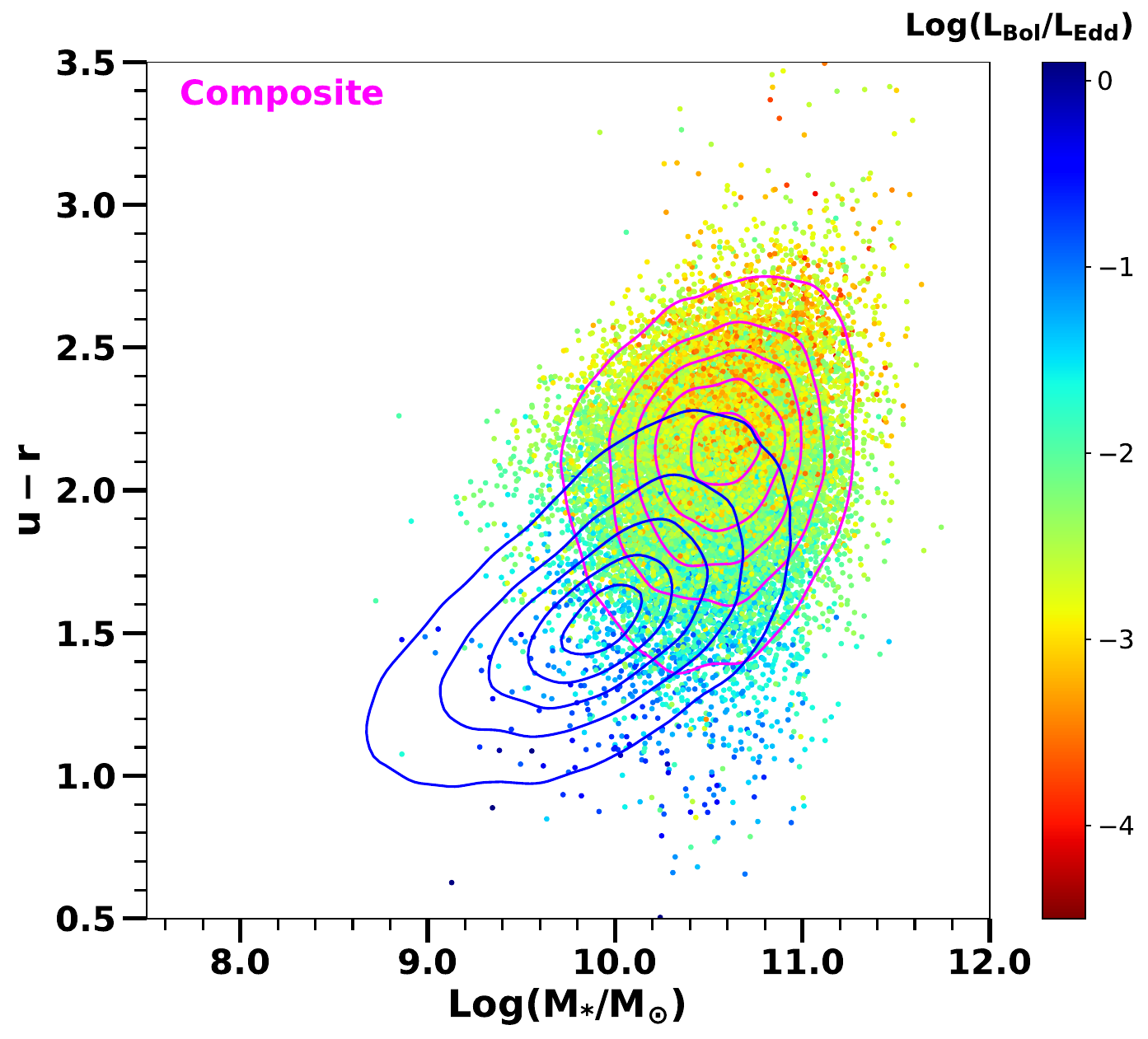}
	\includegraphics[width=0.33\textwidth]{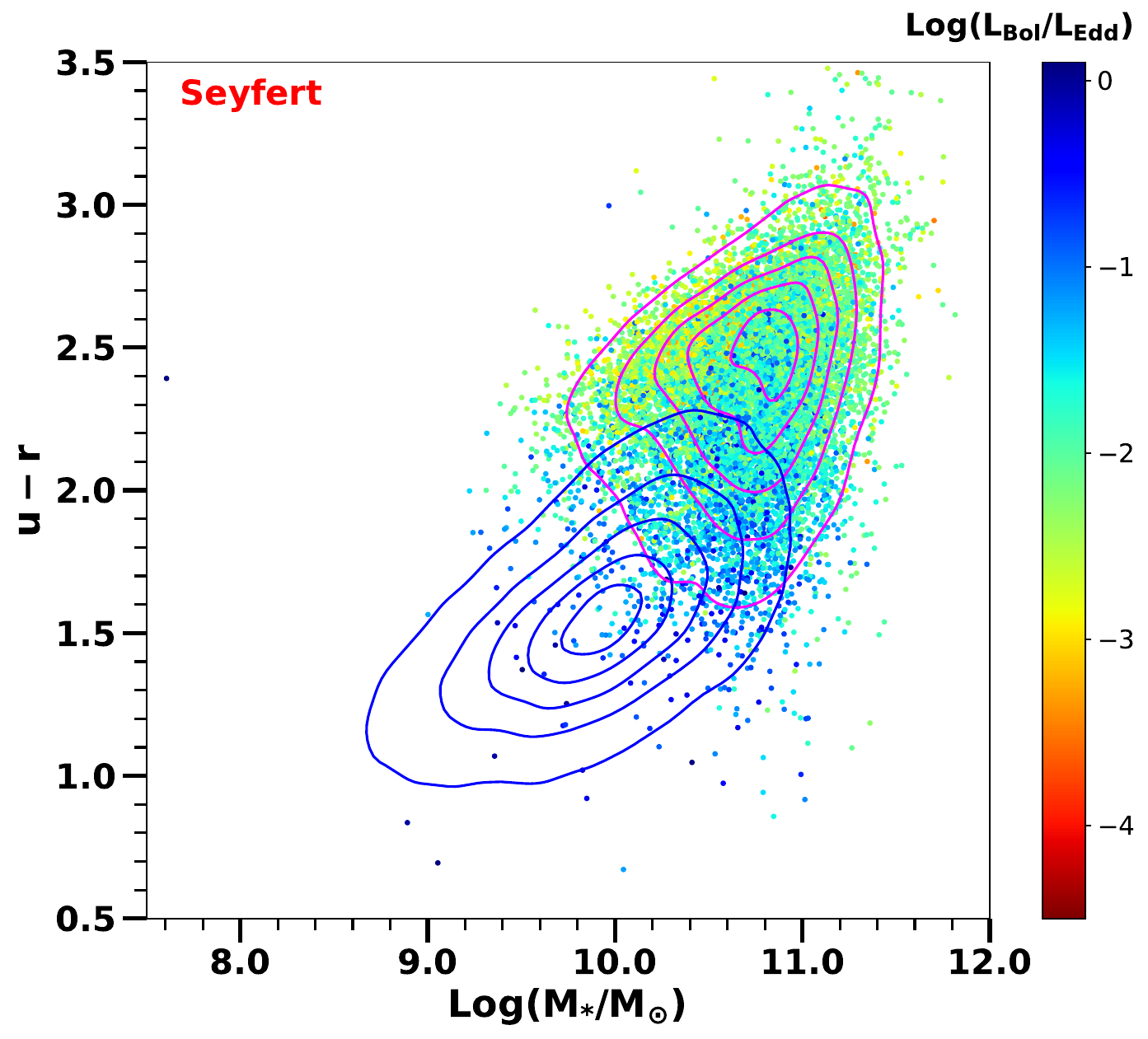}
	\includegraphics[width=0.33\textwidth]{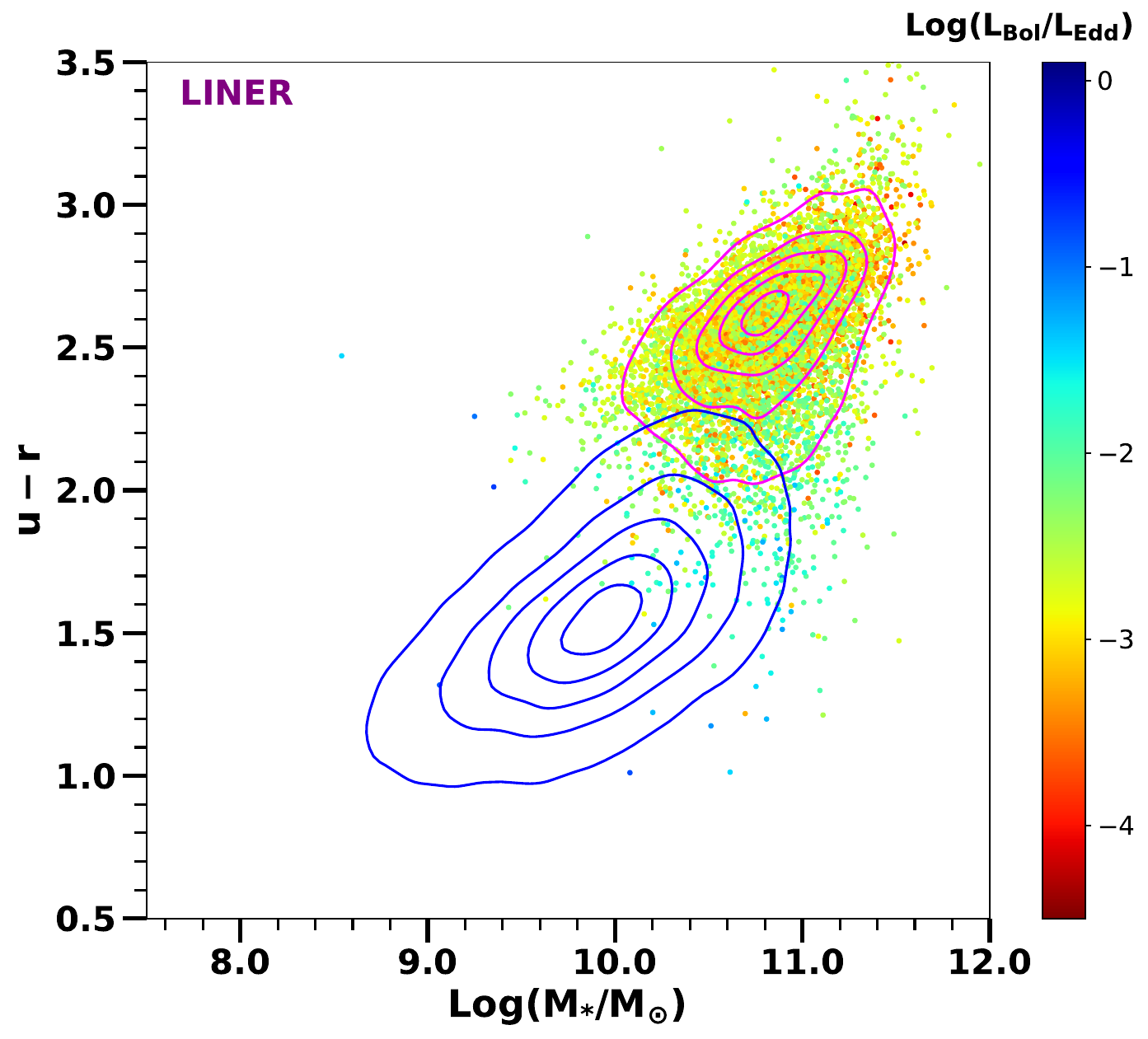}
	\includegraphics[width=0.33\textwidth]{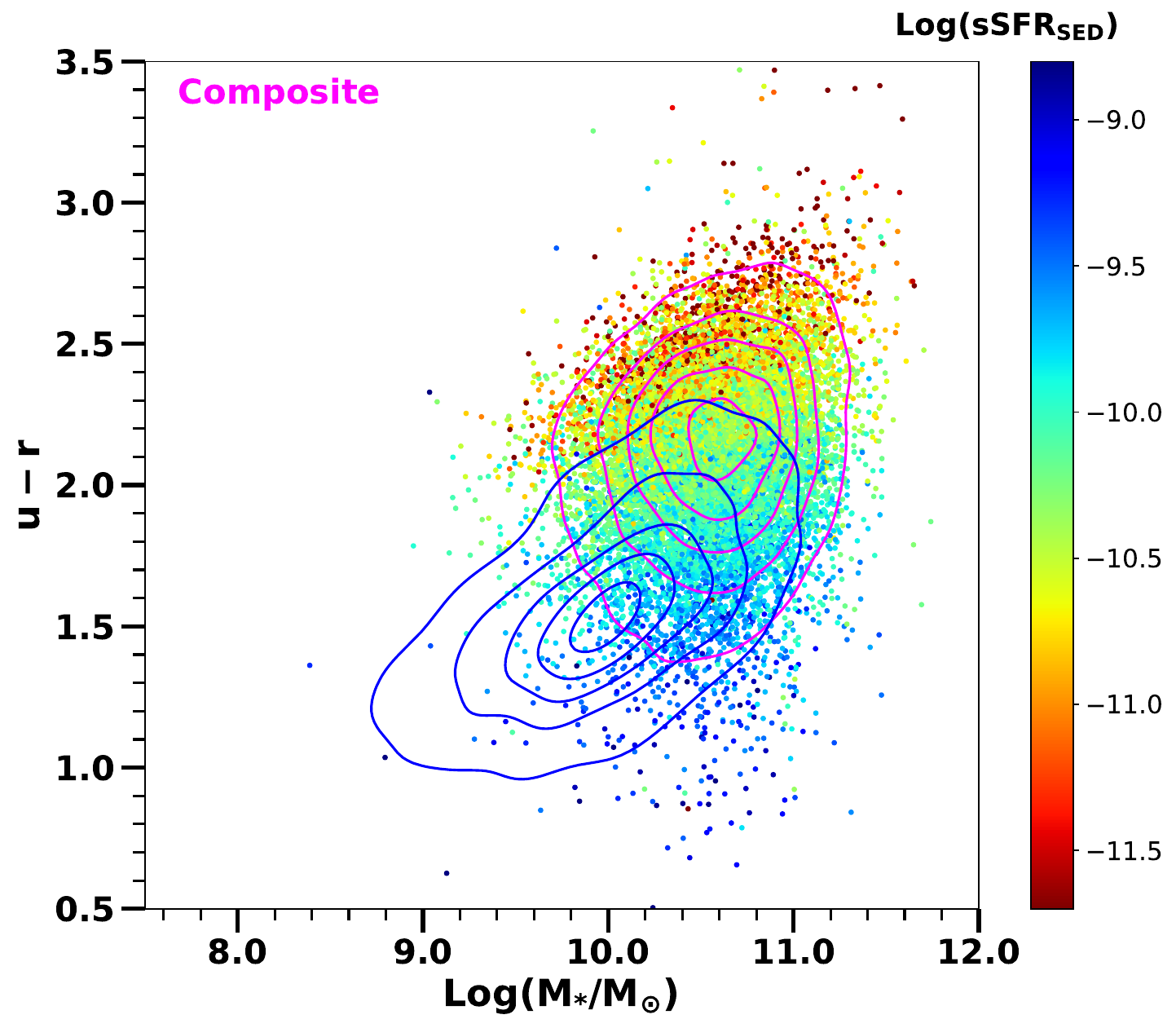}
	\includegraphics[width=0.33\textwidth]{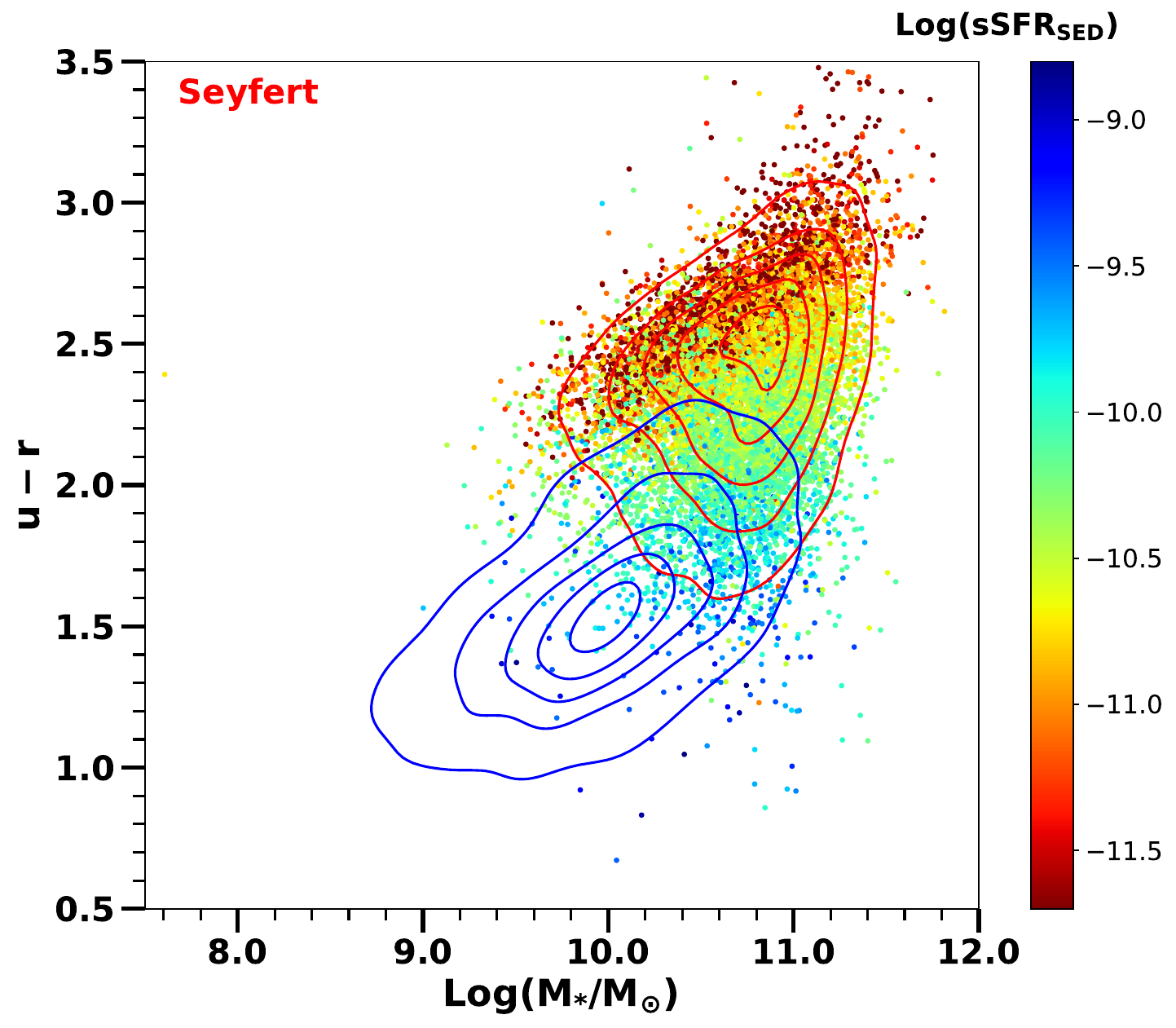}
	\includegraphics[width=0.33\textwidth]{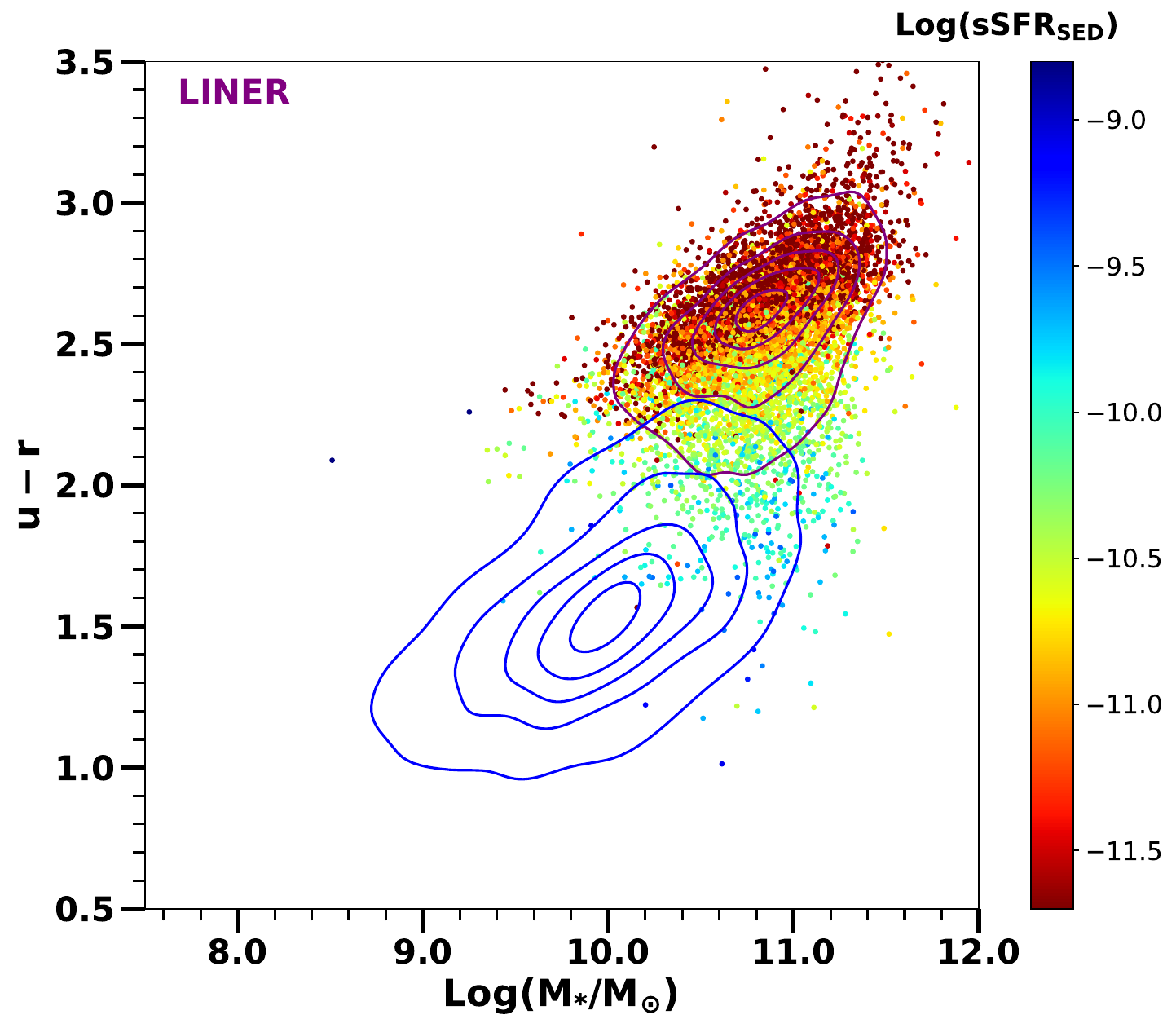}
	\includegraphics[width=0.33\textwidth]{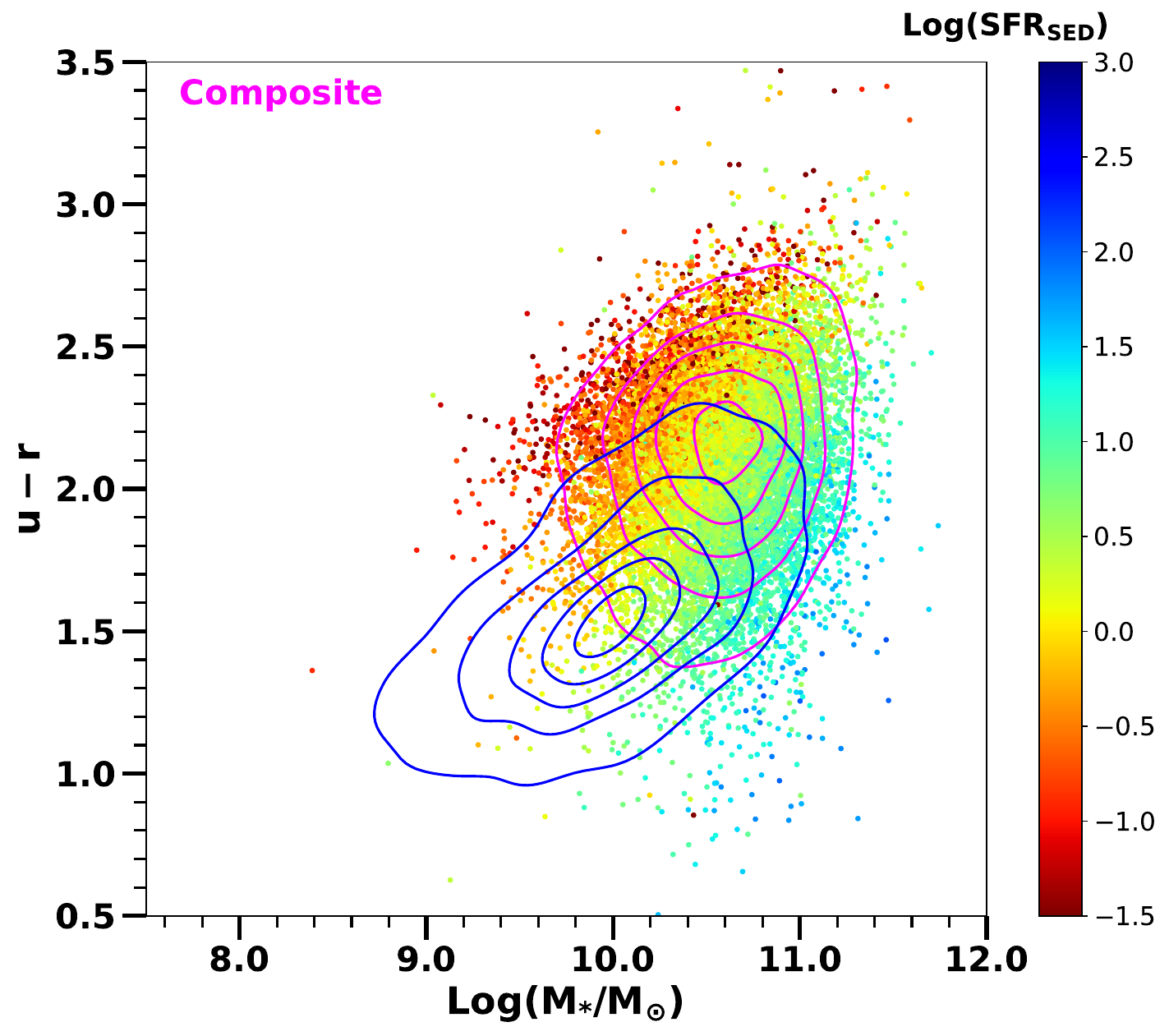}
	\includegraphics[width=0.33\textwidth]{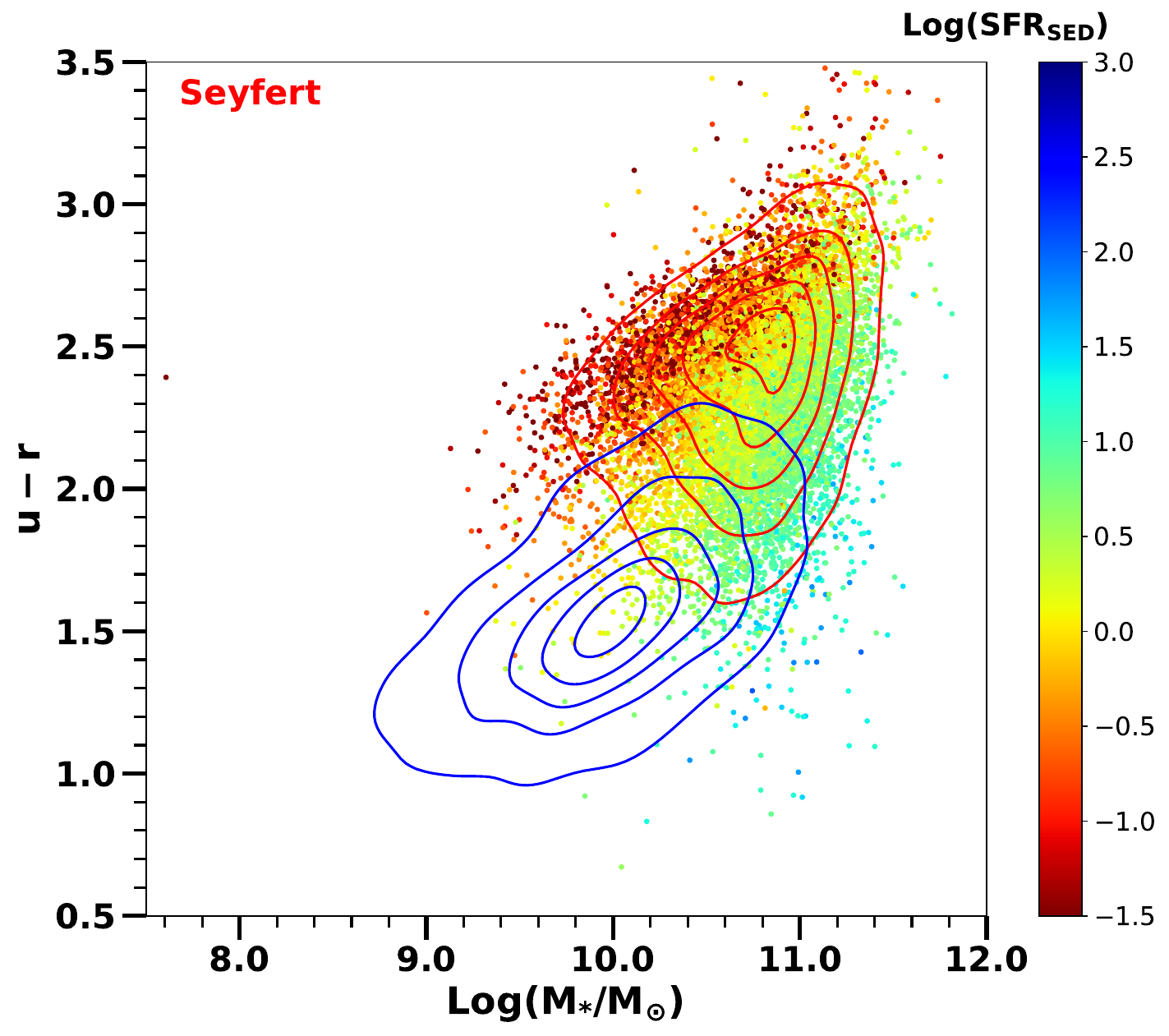}
	\includegraphics[width=0.33\textwidth]{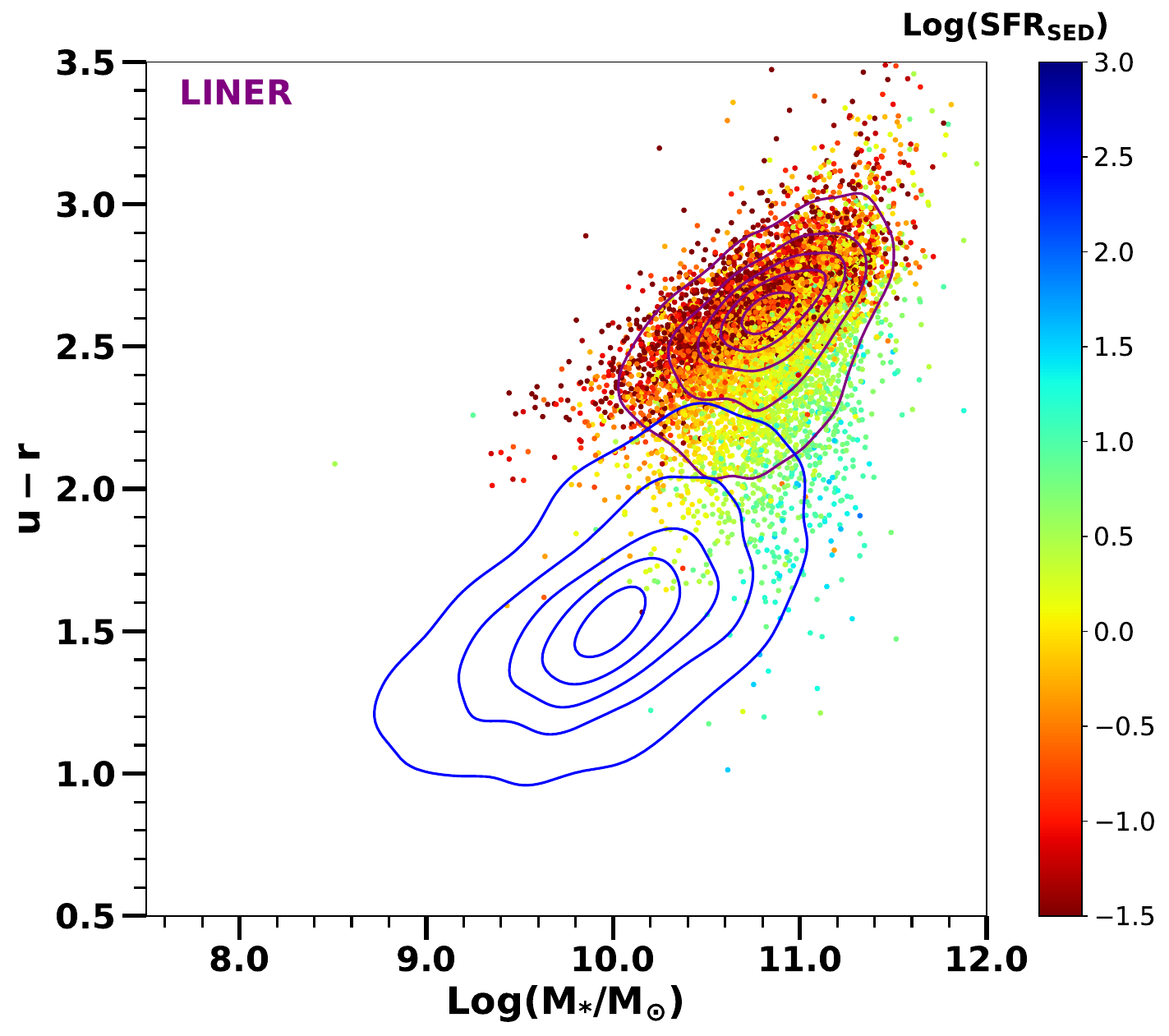}
	\caption{Similar to Figure \ref{fig:color_contour}, but with the color schemes representing Eddington ratio values, $\rm sSFR_{SED}$, $\rm SFR_{SED}$, respectively. 
	\label{fig:color_contour_edd}}
\end{figure*}

In Figure \ref{fig:color_contour_edd}, we display the Eddington ratios, specific SFRs (sSFRs), and SFRs as color scales to examine their variations across different galaxy types in the color-mass diagram for composite, Seyfert, and LINER galaxies. Interestingly, Eddington ratios, SFRs, and sSFRs are continuously distributed across the $\rm (u-r)-M_{*}$ plane, ranging from blue to redder galaxies. The Eddington ratio is high in the region where other galaxy types overlap with blue SF galaxies. As the galaxies transition to a redder color, the Eddington ratio decreases. 

The Eddington ratio values are highest in Seyfert galaxies, followed by composite galaxies, and lowest in LINER galaxies. Interestingly, we observe that in each galaxy type, Eddington ratios are higher when ($\rm u-r$) colors are bluer, and they tend to decrease vertically as ($\rm u-r$) colors become redder, indicating an older stellar population.

The trends in Eddington ratio values are consistent with those observed in sSFRs. We find that sources with blue ($\rm u-r$) colors exhibit high sSFRs, while those with redder ($\rm u-r$) colors have lower sSFRs. The similar trends between sSFRs and Eddington ratios suggest a strong correlation between these two parameters. Specifically, sources with low Eddington ratios tend to have low SF activities, whereas sources with high Eddington ratios are associated with high SF activities. 

Additionally, the SFR trends on the color-mass diagram show variations in SFR values across different galaxy types. These trends differ from the vertical trends of the Eddington ratio and sSFRs. SFRs tend to be larger at higher stellar mass ranges. If we examine vertical slices at constant stellar mass, we observe an upward trend in SFRs as the ($\rm u-r$) color becomes red within each galaxy type. This suggests that sSFRs might be a more appropriate metric than SFRs, as they allow for a fair comparison among sources with similar physical properties (e.g., \citealp{Salim+16}).

\subsection{UV-to-optical Colors versus Eddington Ratio}\label{subsec:color_edd}

Based on the strong correlations observed between the ($\rm u-r$) color, Eddington ratio, and sSFRs, in this part, we aim to explore these relationships in more detail by examining ($\rm u-r$) color as a function of the Eddington ratio. In each galaxy type, we display sSFRs as a color scale to observe their variation in the $\rm (u-r)-L_{Bol}/L_{Edd}$ plane.

\begin{figure*}
\centering
	\includegraphics[width=0.33\textwidth]{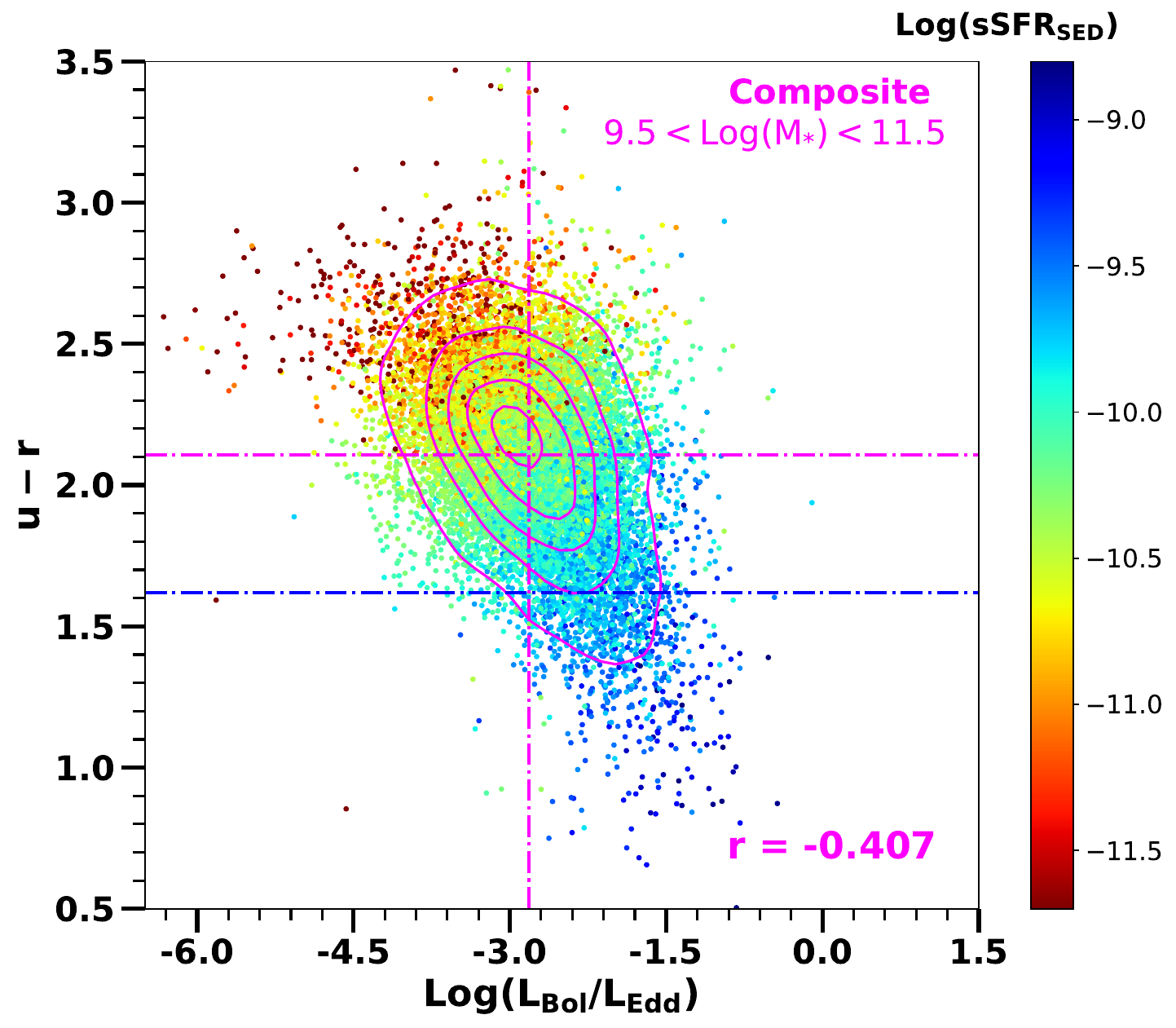}
	\includegraphics[width=0.33\textwidth]{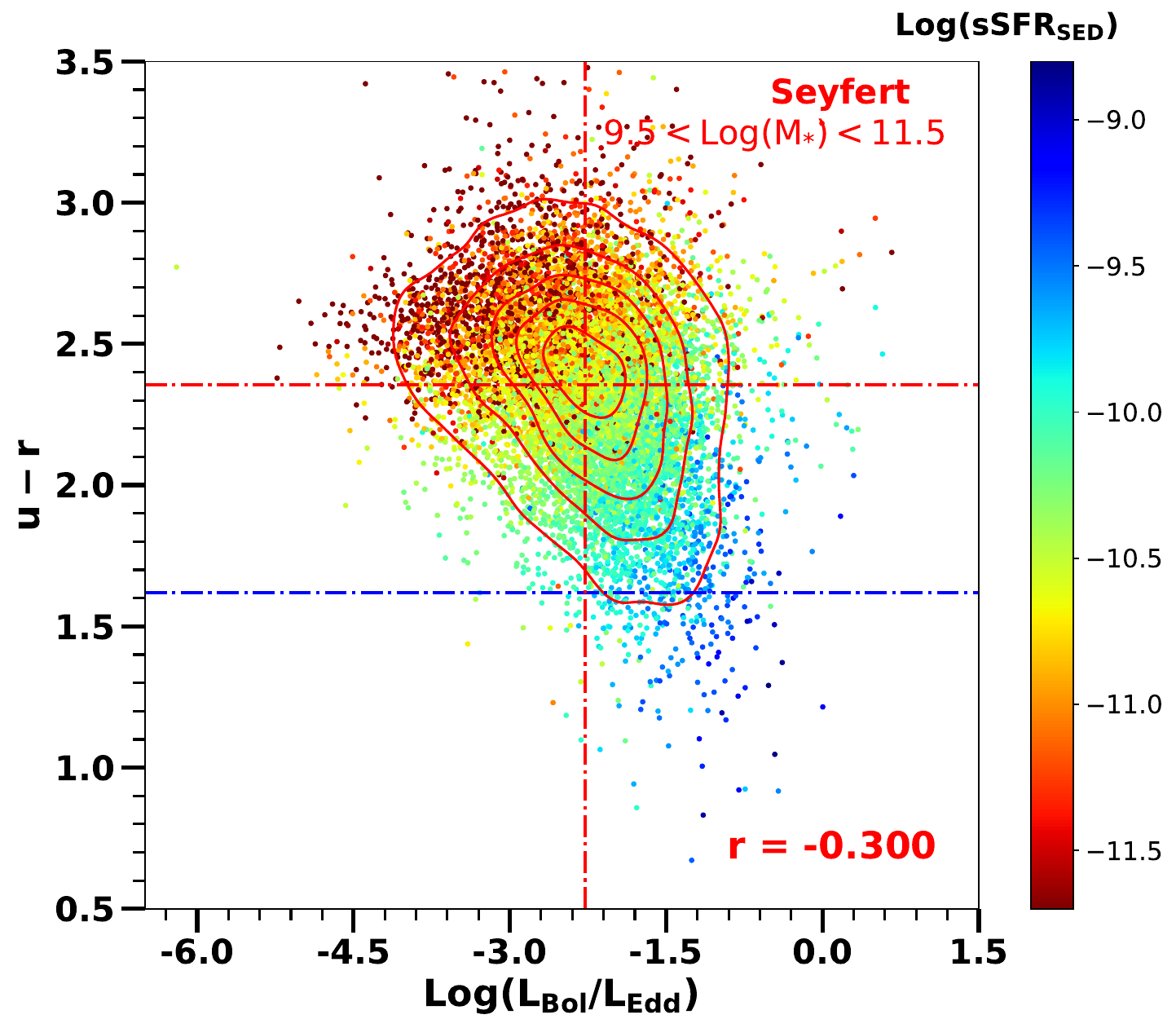}
	\includegraphics[width=0.33\textwidth]{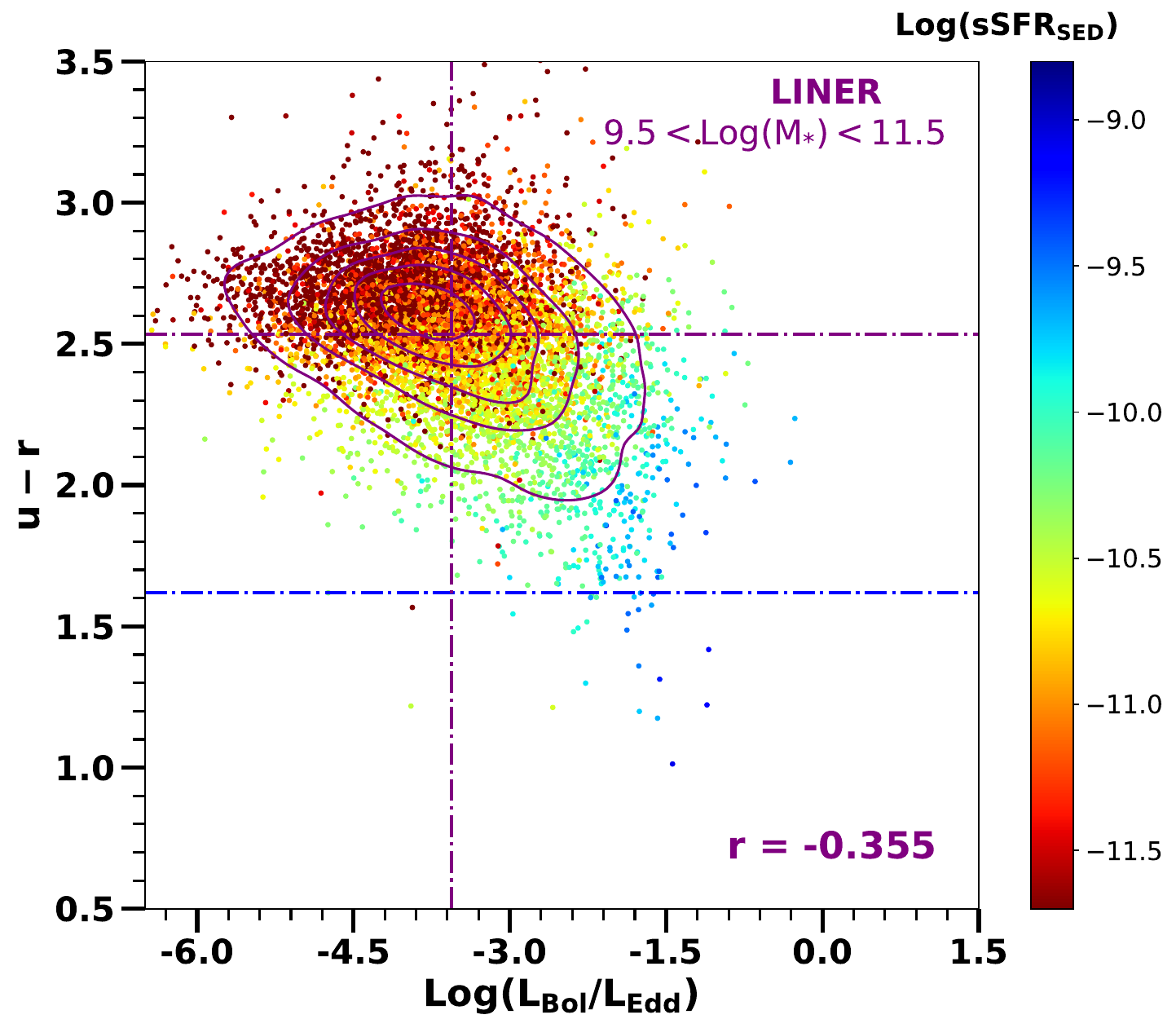}
	\caption{The UV-to-optical ($\rm u-r$) colors as a function of Eddington ratio values. The stellar mass range is from 9.5 $<$ $\rm \log M_{*}$ $<$ 11.5. The color scales represent the $\rm sSFR_{SED}$ values. The $\rm (u-r)-L_{Bol}/L_{Edd}$ space is divided into a grid of 150x150 bins, and contour lines are drawn to represent 10, 30, 50, 70, and 90 percent of the maximum number density. Each panel displays the classifications among galaxies such as composite, Seyfert, and LINER, respectively. Horizontal dashed-dotted lines in blue, pink, red, and magenta correspond to the median UV-to-optical ($\rm u-r$) values for SF, composite, Seyfert, and LINER galaxies, respectively. Vertical dashed-dotted lines indicate the median Eddington ratio values for each galaxy type. The Spearman correlation coefficient (r) is displayed in each panel. 
	\label{fig:color_edd_ssfr}}
\end{figure*}

Figure \ref{fig:color_edd_ssfr} presents $\rm (u-r)$ colors as a function of the Eddington ratio for a stellar mass range of 9.5 $<$ $\rm \log M_{*}$ $<$ 11.5. The $\rm (u-r)$ colors demonstrate a negative correlation with the Eddington ratio. To quantify these correlations, we calculate the Spearman correlation coefficient (r) and report it in each panel. We observe increasing trends in $\rm (u-r)$ colors across SF, composite, Seyfert, and LINER galaxies, respectively. The sSFRs, represented as a color scale, exhibit a strong correlation and smooth transition in the $\rm (u-r)-L_{Bol}/L_{Edd}$ plane. Sources with low Eddington ratios tend to have low sSFRs, while those with high Eddington ratios exhibit high sSFRs. Notably, Seyfert galaxy sources show the highest Eddington ratio values compared to composite and LINER sources, indicating stronger AGN activity in Seyfert galaxies compared to the other types.

\begin{figure*}
\centering
    \includegraphics[width=0.33\textwidth]{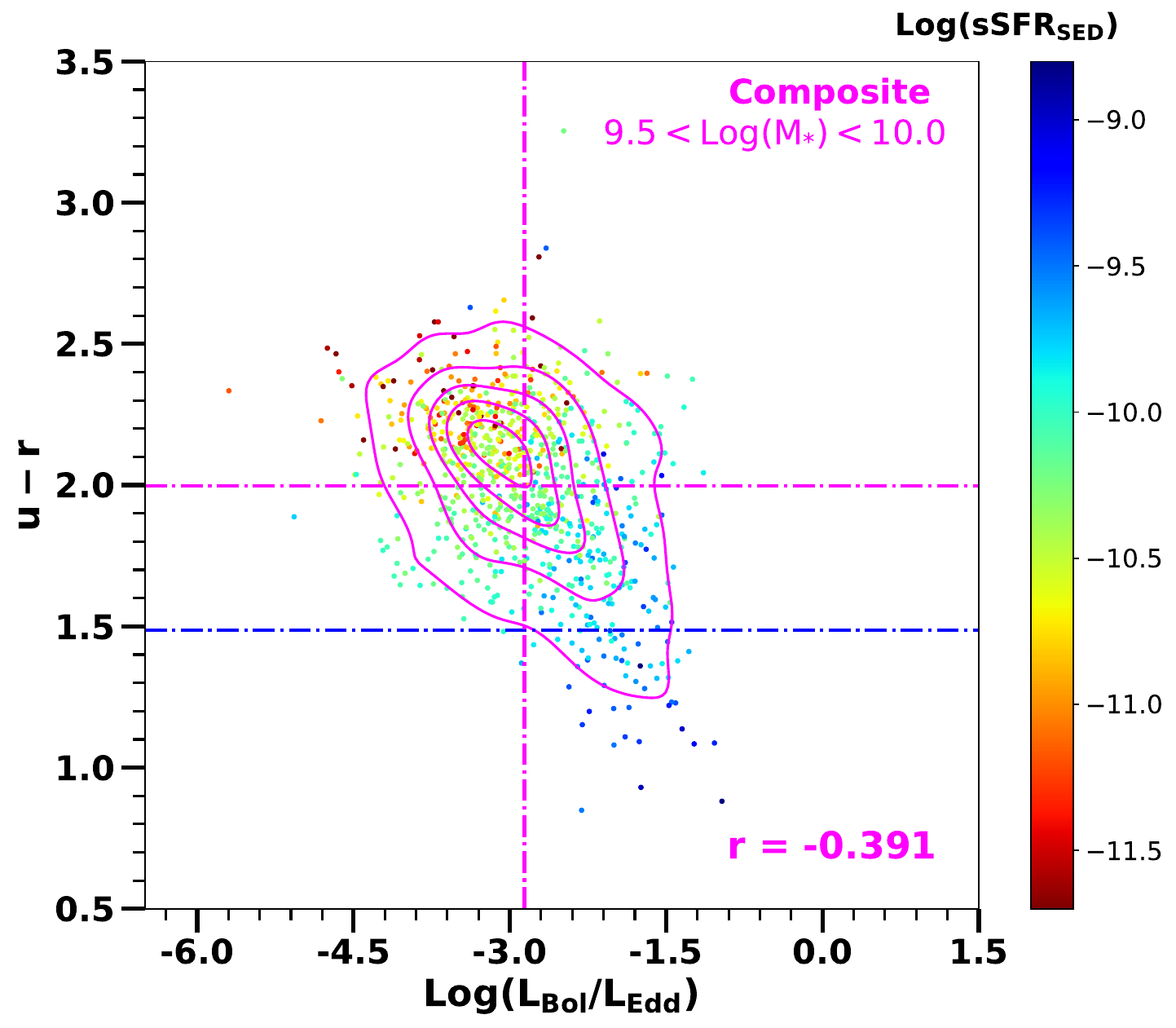}
    \includegraphics[width=0.33\textwidth]{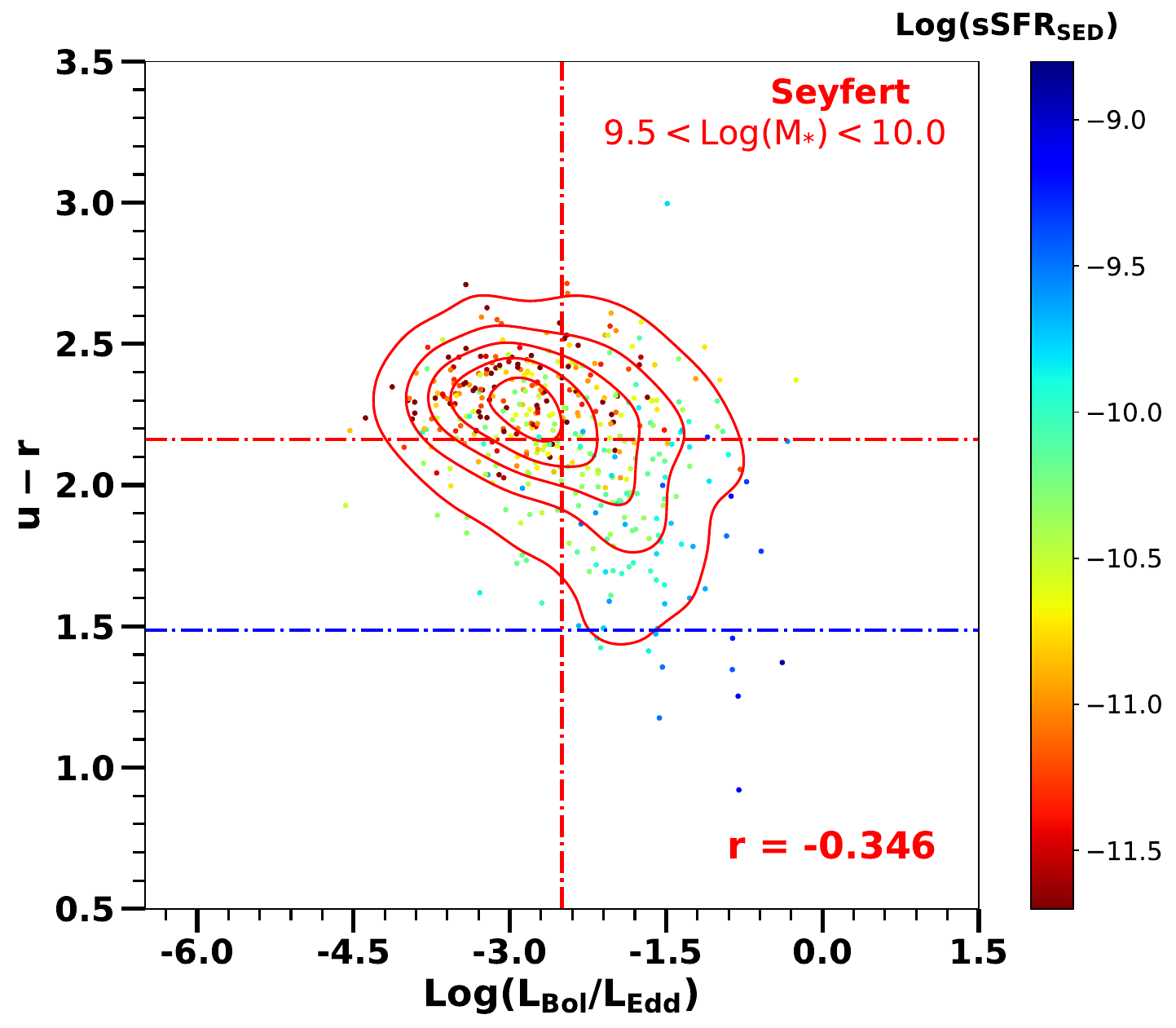}
    \includegraphics[width=0.33\textwidth]{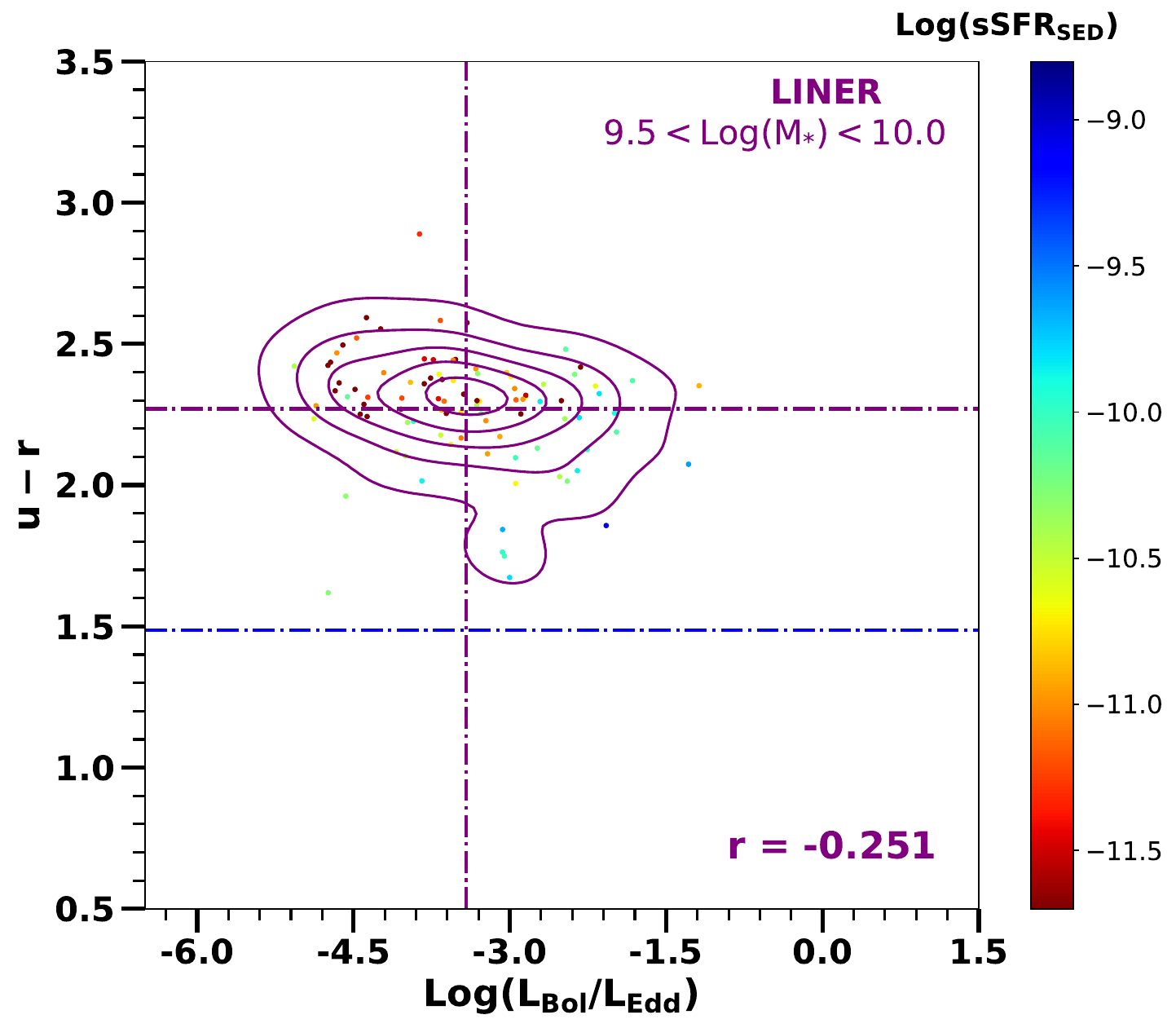}
    \includegraphics[width=0.33\textwidth]{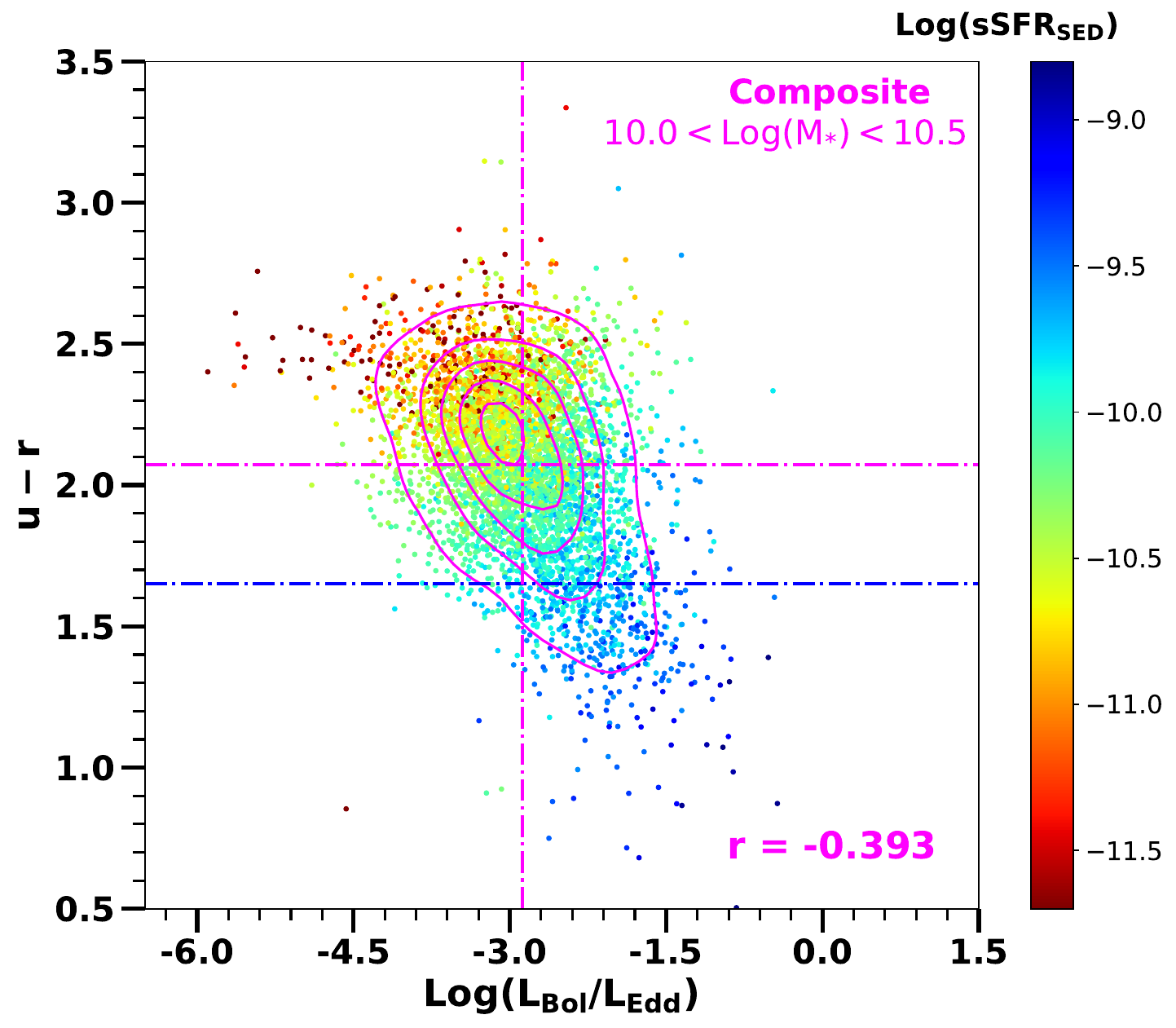}
    \includegraphics[width=0.33\textwidth]{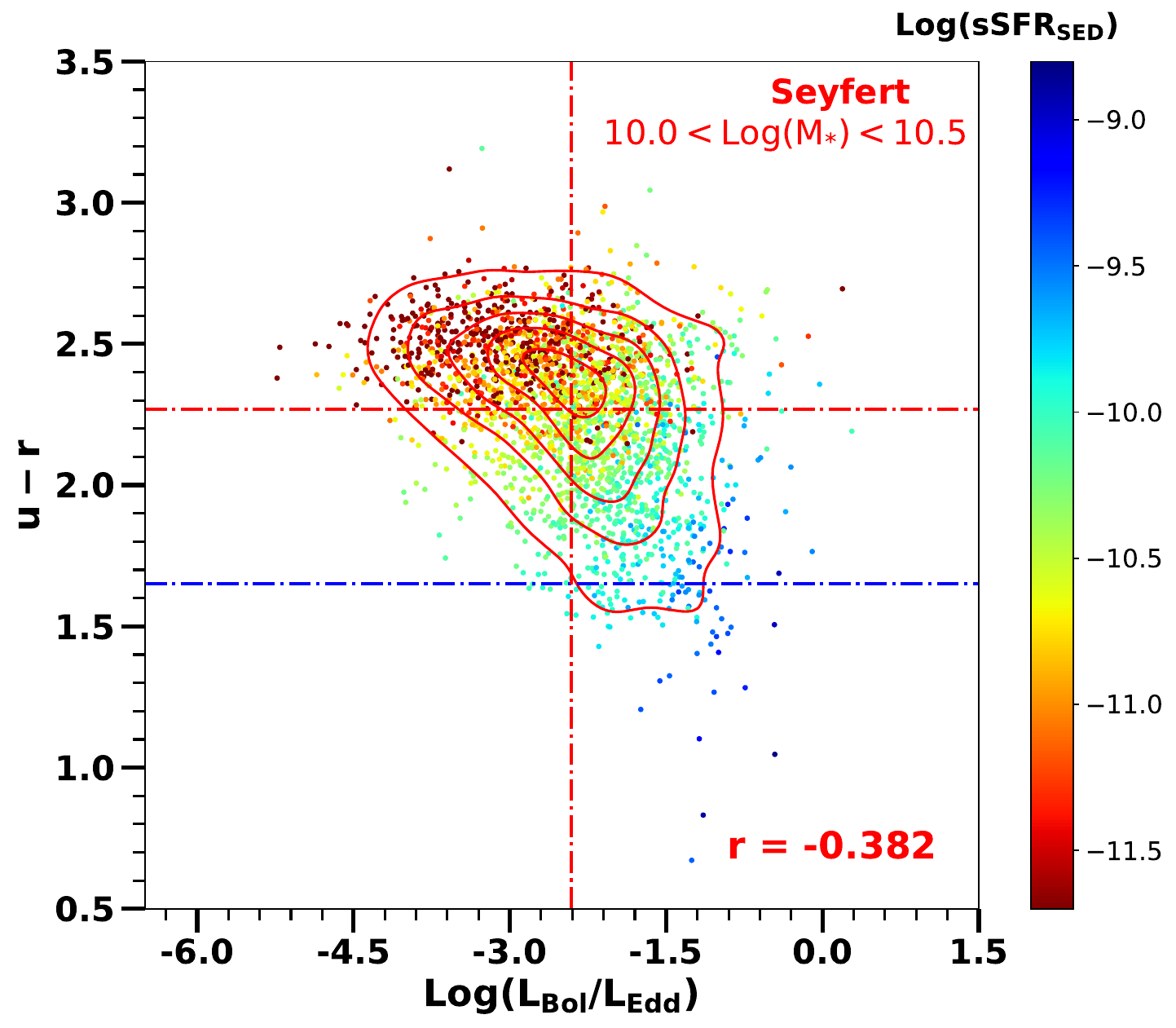}
    \includegraphics[width=0.33\textwidth]{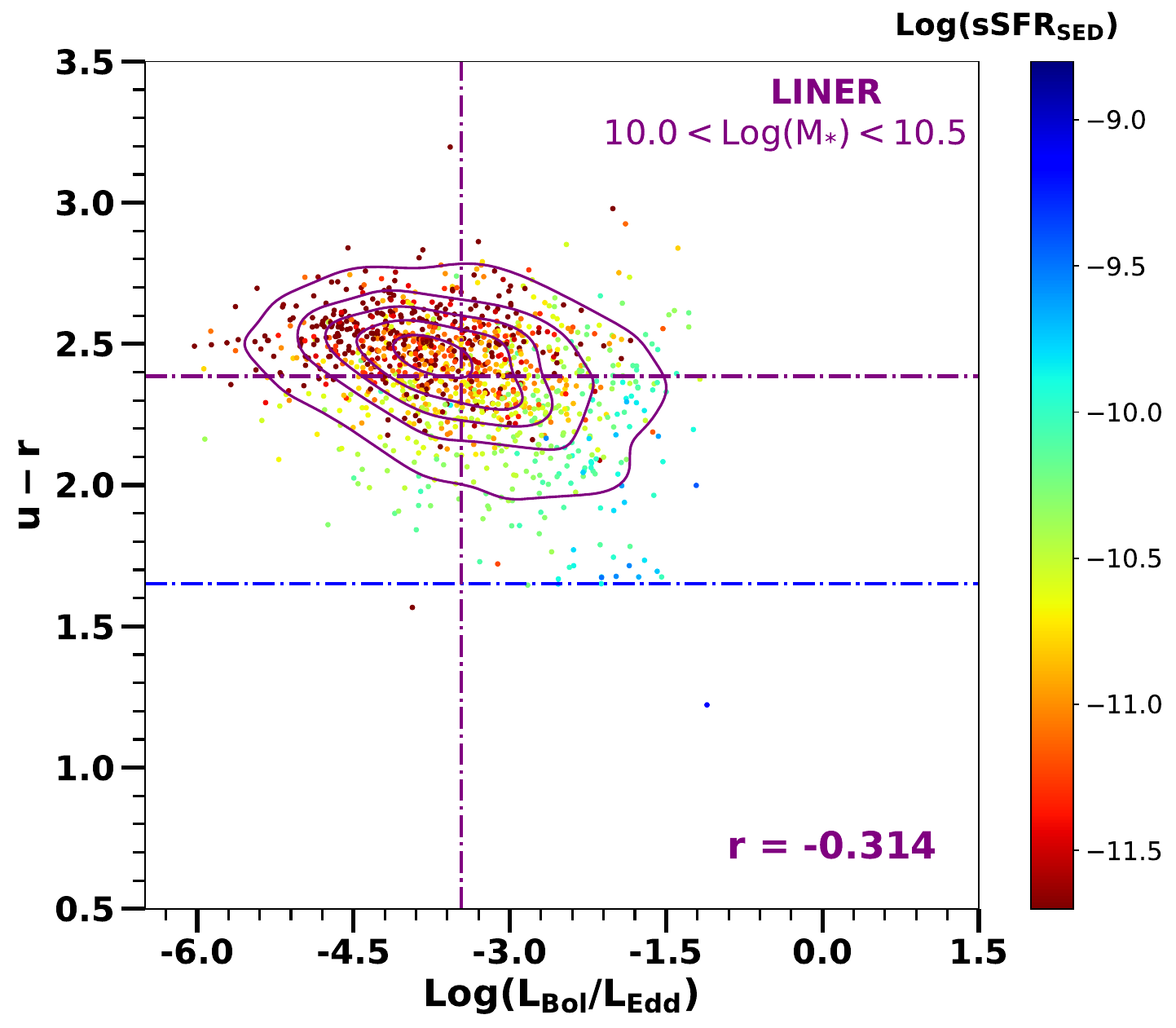}
    \includegraphics[width=0.33\textwidth]{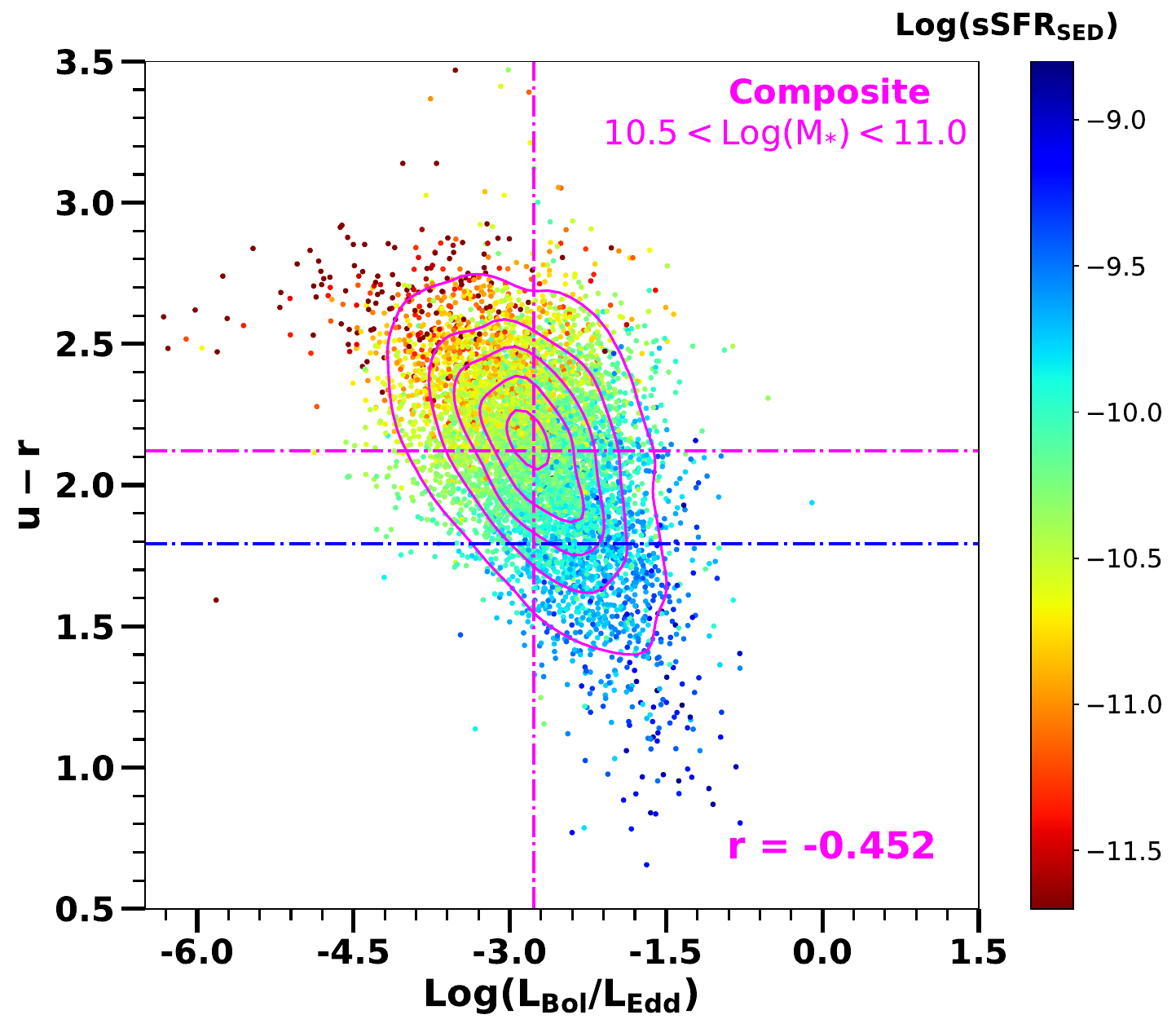}
    \includegraphics[width=0.33\textwidth]{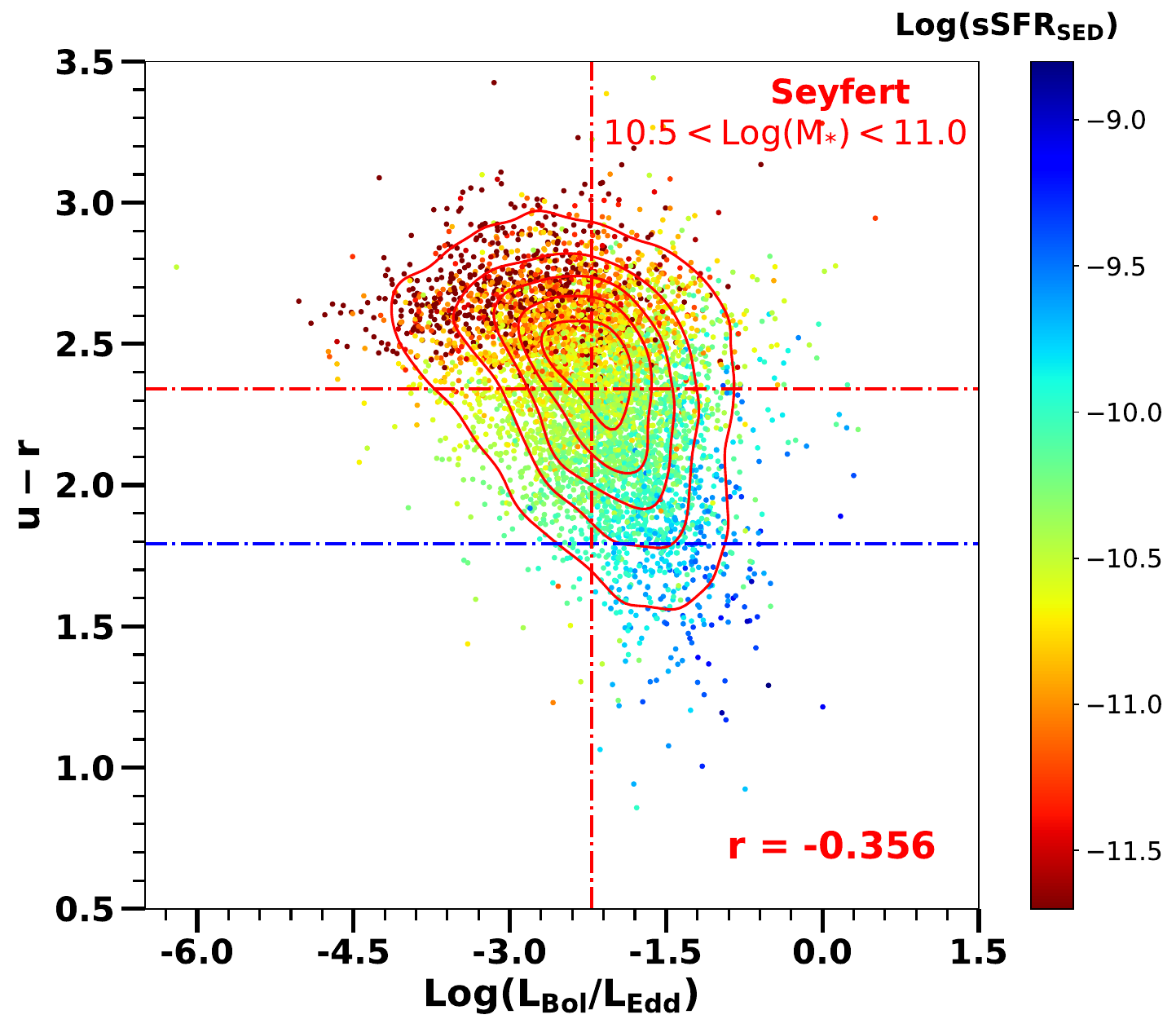}
    \includegraphics[width=0.33\textwidth]{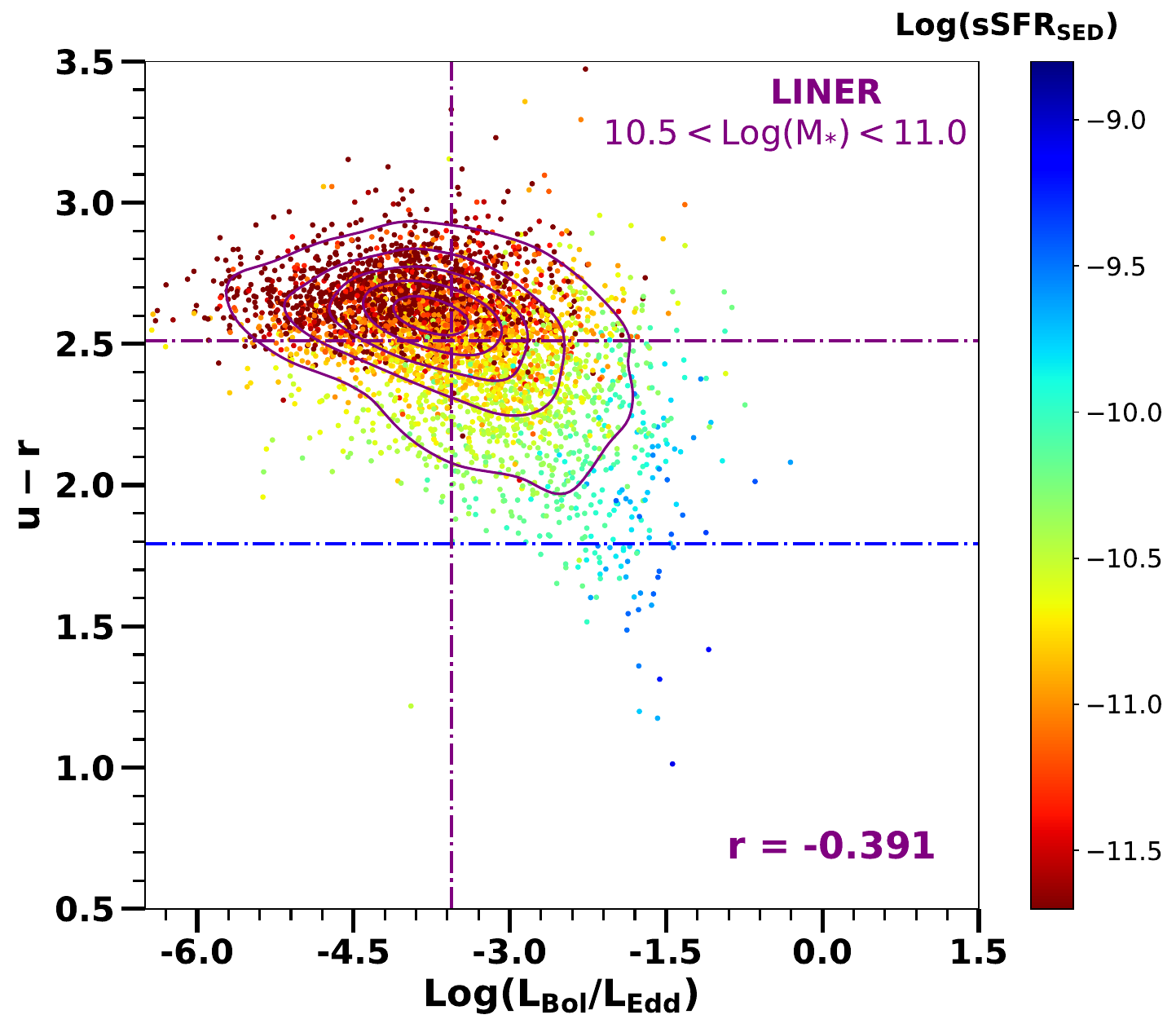}
    \includegraphics[width=0.33\textwidth]{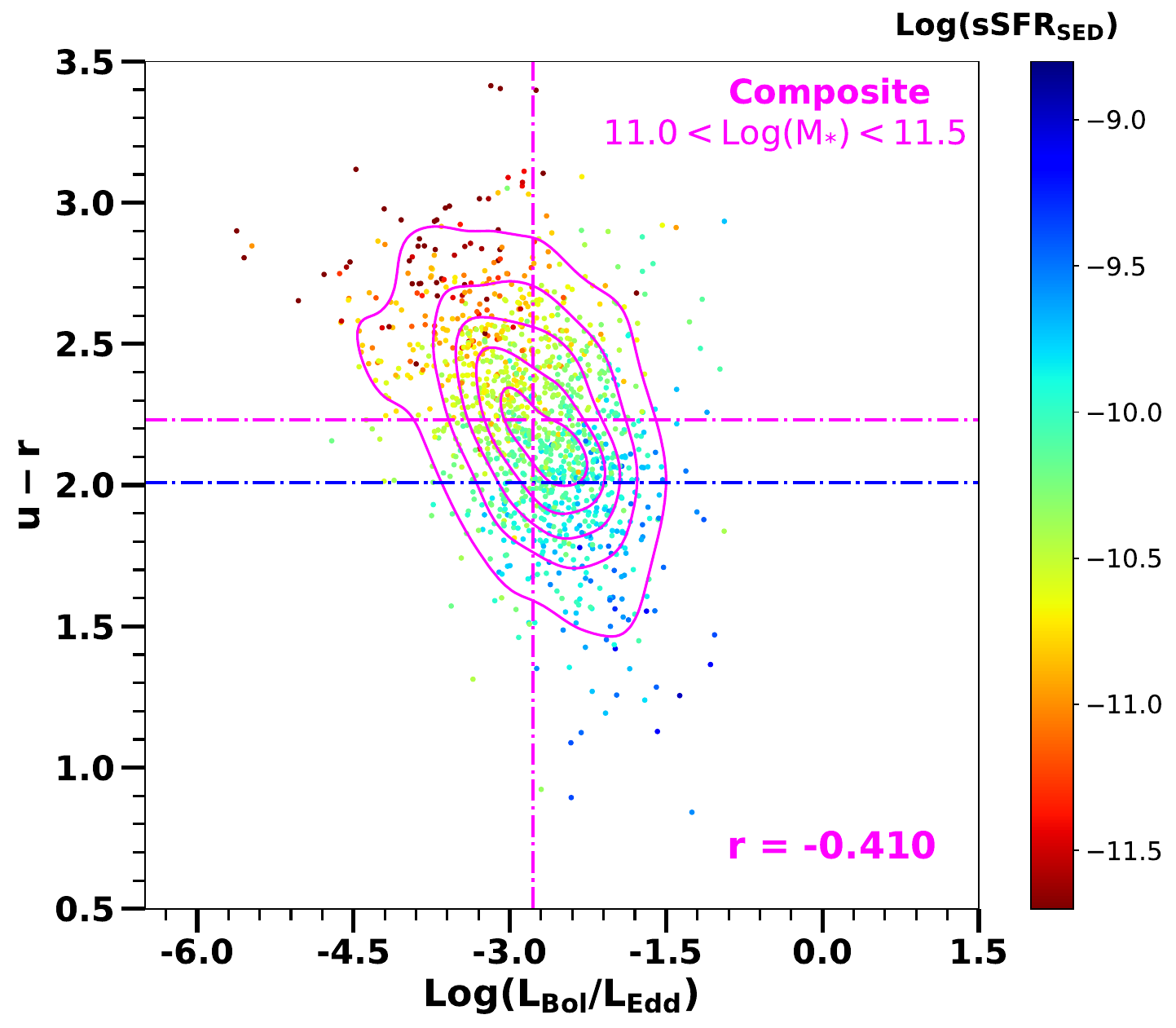}
    \includegraphics[width=0.33\textwidth]{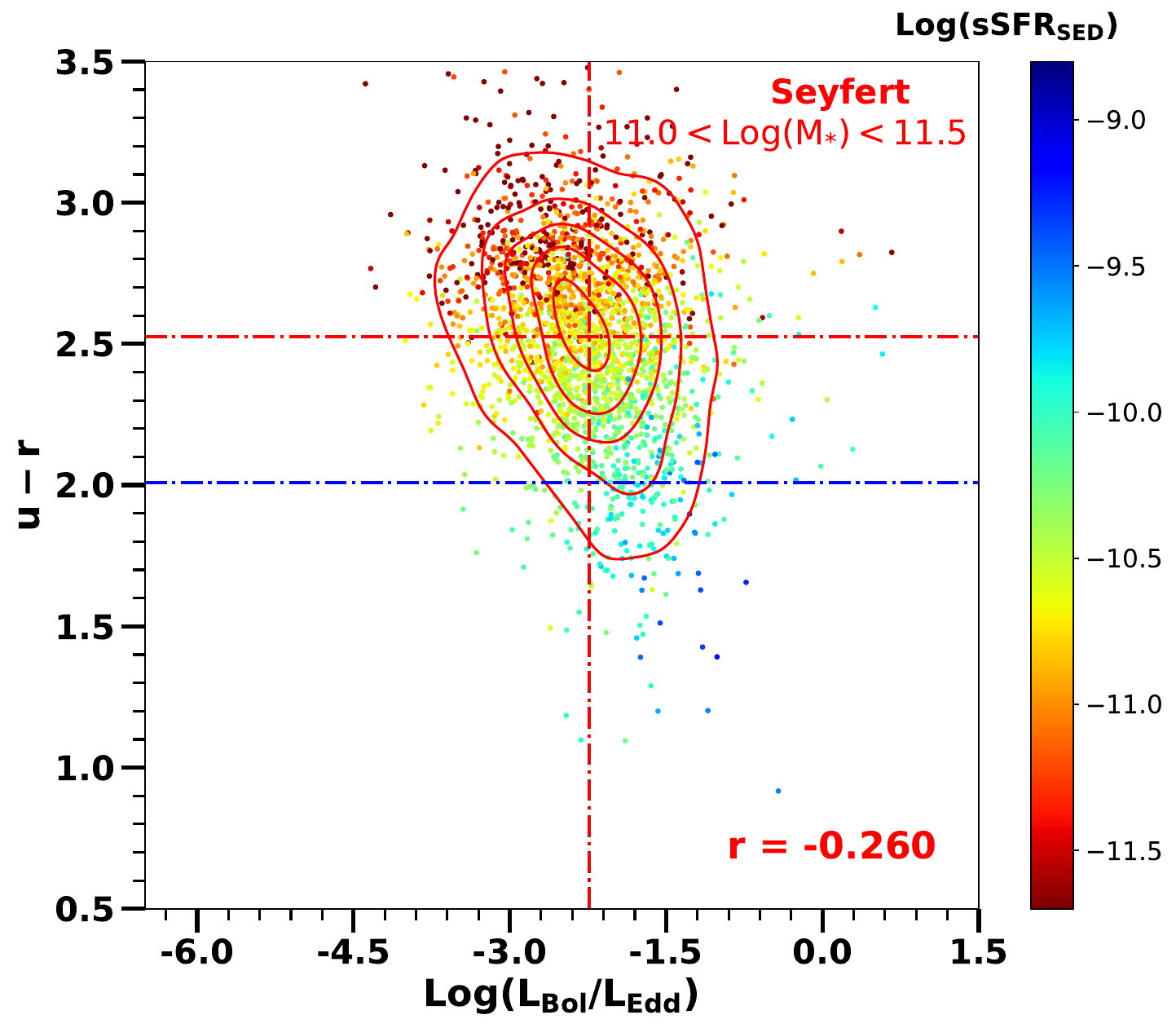}
    \includegraphics[width=0.33\textwidth]{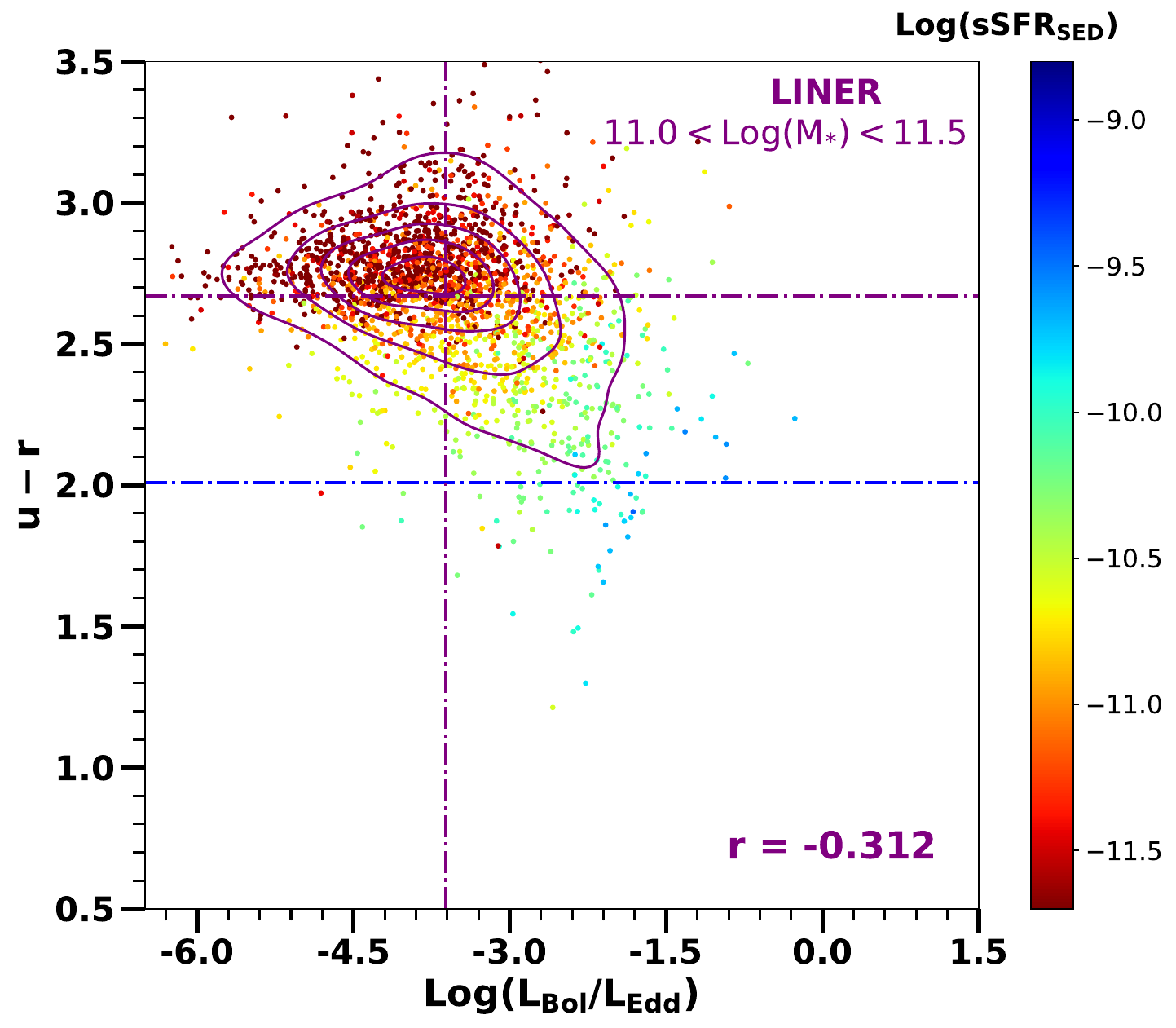}
    \caption{Similar to Figure \ref{fig:color_edd_ssfr}, but with the stellar mass range divided into four bins of $\rm \log M_{*}$ = 9.5$-$10.0, 10.0$-$10.5, 10.5$-$11.0, and 11.0$-$11.5, respectively. 
    \label{fig:color_edd_ssfr_all}}
\end{figure*}

Additionally, it is crucial to examine the $\rm (u-r)-L_{Bol}/L_{Edd}$ plane for mass-matched samples, as this allows us to fully account for AGN activity as a function of stellar mass (e.g., \citealp{Xue+10}). In Figure \ref{fig:color_edd_ssfr_all}, we divide the stellar mass into four different bins, each with a size of $\rm \Delta\log M_{*}$ $=$ 0.5, to examine the correlation between $\rm (u-r)$ colors and the Eddington ratio for galaxies within similar stellar mass ranges. Within each stellar mass bin, we observe consistent correlations between ($\rm u-r$) color and the Eddington ratio across galaxy types. In these correlations, $\rm (u-r)$ colors reach their highest values in LINER galaxies, while Seyfert galaxies exhibit the highest Eddington ratio values. Additionally, sSFRs are elevated in sources with a high Eddington ratio. As stellar mass increases, $\rm (u-r)$ colors tend to become redder, and the Eddington ratio values also increase. Seyfert galaxies consistently show the strongest Eddington ratios compared to composite and LINER galaxies across all stellar mass bins.

\subsection{Radio Luminosity versus Eddington Ratio}\label{subsec:radio_edd}

As mentioned in Section \ref{section:meas}, we cross-matched our sample with the FIRST catalog, identifying 4,877 sources with measured radio luminosity ($\rm L_{1.4GHz}$). These detected radio sources present an intriguing opportunity to study the connection between radio and AGN activities, as well as SF activity. We aim to explore potential relationships between radio luminosity, Eddington ratio, and SFRs in these AGNs.

\begin{figure*}
\centering
	\includegraphics[width=0.23\textwidth]{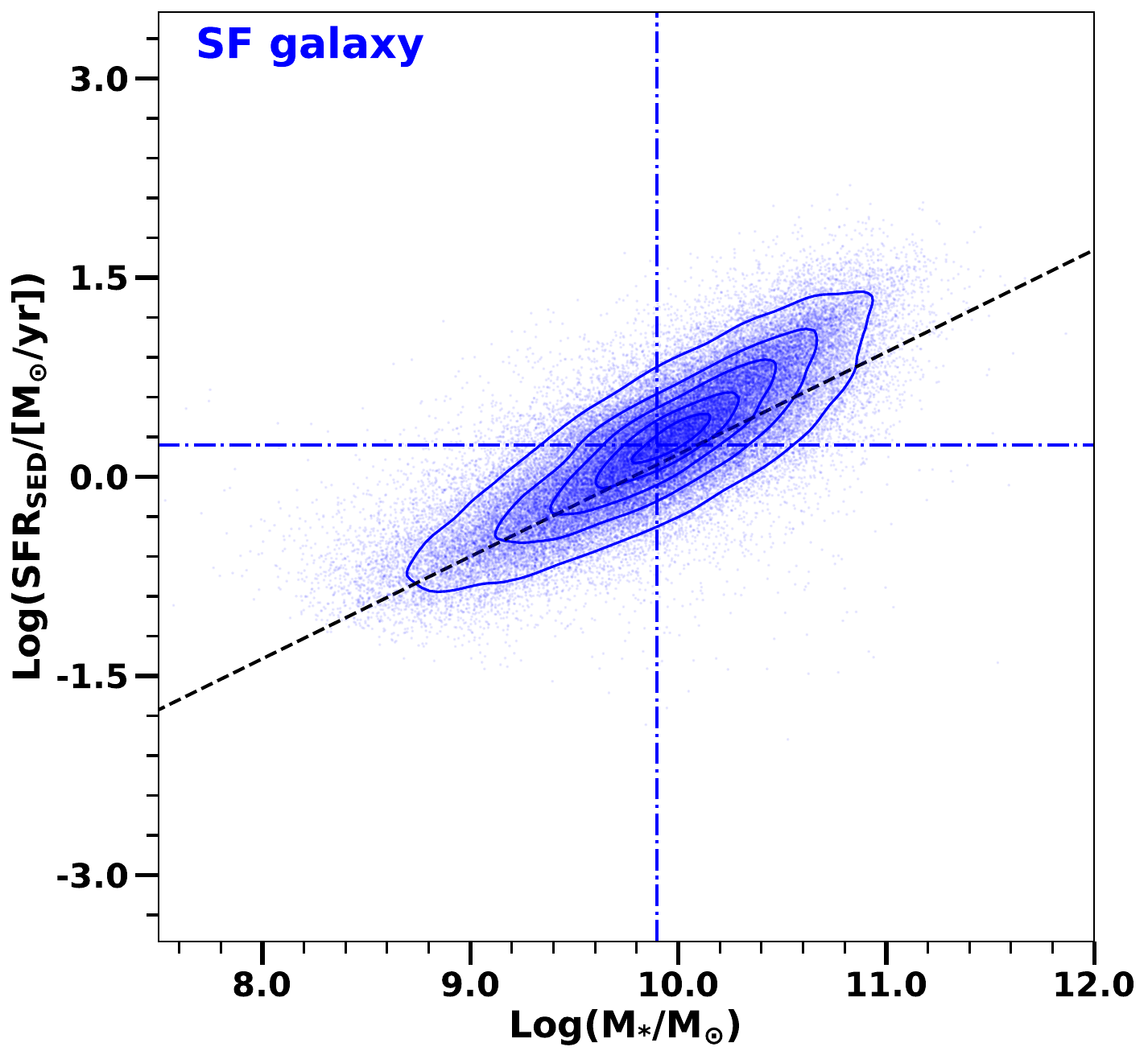}
	\includegraphics[width=0.25\textwidth]{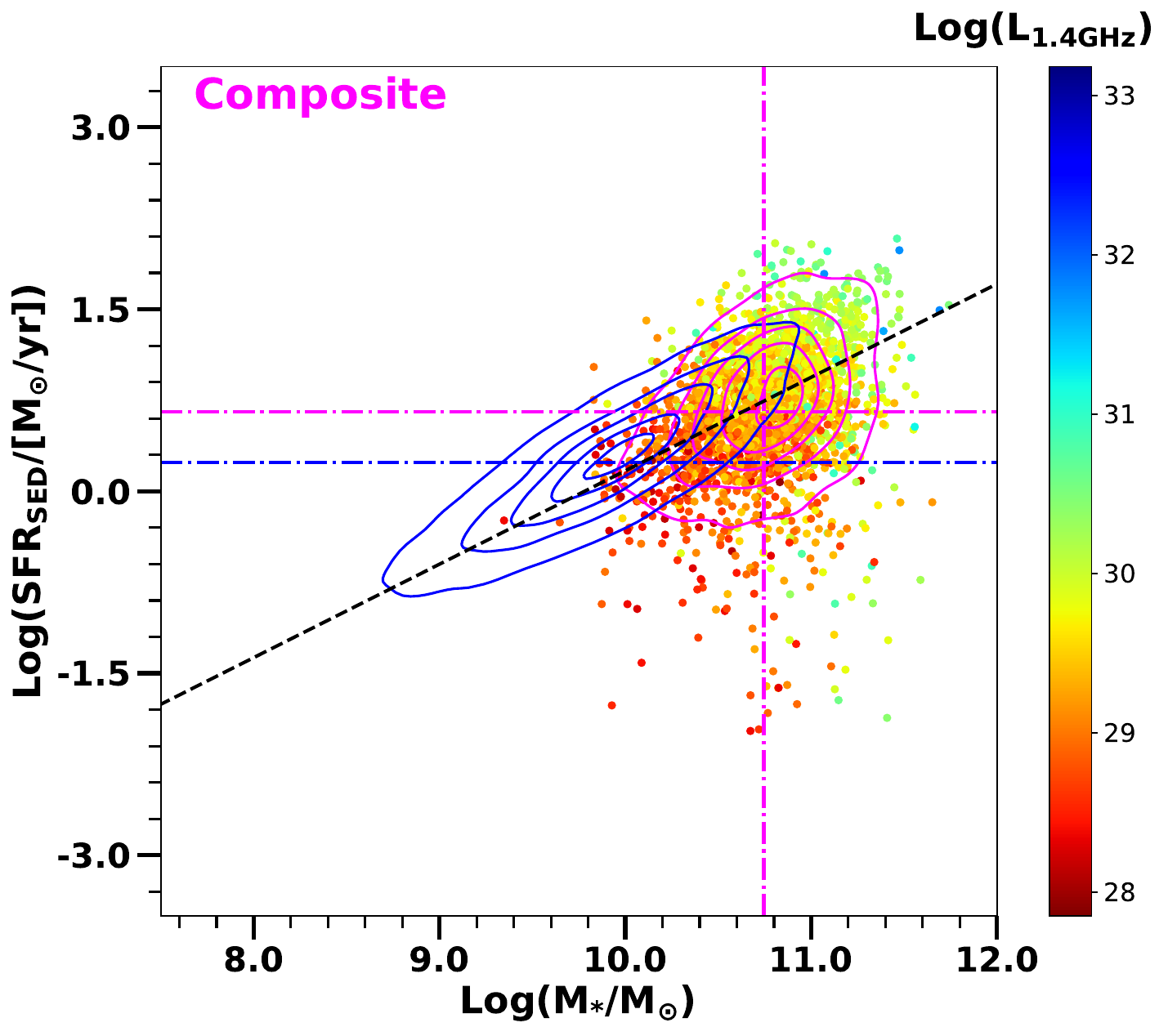}
	\includegraphics[width=0.25\textwidth]{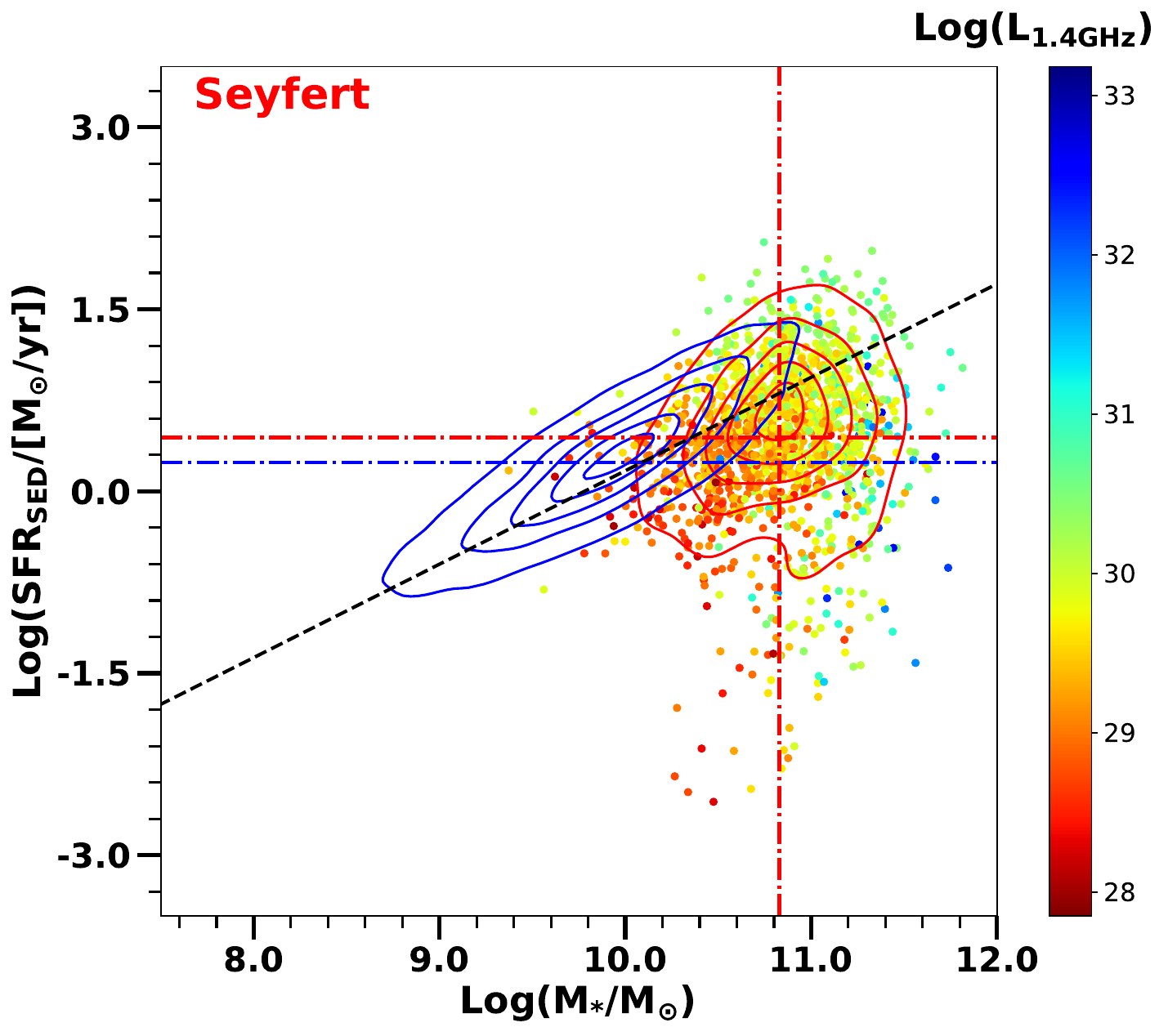}
	\includegraphics[width=0.25\textwidth]{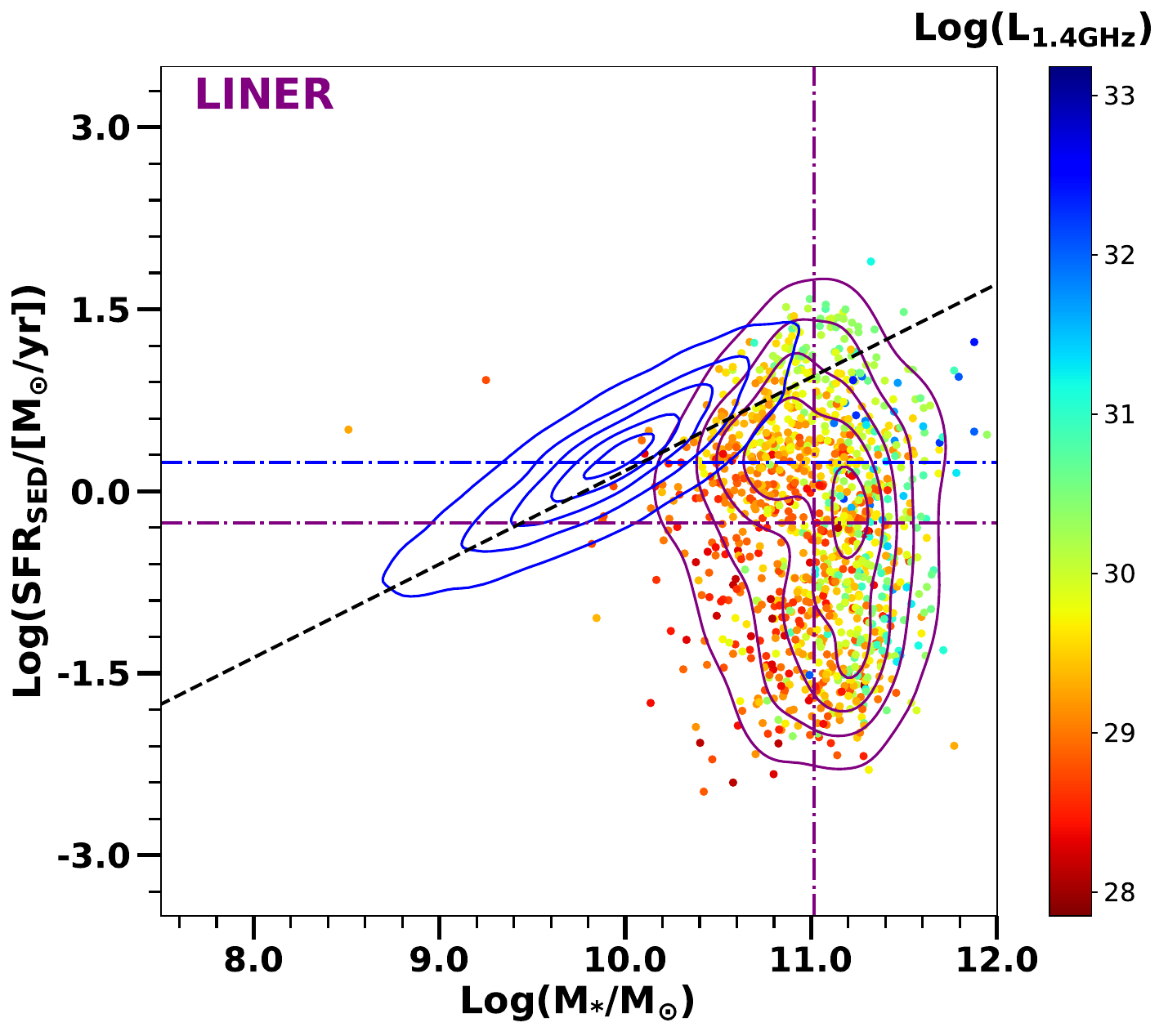}
	\includegraphics[width=0.23\textwidth]{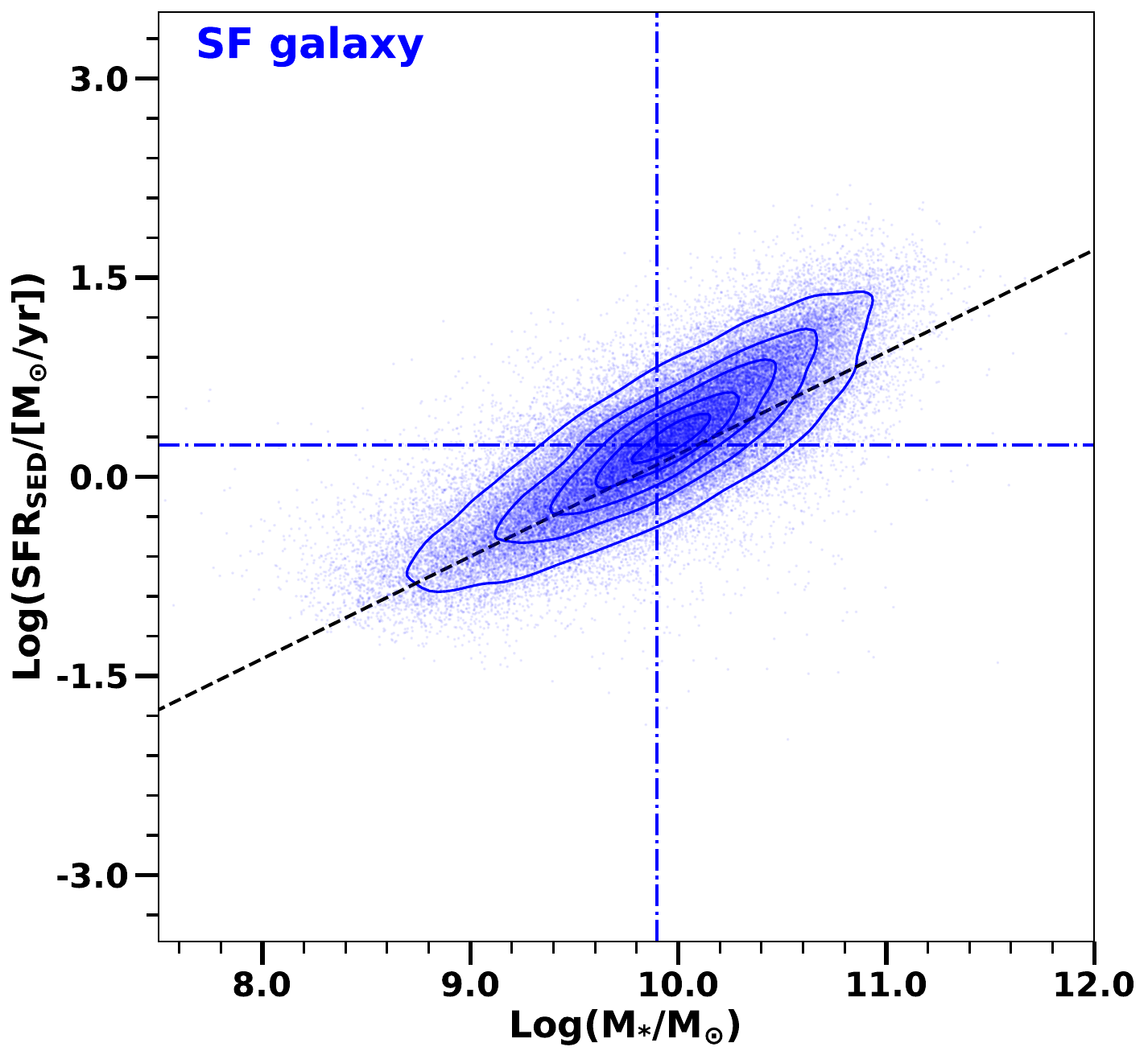}
	\includegraphics[width=0.25\textwidth]{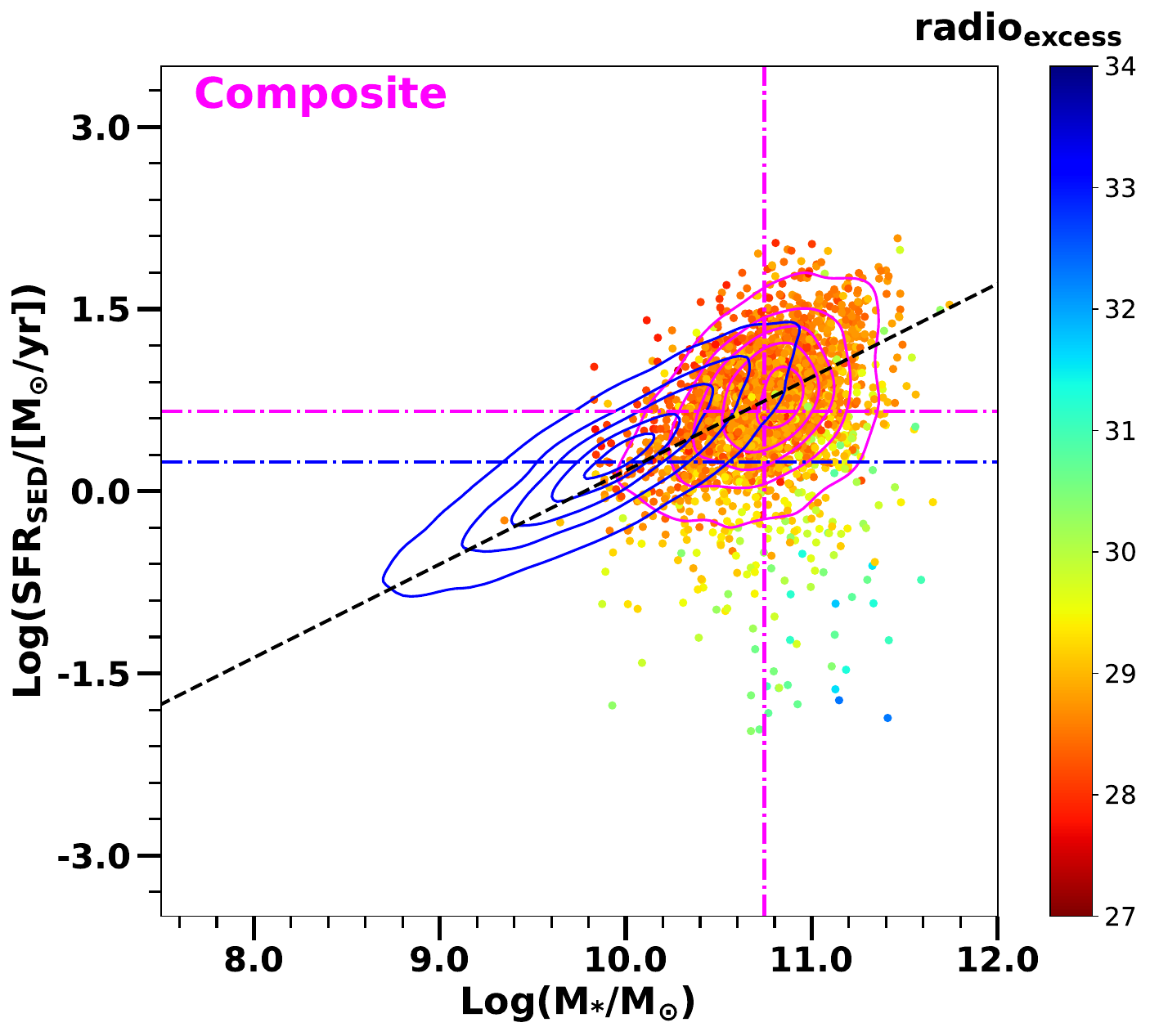}
	\includegraphics[width=0.25\textwidth]{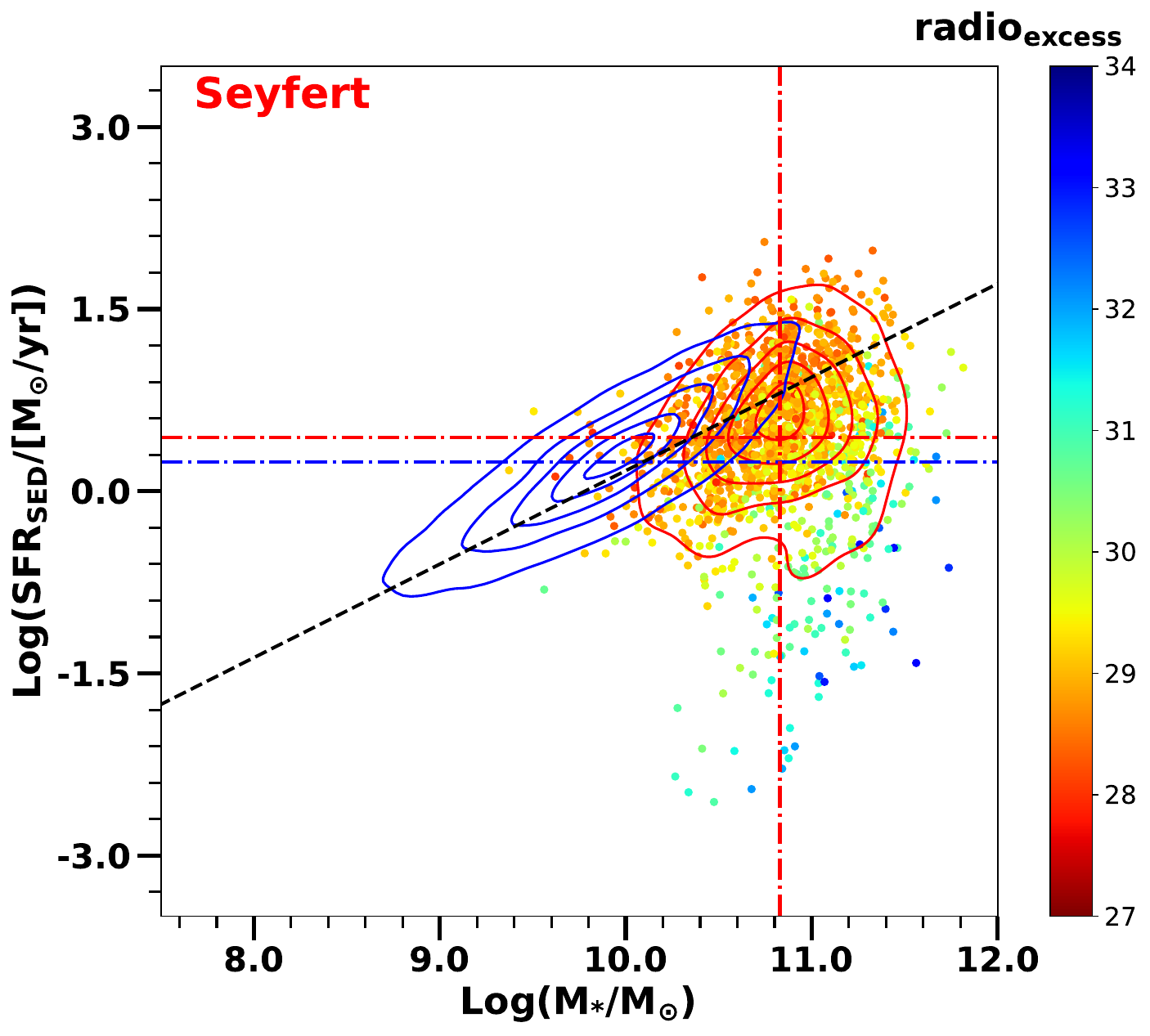}
	\includegraphics[width=0.25\textwidth]{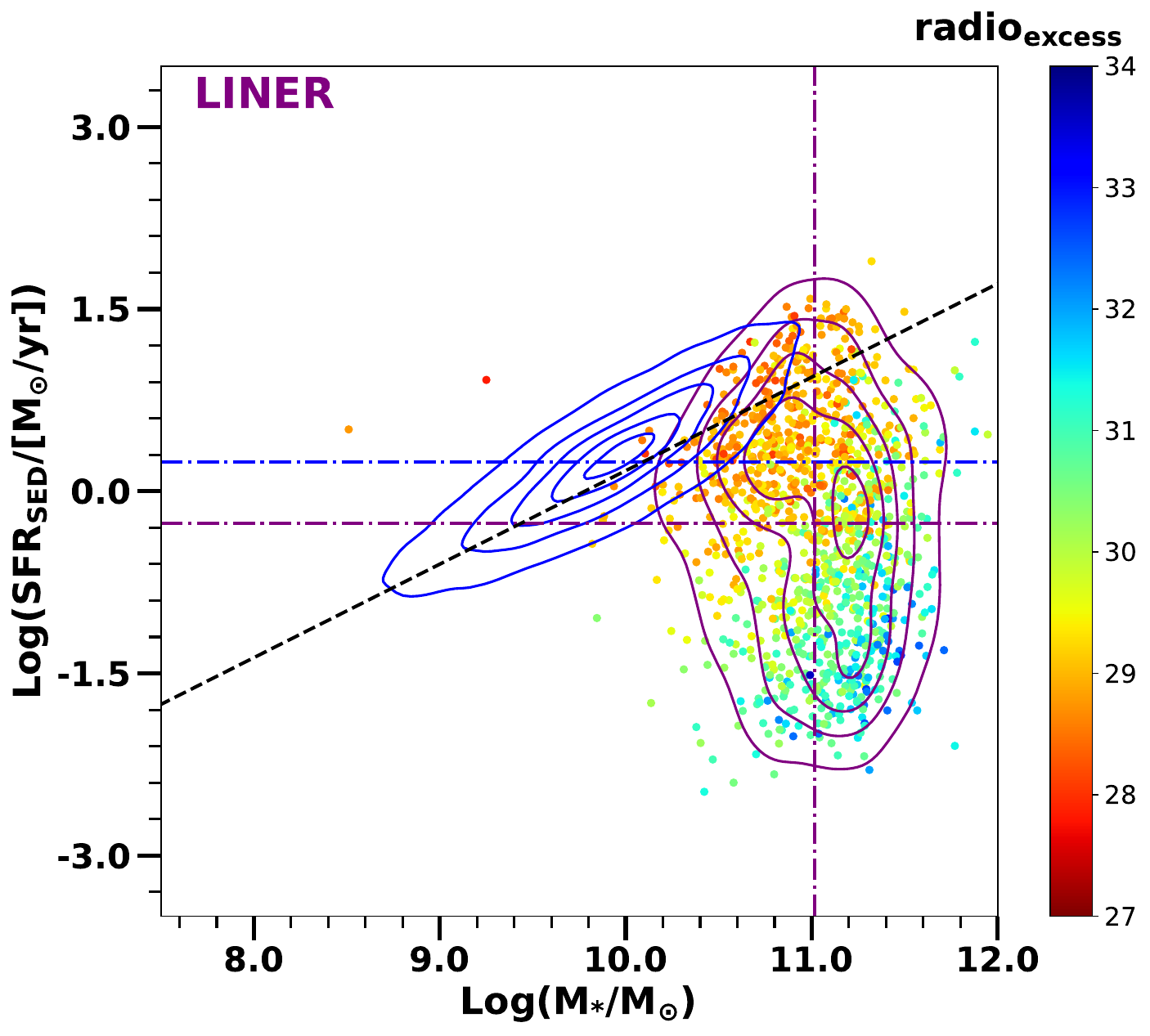}
	\includegraphics[width=0.23\textwidth]{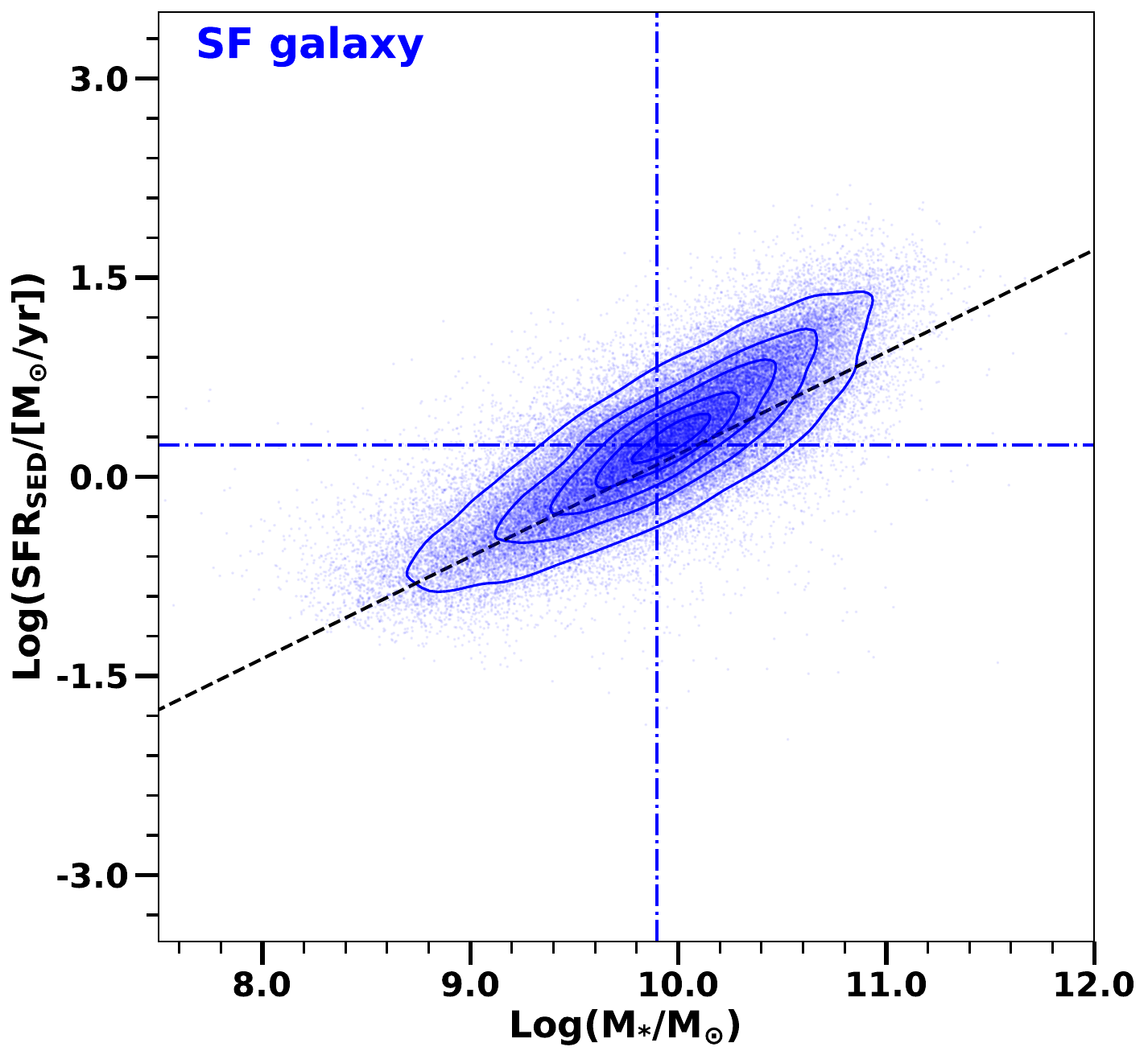}
	\includegraphics[width=0.25\textwidth]{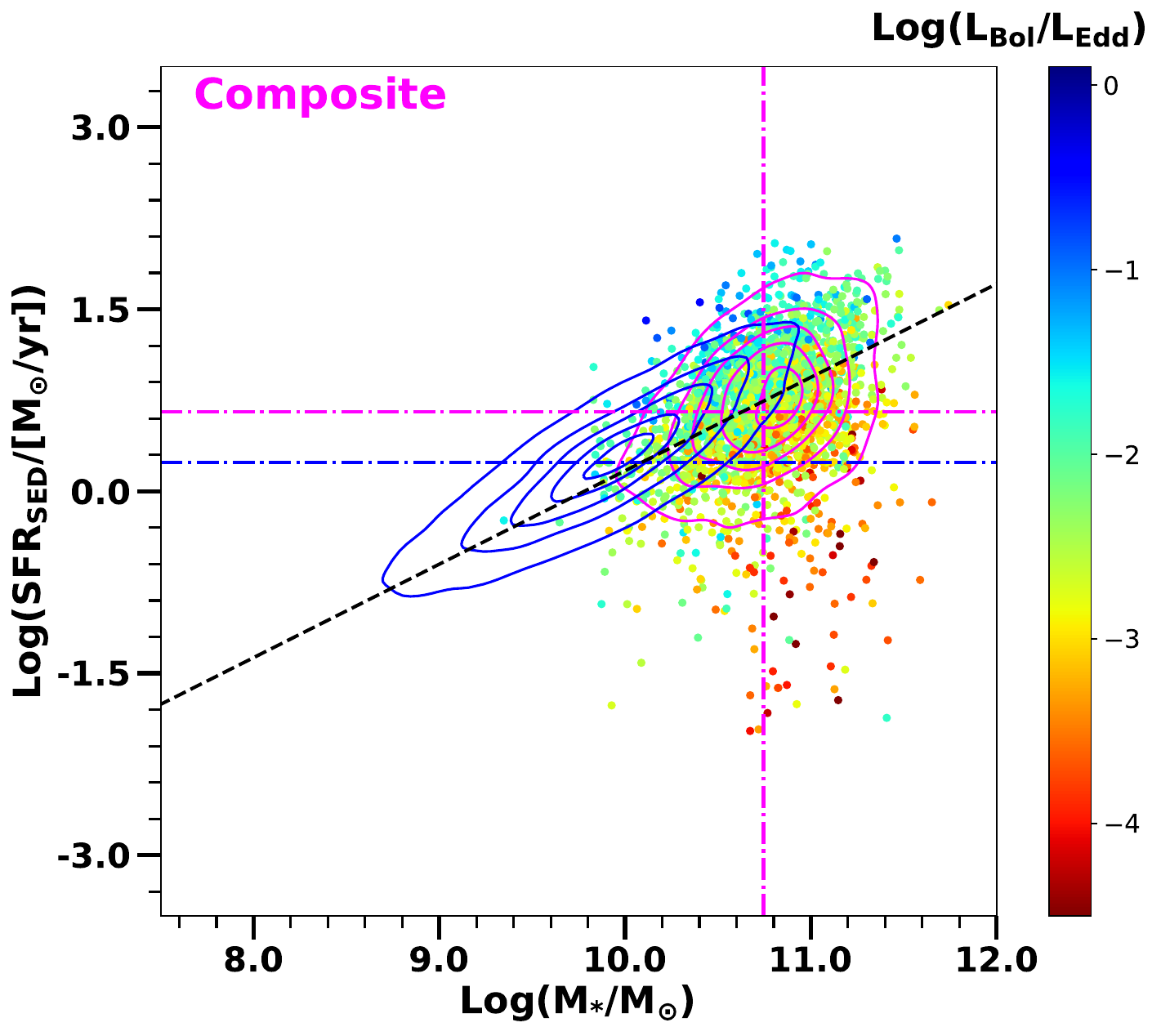}
	\includegraphics[width=0.25\textwidth]{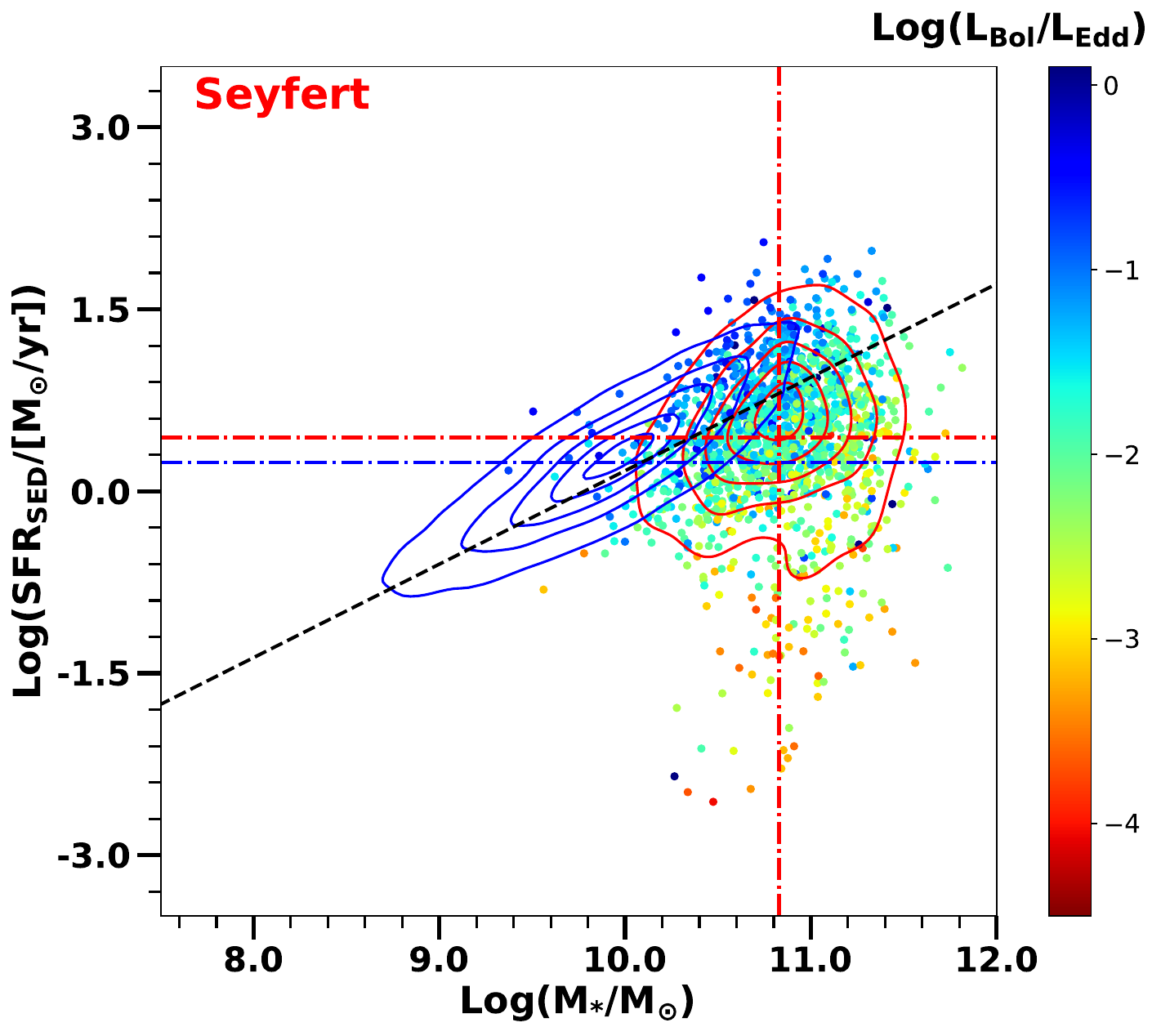}
	\includegraphics[width=0.25\textwidth]{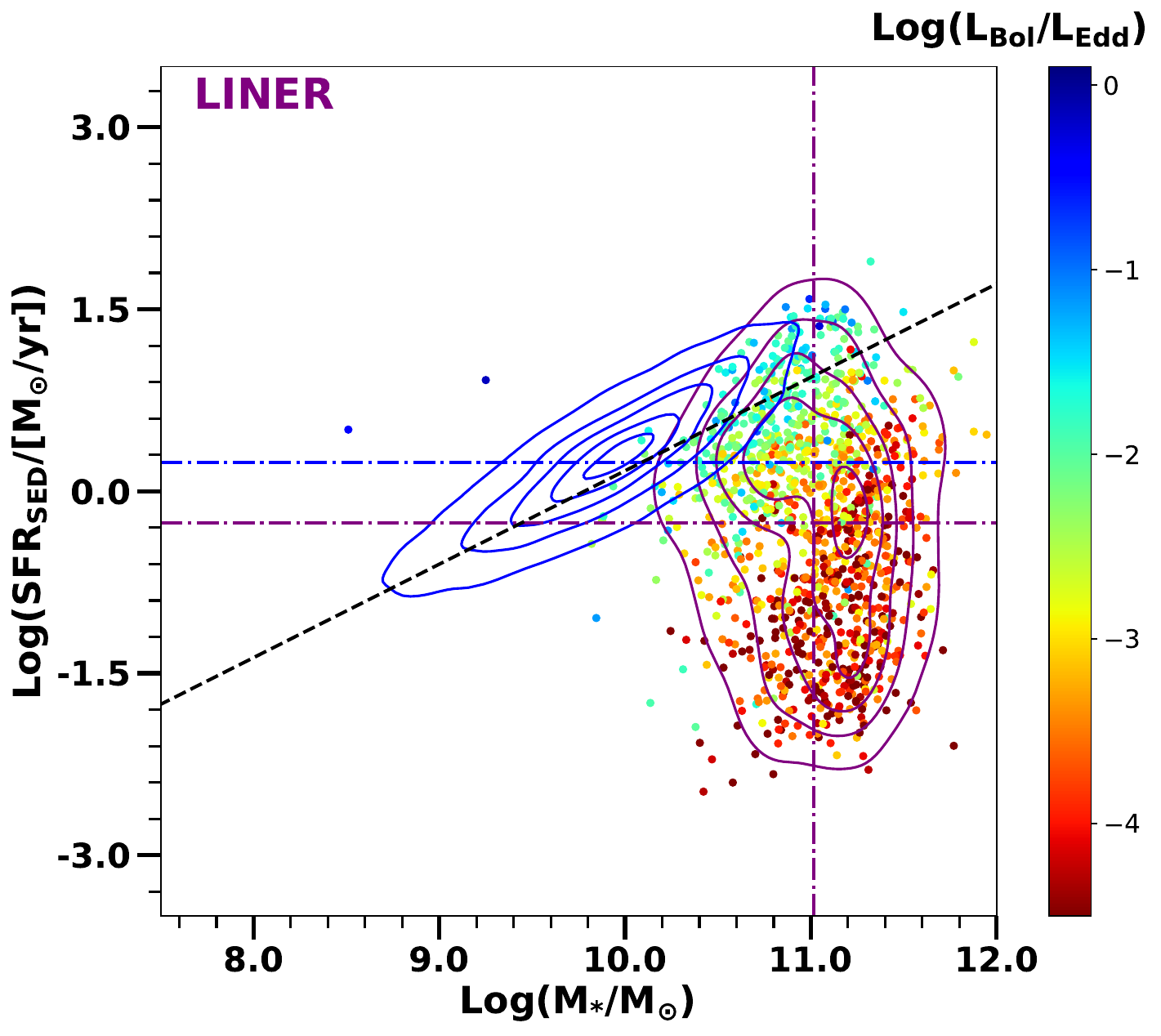}
	\caption{Similar to Figure \ref{fig:sfr_mass_ur}, but the color scales display the radio luminosity $\rm L_{1.4GHz}$ (top panel), the radio excess $\rm L_{1.4GHz}/SFR_{SED}$ (middle panel), and Eddington ratio $\rm L_{Bol}/L_{Edd}$ (bottom panel), respectively. 
	\label{fig:color_contour_radio}}
\end{figure*}

Figure \ref{fig:color_contour_radio} displays the $\rm SFR-M_{*}$ plane for these detected radio sources across different galaxy types. We also analyze the variation trends of radio luminosity ($\rm L_{1.4GHz}$), the radio excess ($\rm L_{1.4GHz}/SFR_{SED}$, e.g., \citealp{Delvecchio+17}) and the Eddington ratio ($\rm L_{Bol}/L_{Edd}$) within the $\rm SFR-M_{*}$ plane, using color scales to illustrate these trends. 

We found that both radio luminosity and Eddington ratios strongly correlate with SFRs. However, their variation trends on the $\rm SFR-M_{*}$ plane differ somewhat. The Eddington ratio remains high for sources lying along the MS line but decreases for sources moving vertically downward from the MS line, where SFRs are also lower. In contrast, radio luminosity tends to increase along the MS line and with increasing stellar mass. Sources with lower stellar mass exhibit low radio luminosity, while those with higher stellar mass have higher radio luminosity. Interestingly, the distribution of radio excess on the $\rm SFR-M_{*}$ plane shows an increasing trend perpendicular to the MS line, with radio excess being low for sources along the MS line and increasing for sources moving vertically downward, where SFRs are lower.

\begin{figure*}[h]
\centering
	\includegraphics[width=0.33\textwidth]{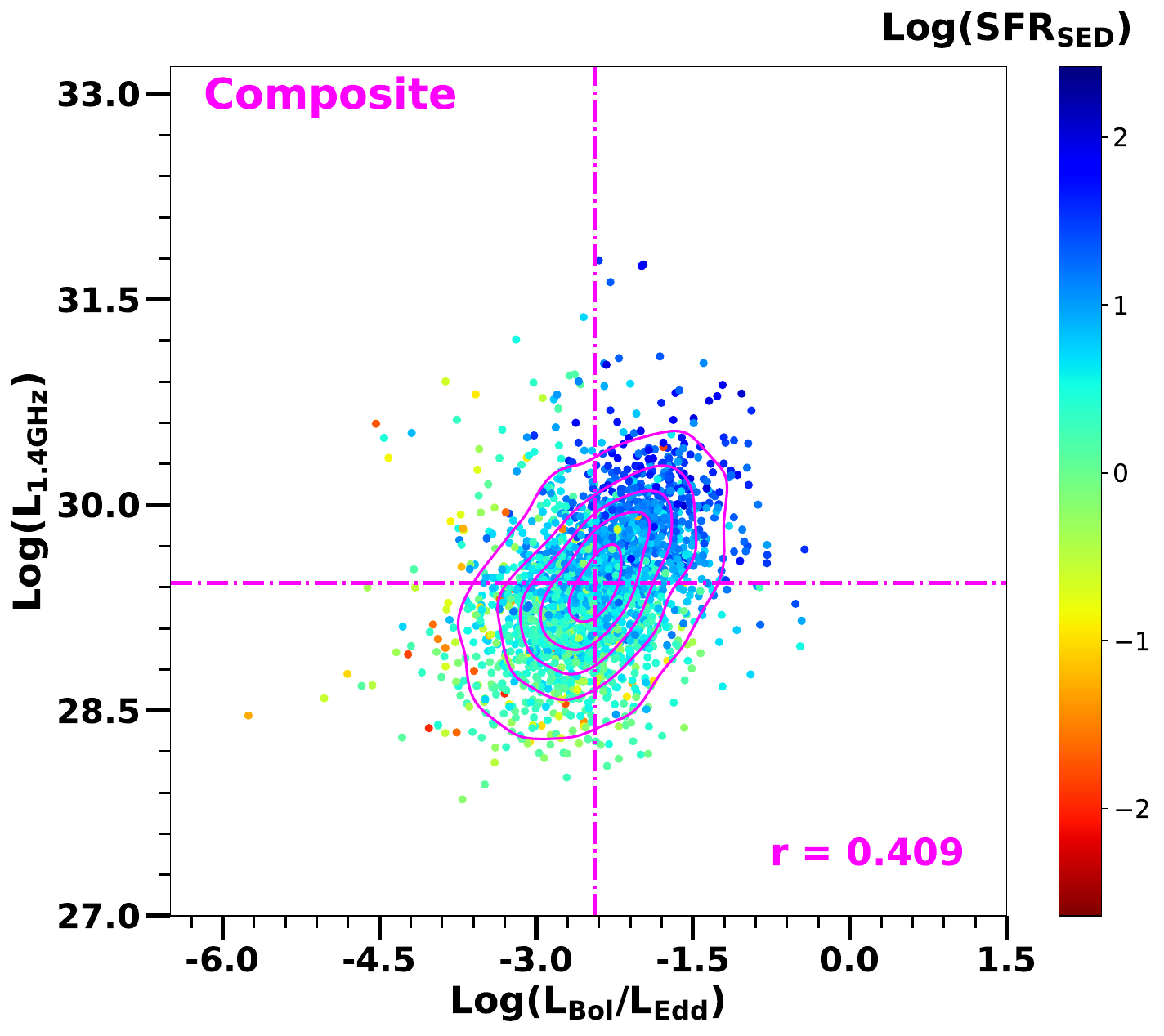}
	\includegraphics[width=0.33\textwidth]{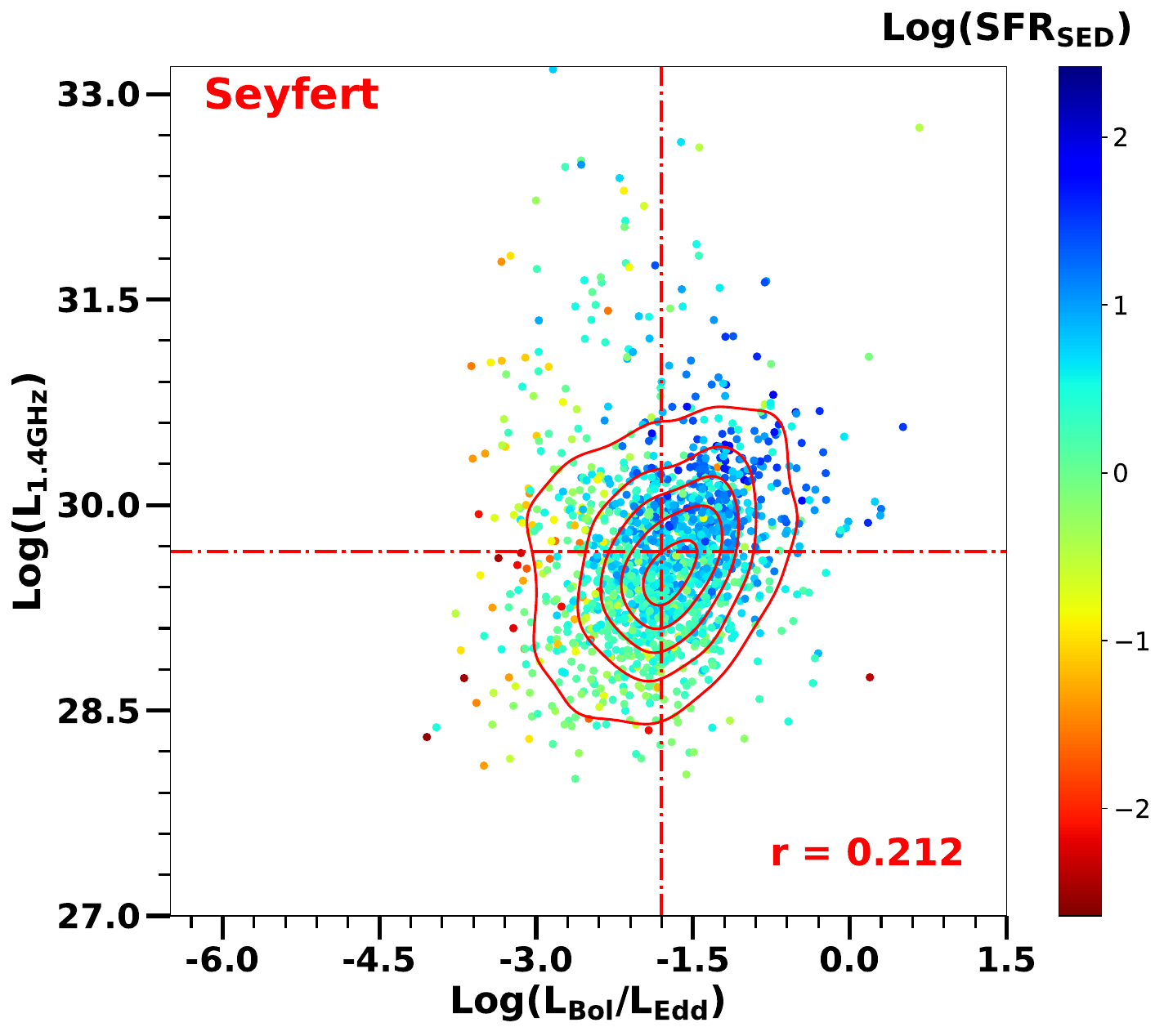}
	\includegraphics[width=0.33\textwidth]{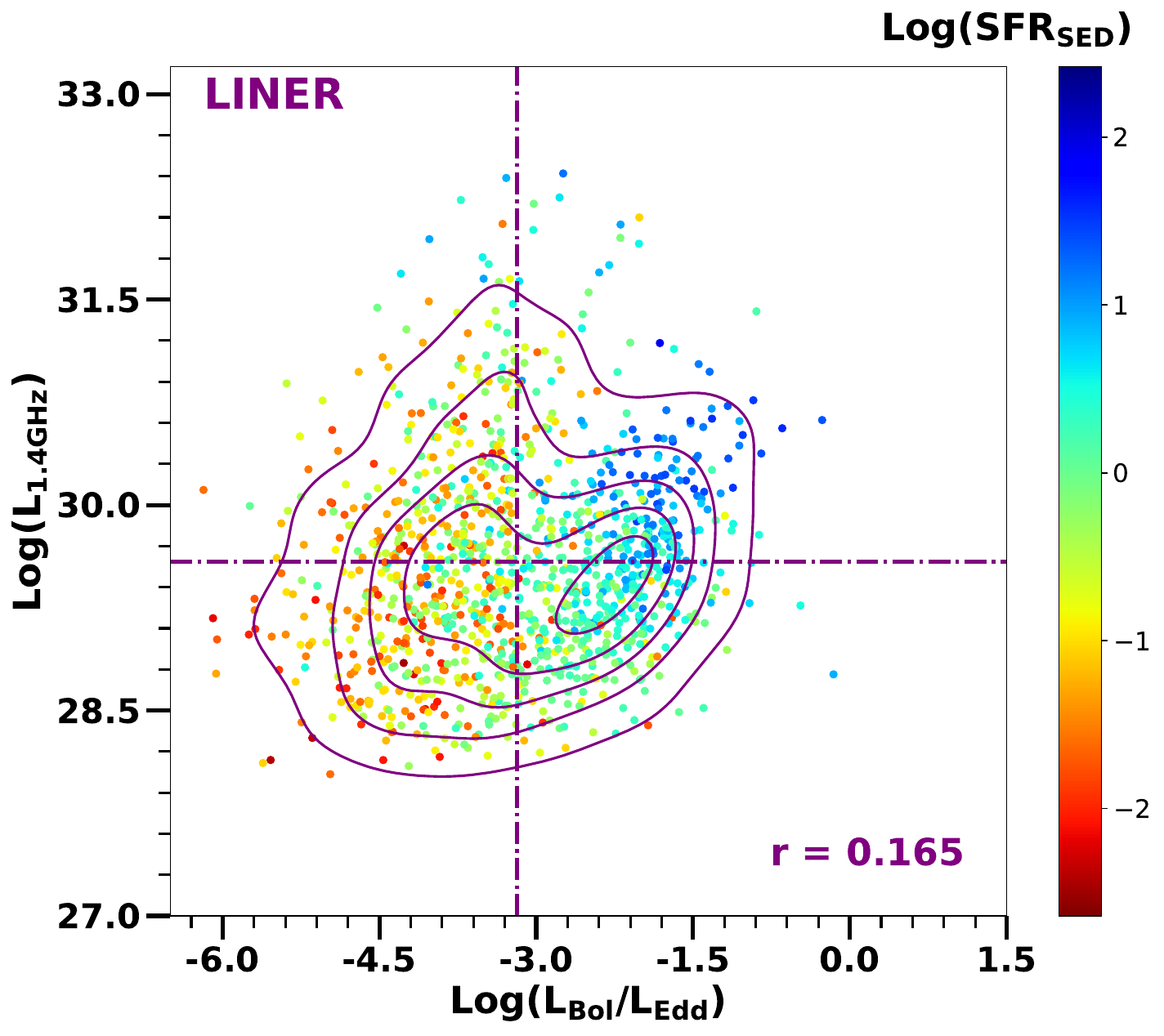}
	\includegraphics[width=0.33\textwidth]{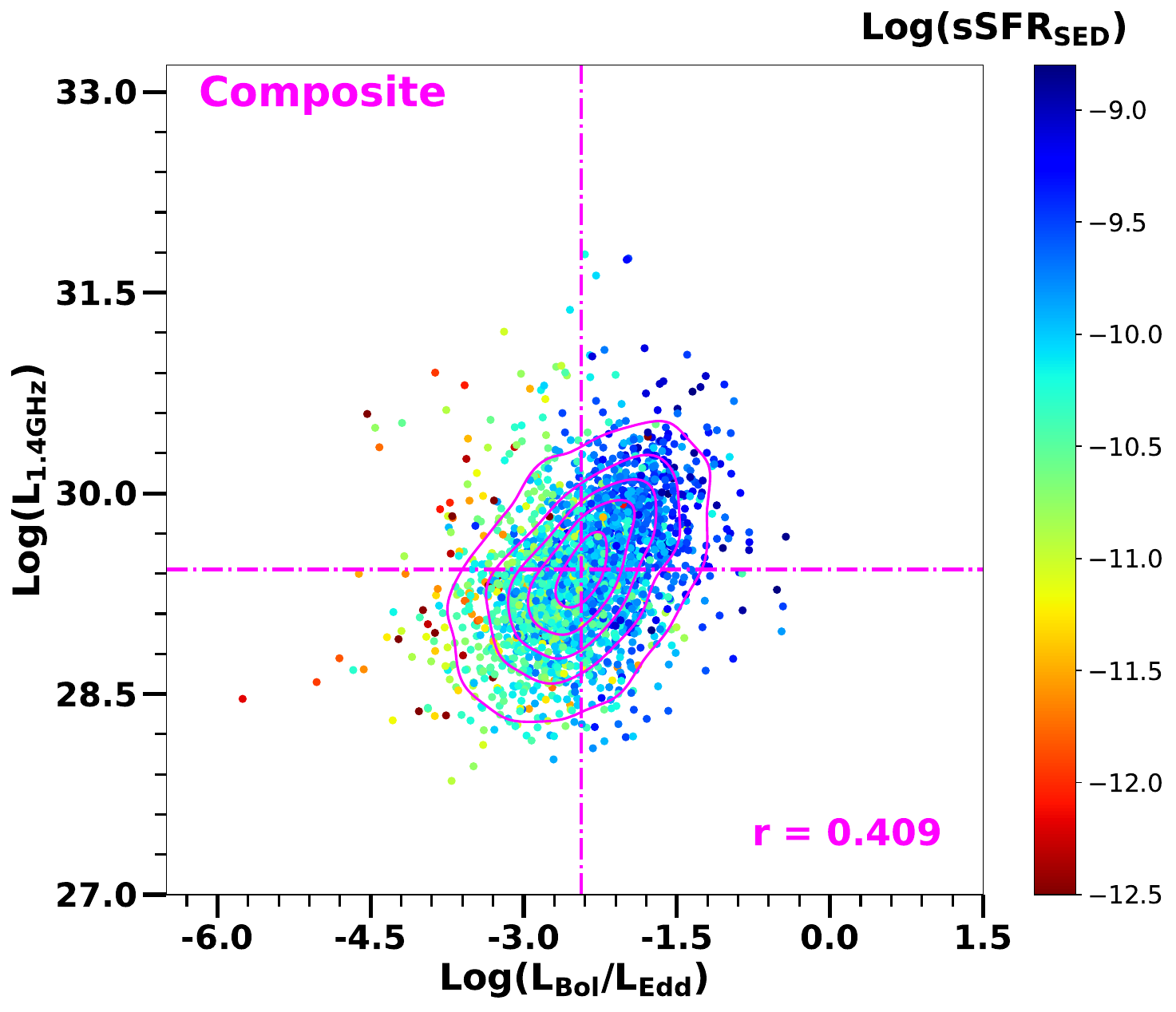}
	\includegraphics[width=0.33\textwidth]{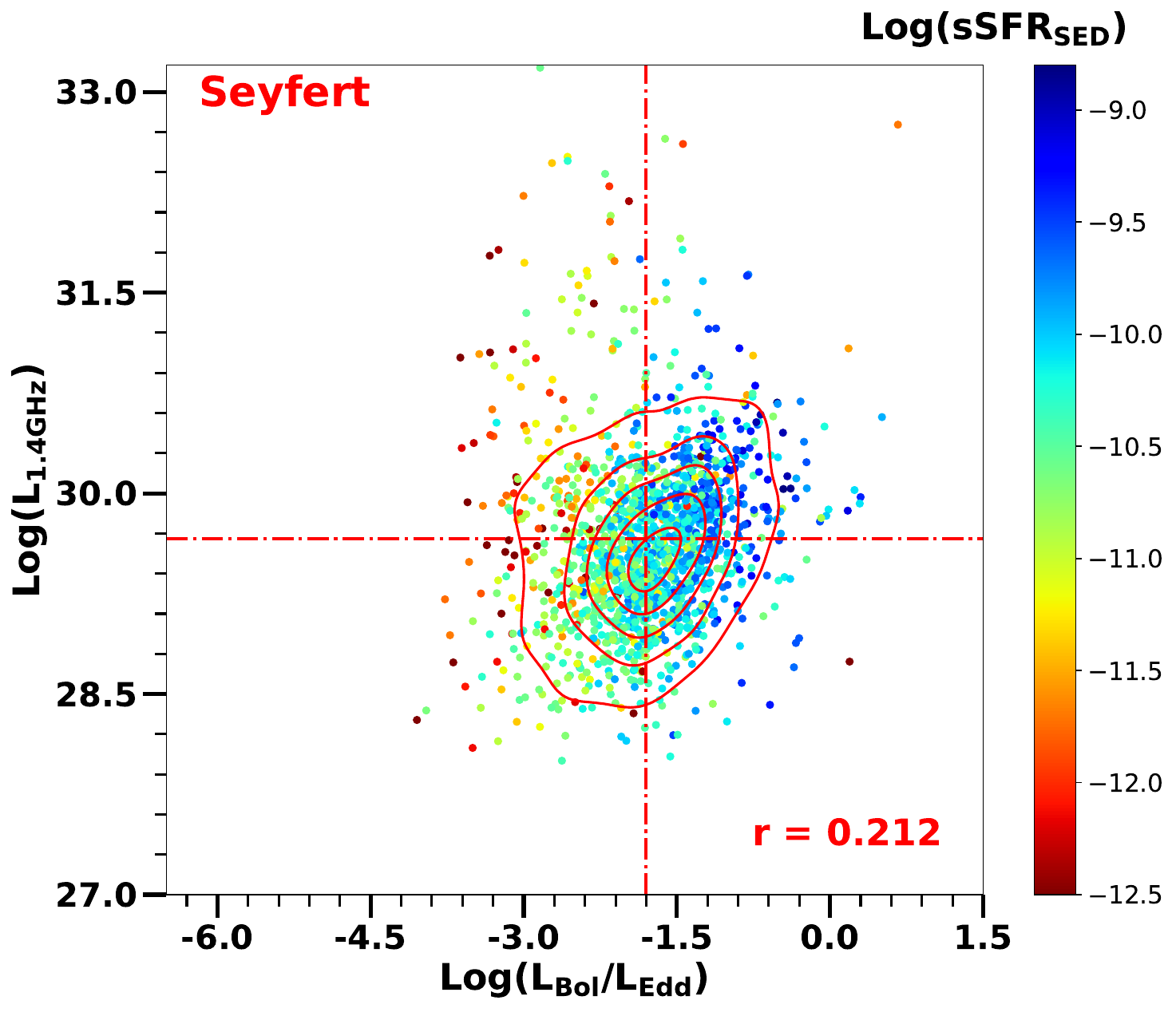}
	\includegraphics[width=0.33\textwidth]{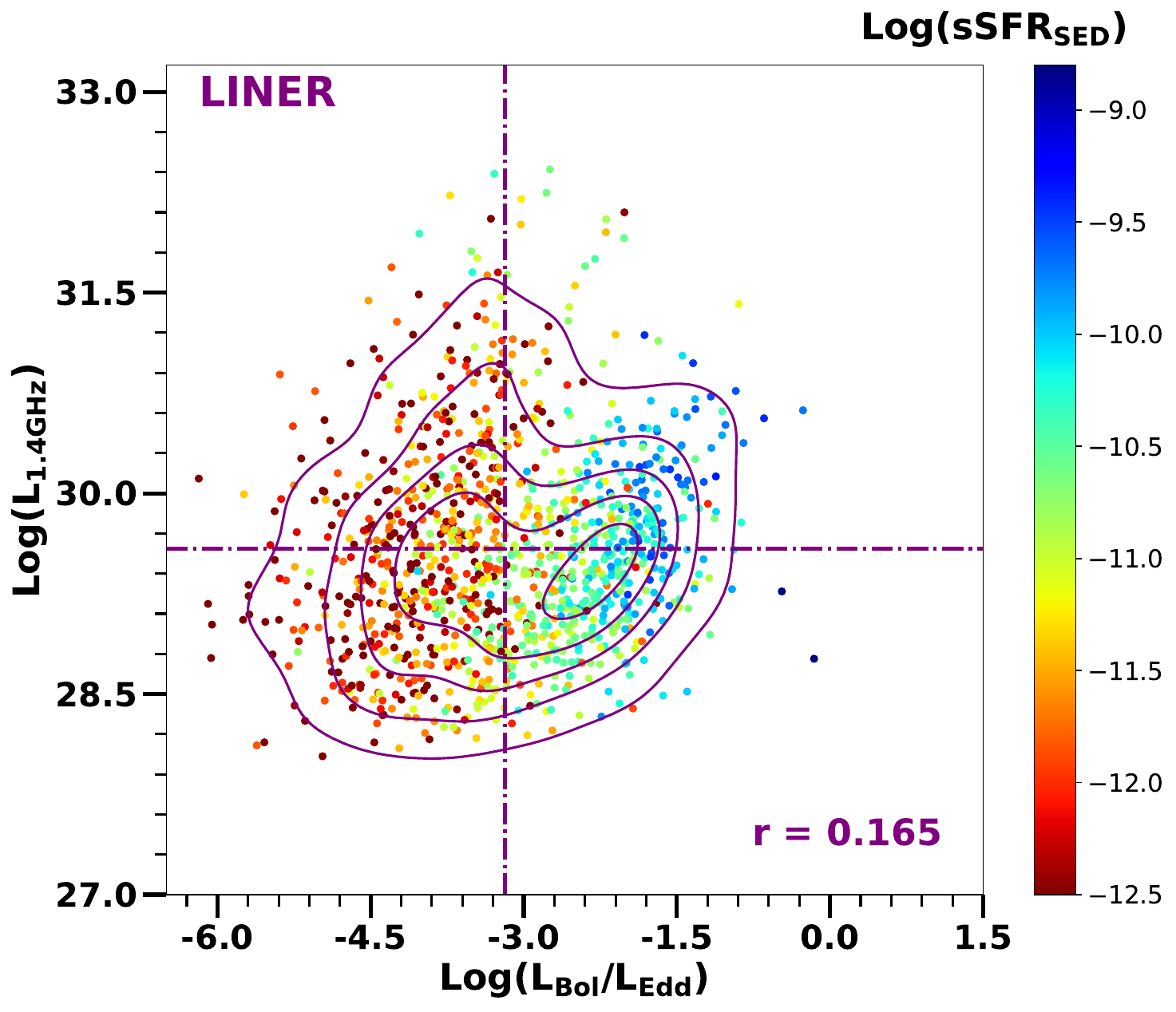}
	\includegraphics[width=0.33\textwidth]{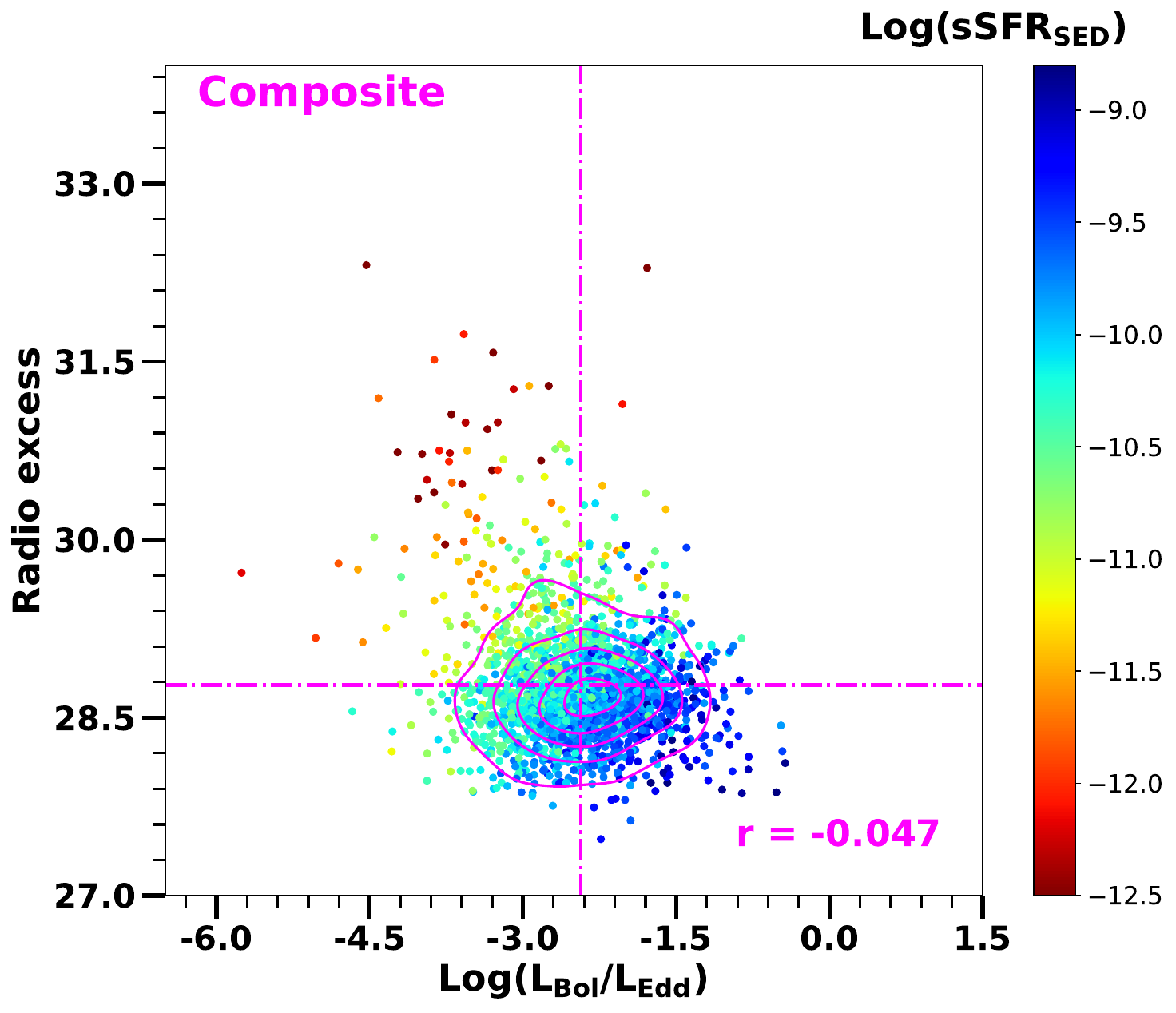}
	\includegraphics[width=0.33\textwidth]{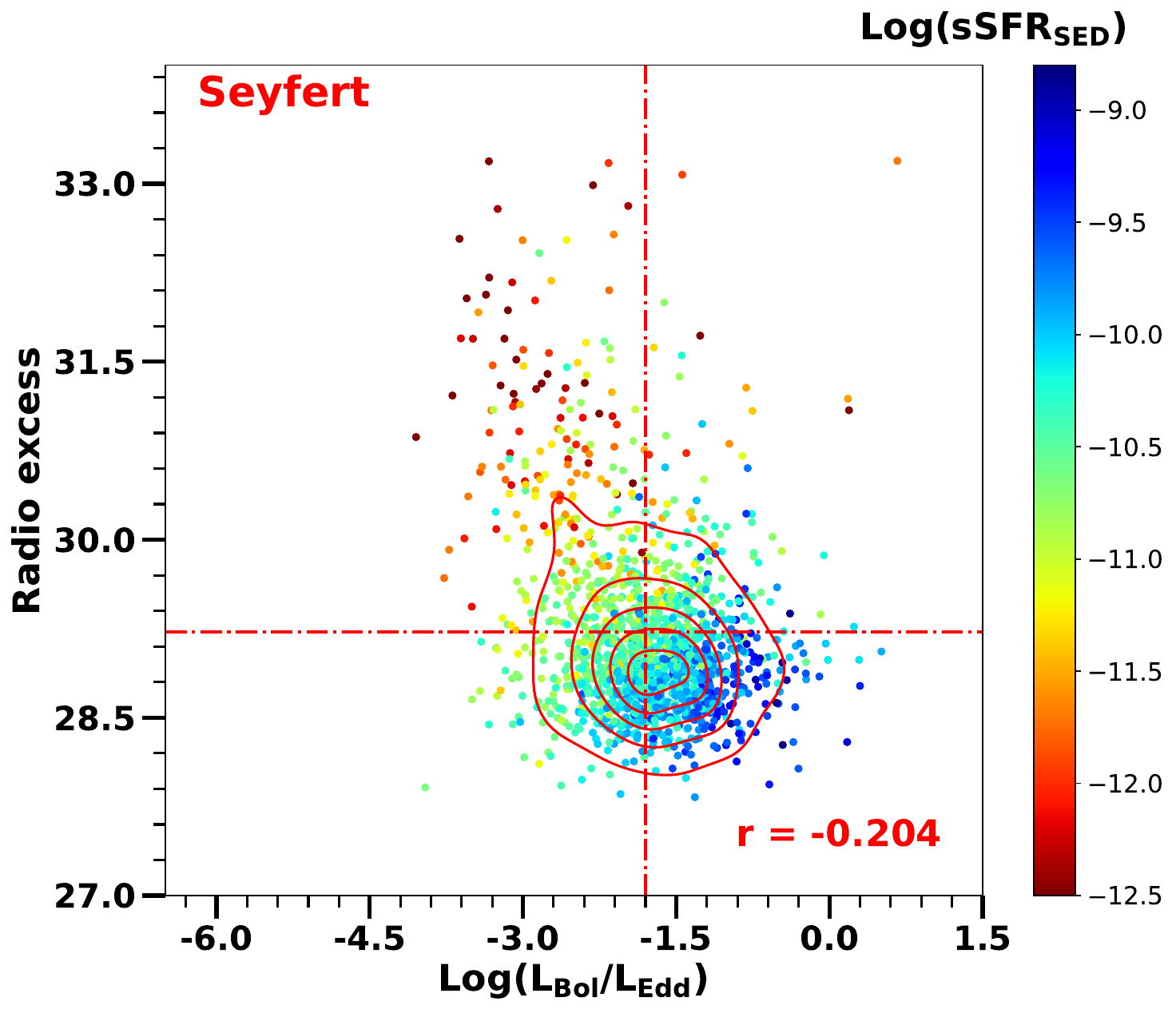}
	\includegraphics[width=0.33\textwidth]{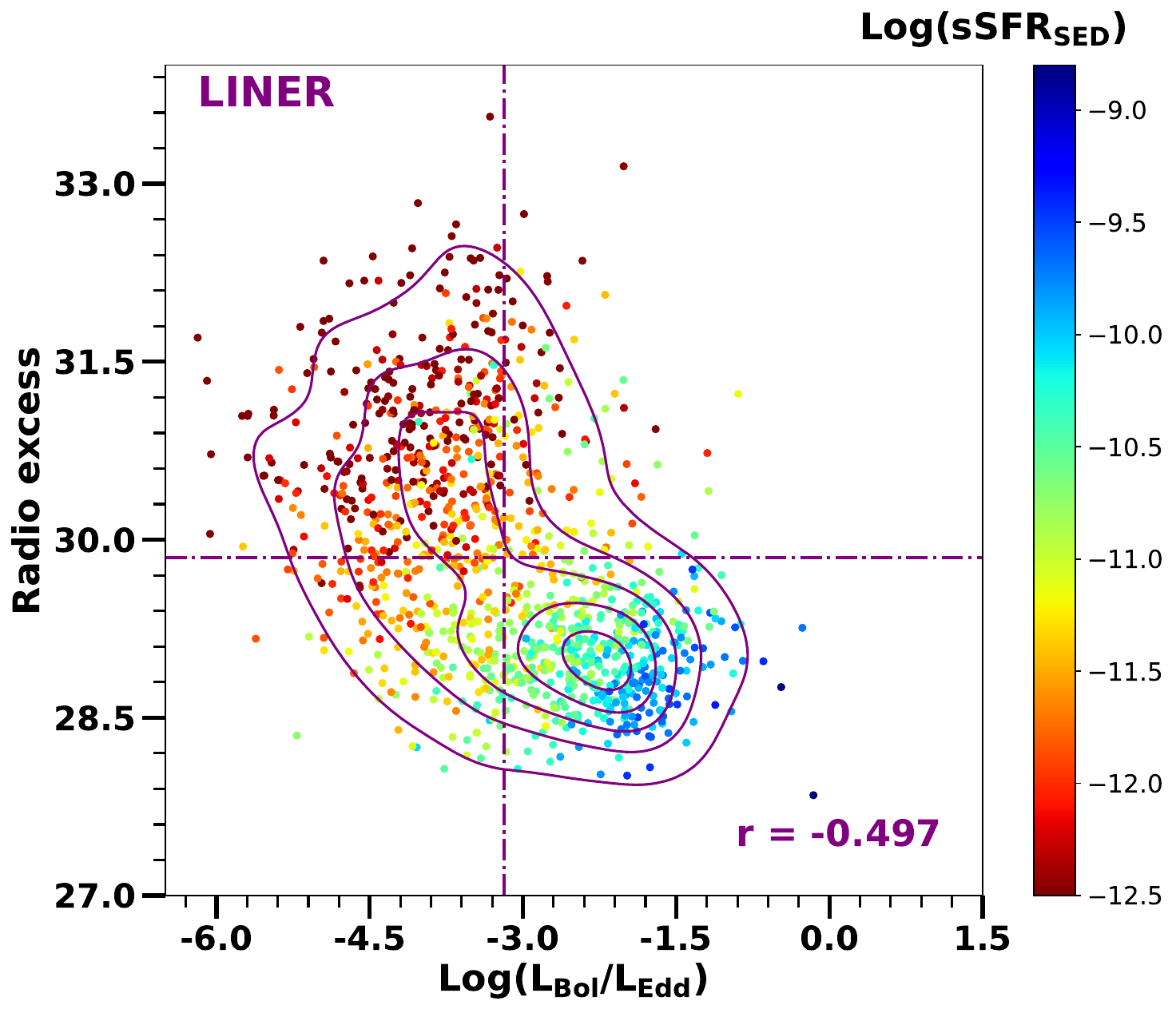}
	\caption{Radio luminosities $\rm L_{1.4GHz}$ as a function of Eddington ratios $\rm L_{Bol}/L_{Edd}$ (top and middle panels). Radio excess $\rm L_{1.4GHz}/SFR_{SED}$ as a function of Eddington ratios (bottom panel). The color scheme in each panel presents $\rm SFR_{SED}$ or $\rm sSFR_{SED}$. The Spearman correlation coefficient (r) is displayed in each panel. The $\rm L_{1.4GHz}-L_{Bol}/L_{Edd}$ space is divided into a grid of 150x150 bins, and contour lines are drawn to represent 10, 30, 50, 70, and 90 percent of the maximum number density. The vertical and horizontal dashed-dotted lines display the medium values of $\rm L_{Bol}/L_{Edd}$, $\rm L_{1.4GHz}$, and radio excess of each galaxy type, respectively.
	\label{fig:color_edd_radio}}
\end{figure*}

\section{Discussions}\label{section:discuss}

\subsection{Roles of Gas Supply and AGN Feedback in the Evolutionary Stage of Galaxies}

Our results are consistent with the previous work of \citet{Leslie+16}, who investigated the distribution of different galaxy types on the $\rm SFR-M_{*}$ plane using a large sample from the SDSS data. They proposed an evolutionary pathway for galaxies that progress from active star-forming galaxies to composite, Seyfert, and LINER galaxies, illustrating the transition from \emph{blue to red (active to quenching)} states. \citet{Leslie+16} emphasized the significant role of AGNs in quenching SF within this evolutionary pathway, a conclusion that aligns with earlier studies using optical and X-ray selected samples (e.g., \citealp[]{Salim+07, Schawinski+07, Shimizu+15}).



Motivated by previous works (e.g., \citealp{Leslie+16}), this study aims to examine in detail the evolutionary stages of SF, composite, Seyfert, and LINER galaxies. Using a large sample of $\sim$113,000 local sources with high S/N emission lines from the SDSS data, we analyze their physical properties, including UV-to-optical colors ($\rm u-r$), SFRs, $\rm L_{[OIII]}/\sigma_{*}^{4}$, and Eddington ratios in the $\rm SFR-M_{*}$ plane and the color-mass diagram. Our results show that SF galaxies exhibit blue colors and span a broad range of stellar masses. In contrast, composite, Seyfert, and LINER galaxies are located in the higher stellar mass range ($\rm \log M_{*} \gsim 10^{10}$) and exhibit progressively redder colors, with colors transitioning from composite to Seyfert and then to LINER galaxies. 

Additionally, SF activity follows similar trends, with high sSFRs comparable to blue SF galaxies, followed by a smooth decrease in composite, Seyfert, and LINER galaxies. The Eddington ratio correlates well with sSFRs: high Eddington ratio sources show strong SF activity and blue colors, while redder sources with low SFRs exhibit low Eddington ratio values. The Eddington ratio and $\rm L_{[OIII]}/\sigma_{*}^{4}$ peak in Seyfert sources, indicating strong AGN activity in Seyfert galaxies.

Our results may support the scenario of an evolutionary pathway where galaxies transform from blue SF galaxies in an active SF phase to transitional composite and Seyfert galaxies, ultimately culminating in red LINER galaxies in a quenching SF phase. In this sequence, the gas supply plays a primary role. We can assume that when the gas supply is abundant, both sSFRs and Eddington ratios are high, keeping galaxies in an active SF phase as blue SF galaxies. As the gas supply diminishes, both sSFRs and Eddington ratios decrease. Once the gas supply is nearly depleted, galaxies transition to a red state, quenching SF and becoming LINER galaxies. Additionally, AGN feedback may play a secondary role in this evolutionary sequence. We observe that the Eddington ratio and $\rm L_{[OIII]}/\sigma_{*}^{4}$ are higher in Seyfert galaxies compared to composite and LINER galaxies, suggesting that the Seyfert stage represents the most active phase of AGN activity. During this process, AGN feedback, in conjunction with the gas supply, may contribute to triggering SF activities and beginning the quenching of SFRs. Finally, in the LINER stage, both the gas supply and AGN activity weaken, resulting in lower Eddington ratios and SFRs.

Recent studies have found that SF, composite, Seyfert, and LINER galaxies exhibit trends of increasing stellar ages, higher black hole masses, and redder colors \citep[e.g.,][]{Schawinski+07, Vitale+13}, consistent with our results. An evolutionary sequence has been proposed, progressing from blue SF galaxies, through the transitional stages of Seyfert and composite galaxies, and culminating in red LINERs. For example, \citet{Schawinski+07} analyzed a large sample of 16,000 early-type galaxies ($z < 0.1$) from the SDSS and identified a sequence from SF to quiescent galaxies driven by nuclear activity. They estimated that this transition process lasts $\sim$1 Gyr, with peak AGN activity occurring $\sim$0.5 Gyr after the starburst. Alternatively, using $\sim$50,000 optically selected galaxies ($z \sim 0.1$), \citet{Salim+07} proposed an evolutionary pathway linking normal star-forming galaxies to quiescent galaxies, mediated by strong and weak AGNs. Moreover, using 119 radio galaxies at intermediate redshifts ($0.04 < z < 0.4$), \citet{Vitale+15} found evidence of spectral evolution across SF, composite, Seyfert, and LINER galaxies. They observed a flattening trend in the spectral index for SF, composite, and Seyfert galaxies on the BPT diagram. This flattening is thought to result from high-energy outputs driven by central nuclear activity or radio jet emission. The authors proposed that the transition from blue active SF galaxies to red passive elliptical galaxies is significantly influenced by AGN activity, which plays a key role in quenching star formation during this evolutionary sequence.

Additionally, our results are also consistent with the simulated findings of \citet{Chen+10}, who used evolutionary population synthesis (EPS) models to track the stellar evolution of optical spectra across different galaxy types, including SF, composite, Seyfert 2, LINER, and early-type galaxies from the SDSS. Their simulations revealed an age sequence among galaxy types, where the fraction of young stellar populations decreases significantly from SF and composite galaxies, through Seyfert and LINER galaxies, to early-type galaxies.

Recent works \citep[e.g.,][]{Schawinski+07, Vitale+13, Leslie+16} have proposed that AGN feedback, where the suppression of SF by nuclear activity plays a significant role, is a key factor in the evolutionary path of galaxies across SF, composite, Seyfert, and LINER stages. While our results align with these findings, we propose an alternative perspective on the evolutionary sequence, suggesting that gas supply may play a primary role, with AGN feedback as a secondary factor. In our observations, we find a strong correlation between AGN and SF activities, as indicated by a positive relationship between Eddington ratios and SFRs. This may indicate that the central engine and the host galaxy share a common gas reservoir. It is natural to expect that gas supply fuels both intense nuclear activity and enhanced star formation processes. We note that a strong positive correlation between SFRs, black hole accretion rates, and bolometric luminosity has also been reported by \citet{Zhuang+20}. Moreover, we propose that following the primary role of gas supply, AGN feedback may play a secondary role by interacting with the gas supply process to both trigger SF activity and initiate the quenching of SFRs during the peak of AGN activity in Seyfert galaxies.

It is important to note that AGN feedback is a complex process, and the relationship between AGN activity and SFRs is not simply characterized by a positive or negative correlation. This complexity arises, in part, because we are comparing AGN activity on small scales (e.g., Eddington ratios) with star formation activity on larger scales in the host galaxy. Careful considerations are needed to ensure a fair comparison between these two physical quantities. For example, using a large sample from the SDSS, \citet{Woo+20} proposed that AGN feedback operates on a delayed timescale, requiring a dynamical time for AGN activity to significantly influence the star-forming regions in the host galaxy. Furthermore, both positive and negative feedback effects have been observed in NGC 5728, as reported by \citet{Shin+19}. The authors found enhanced SFRs in the encounter region, likely caused by compression of the ISM driven by AGN activity. In contrast, the outer regions of the spiral arms showed no signs of star formation activity. Morever, using model simulations, \citet{Yuan+18} found that AGN feedback can have either positive or negative effects on star formation, depending on the location within the galaxy. In the inner regions of the galaxy ($\rm r \lsim 15\ kpc$), star formation is suppressed (negative feedback), whereas in the outer regions ($\rm r \gsim 15\ kpc$), star formation is enhanced (positive feedback). Furthermore, the authors showed that, on a galaxy-scale, star formation can be alternately suppressed or enhanced over time. They also noted that whether AGN feedback is positive or negative depends not only on the location but also on the timescale.

In summary, based on our results, we tentatively propose an evolutionary pathway for galaxies transitioning through the SF, composite, Seyfert, and LINER stages. Gas supply may play a primary role, while the complex AGN feedback mechanisms may serve as a secondary factor in this evolutionary process. However, our findings are limited by the sample, which is selected based on optical emission lines and primarily consists of galaxies at local redshifts. A larger, more robust sample that includes galaxies at higher redshifts is crucial to further test our results. Also, it is important to note that our sample was selected based on the BPT diagram. This method poses the risk of misclassifying weak AGNs with low Eddington ratios as SF galaxies with high SFRs, as their emission lines may be strongly influenced by star formation processes. To address this potential limitation in our sample selection, a similar analysis incorporating X-ray data is necessary. Furthermore, the emission lines observed in composite galaxies may originate not only from SF or AGN activity but also from shock excitation. Similarly, the emission lines in LINER galaxies could be driven by shocks or low-luminosity AGNs (e.g., \citealp{Ho+93}, \citealp{Ho08}). Additionally, our findings align with the results of \citet{Xue+10}, who used deep X-ray observations from the Chandra Deep Field-North (CDF-N) and Chandra Deep Field-South (CDF-S) surveys (see \citealp{Xue17} for a review on the Chandra Deep Fields). Their study revealed that most AGNs tend to reside in massive host galaxies and exhibit redder colors as stellar mass increases.

\subsection{Connection between Radio and AGN Activities in Relation to SFRs?}

The relationship between radio activity and nuclear activity has garnered significant attention in recent years. Utilizing 4,877 sources, including 2,107 SF, 111 composite, 1,496 Seyfert, and 1163 LINER galaxies cross-matched with the FIRST catalog, we aim to explore the correlation between radio activity, represented by $\rm L_{1.4GHz}$, and AGN activity, indicated by the Eddington ratio $\rm L_{Bol}/L_{Edd}$, particularly in their connection to SF activity. In Figure \ref{fig:color_contour_radio}, we presented the color scales of radio luminosity $\rm L_{1.4GHz}$, the radio excess $\rm L_{1.4GHz}/SFR_{SED}$, and the Eddington ratio $\rm L_{Bol}/L_{Edd}$ on the $\rm SFR-M_{*}$ plane.

Notably, the high and low Eddington ratios are directly linked to the high and low SFRs on the $\rm SFR-M_{*}$ plane. This variation is almost vertical to the MS line, where high Eddington ratio sources align with the MS line, characterized by blue and high SFR sources. This trend exhibits a smooth decrease towards lower Eddington ratios associated with red and low SFR sources. Additionally, we found that the radio excess increases perpendicular to the MS line. Sources on the MS line exhibit low radio excess, while sources below the MS line show higher radio excess that may indicate contributions from AGN and/or jet emission that suppresses SFRs in these sources.

Furthermore, the trends of radio luminosity on the $\rm SFR-M_{*}$ plane show a distinct directional variation compared to Eddington ratios. Specifically, $\rm L_{1.4GHz}$ tends to vary along the MS line, increasing with stellar mass. Our results align with previous studies on SF galaxies, showing that radio luminosity is strongly correlated with stellar mass and SFR \citep[e.g.,][]{Smith+21, Delvecchio+21}.

Our findings may indicate that radio luminosity and Eddington ratios play somewhat different roles in the $\rm SFR-M_{*}$ plane. Eddington ratios are strongly associated with the blue and red sources on the $\rm SFR-M_{*}$ plane, suggesting a direct connection between Eddington ratios and the active or quenching phases of SF activity. In contrast, radio luminosity tends to follow the MS line, increasing as both SFR and stellar mass increase. 

Figure \ref{fig:color_edd_radio} presents radio luminosities as a function of Eddington ratios. In each galaxy type, to quantify the correlation between radio luminosities and Eddington ratios, we calculate the Spearman correlation coefficient (r). Radio luminosities and Eddington ratios in composite galaxies show a clear correlation (r $=$ 0.407). In comparison, there is no correlation between the two parameters in Seyfert (r $=$ 0.210) and LINER (r $=$ 0.158) galaxies. We also present the $\rm SFR_{SED}$ and $\rm sSFR_{SED}$ as color schemes to visualize their variation in the relations between radio luminosities and Eddington ratios. The low (red) and high (blue) sSFR sources are directly linked to the low and high Eddington ratio sources, though such trends are not seen in comparison to radio luminosities. The radio excess exhibits a negative correlation with Eddington ratios, particularly in LINER galaxies. This negative correlation may result from the inclusion of negative SFR values in the calculation of radio excess or could indicate the influence of jet emission in quenching star formation activity.

Numerous studies have explored the relationship between radio luminosity and AGN activity, such as gas kinematic outflows \citep[e.g.,][]{Woo+16, Le+17b, Rakshit+18, Ayubinia+23}. \citet{Woo+16} and \citet{Rakshit+18} found no correlation between the gas outflow kinematics, as indicated by the dispersion velocities of \OIII\ lines, and radio luminosity ($\rm L_{1.4GHz}$) for type 1 and type 2 AGNs, using large local samples from the SDSS. Additionally, \citet{Ayubinia+23} identified a significant difference in \OIII\ gas outflow kinematics for low and high $\rm L_{1.4GHz}/L_{Edd}$. Our results align with these findings, indicating no direct connection between radio luminosity and Eddington ratios. Also, Eddington ratios play a significant role, strongly linking to the active (blue) or quenching (red) phases of SF activities in AGNs.

Additionally, the radio sources in our sample are incomplete because we only cross-matched the sample with the FIRST catalog. Operating at lower frequencies ($<200$ MHz) than the FIRST survey, the International LOFAR telescope enables the detection of lower-energy synchrotron emission and observes fainter SF galaxies, lower-luminosity AGNs, and more extended radio sources, e.g., the LOFAR Two-Metre Sky Survey (LoTSS-DR2, e.g., \citealp{Drake+24}). Incorporating a larger sample, such as those detected by LoTSS-DR2, will significantly enhance the completeness of our sample, particularly for faint radio source populations.

Moreover, some contexts are still missing in the analysis of the radio sources in our sample. For example, \citet{Smolvic09} analyzed a sample of $\sim$500 radio AGNs ($z < 0.1$) with available optical spectroscopy to discuss different phases of weak and powerful radio AGNs in the \textit{blue to red} galaxy evolution, by distinguishing between low-excitation ($\rm L_{1.4GHz} < 10^{25}\ W\ Hz^{-1}$) and high-excitation ($\rm L_{1.4GHz} > 10^{25}\ W\ Hz^{-1}$) in their sample. Similarly, using a local sample of 2215 radio-loud AGNs ($z < 0.3$), \citet{Best+05} investigated the different host properties of radio and optical AGNs by separating \OIII\ luminosity at $\rm L_{[OIII]} = 10^{7}\ L_{\odot}$. These findings provide valuable insights that could enhance our analysis. However, within the scope of our statistical study focusing on the comparison of radio luminosity and Eddington ratio using our large sample, we aim to incorporate the results of \citet{Smolvic09} and \citet{Best+05} in future work, potentially by combining our data with larger samples such as the LOFAR.

\section{Summary}\label{section:sum}

\begin{figure*}
\centering
	\includegraphics[width=0.7\textwidth]{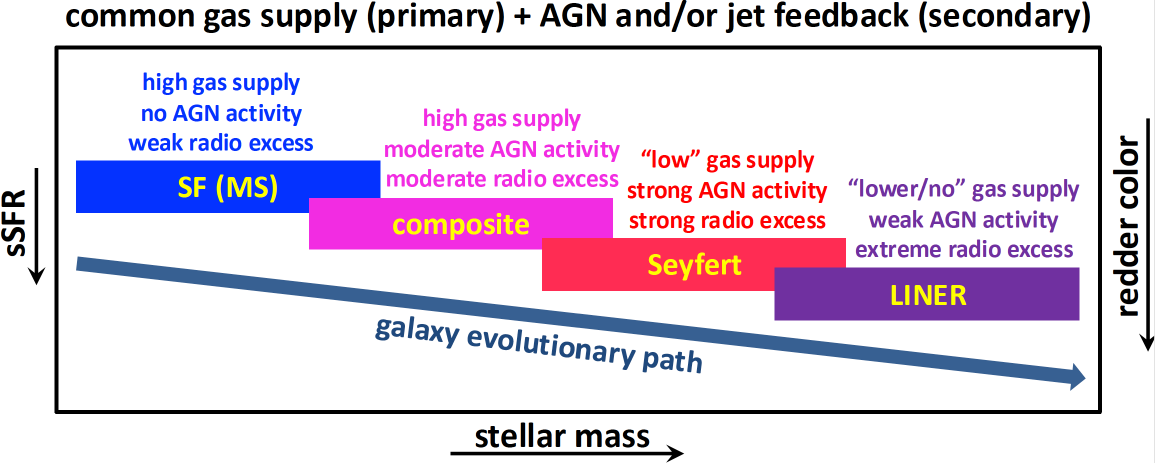}
	\caption{Cartoon illustrating the evolutionary path of galaxies, where they transition from blue SF galaxies to red LINERs, passing through composite and Seyfert phases. 
	\label{fig:cartoon}}
\end{figure*}

Using a large sample of $\sim$113,000 galaxies ($\rm z$ $<$ 0.3) characterized by high S/N emission lines in both UV and optical bands, we divided the sample into four types of galaxies: SF, composite, Seyfert, and LINER galaxies. Using these galaxies, we study the relationships between UV-to-optical colors ($\rm u-r$), SFRs, and Eddington ratios among them. Additionally, by cross-matching with the FIRST catalog, we obtained 4,877 radio-detected sources in which we study the connection between radio activity and Eddington ratio on the $\rm SFR-M_{*}$ plane. Figure \ref{fig:cartoon} presents a visual summary of our key findings. We summarize our main results as follows:

\smallskip
(1) SF galaxies predominantly feature young, blue stars on the MS line, while composite, Seyfert, and LINER galaxies deviate from the MS line, showing decreasing SFRs and older stellar populations. UV-to-optical colors shift from blue in SF galaxies to progressively redder hues in composite, Seyfert, and LINER galaxies, with LINERs being the reddest. 

\smallskip
(2) The sSFRs exhibit a strong correlation and smooth transition in the $\rm (u-r)-L_{Bol}/L_{Edd}$ plane. Sources with low Eddington ratios tend to have low sSFRs, while those with high Eddington ratios exhibit high sSFRs. Eddington ratios are highest in Seyfert galaxies, moderate in composite galaxies, and lowest in LINER galaxies, with higher values corresponding to bluer ($\rm u-r$) colors, indicating younger stellar populations. These trends mirror those of sSFRs, where blue ($\rm u-r$) colors and high sSFRs correlate with high Eddington ratios, suggesting a strong link between Eddington ratios and SF activity. 

\smallskip
(3) $\rm L_{[OIII]}/\sigma_{*}^{4}$ shows high values for sources lying on the MS line. In contrast, sources below the MS line show lower values.

\smallskip
(4) The radio excess increases perpendicular to the MS line, with low values on the line and higher values below, potentially indicating AGN and/or jet contributions suppressing SFRs.

\smallskip
(5) Gas supply may play a crucial role in the strong correlations between SF and AGN strength activities. When the gas supply is high, we observe a strong Eddington ratio and high star-formation activity. Conversely, SF and AGN strength activities decrease when the gas supply decreases. 

\smallskip
(6) There may be an evolutionary pathway in which galaxies progress from blue SF galaxies with active SF activities to composite, Seyfert, and ultimately to red LINER galaxies with quenched SF activities. Gas supply may play a primary key role in this evolutionary stage, while AGN feedback may play a secondary impact. 

\smallskip
(7) Both radio luminosity and Eddington ratios correlate with SFRs, but their trends on the $\rm SFR-M_{*}$ plane differ. The Eddington ratio is highest along the MS line and decreases as SFRs drop vertically from the MS line. In contrast, radio luminosity increases along the MS line and with rising stellar mass, with lower luminosity in lower-mass sources and higher luminosity in higher-mass sources. There may be no direct connection between radio luminosity and Eddington ratios. Eddington ratios play a significant role in the active (blue) or quenching (red) phases of SF activities in AGNs. \\

We thank the referee for valuable suggestions and comments that significantly improved the presentation and clarity of this paper. This work has been supported by National Key R\&D Program of China No. 2022YFF0503401, the National Natural Science Foundation of China (NSFC-12473014, NSFC-11890693), and the science research grants from the China Manned Space Project with NO. CMS-CSST-2021-A06. We thank Prof. Jong-Hak Woo and Dr. Hyun-Jin Bae for generously sharing their catalog. We thank Dr. Wang Yijun and Dr. Xiaozhi Lin for their thoughtful discussion.





\end{CJK*}

\end{document}